\documentclass[notitlepage,a4paper,aps,prd,onecolumn,superscriptaddress,nofootinbib,groupedaddress]{revtex4}
\usepackage{amsmath}
\usepackage{amsfonts,color}
\usepackage{amsmath, mathtools, bm}
\usepackage{accents}
\usepackage{amsthm}
\usepackage{relsize}

\usepackage{amsfonts,color}
\usepackage{amssymb,float}
\usepackage{ytableau}
\ytableausetup
{mathmode, aligntableaux = center}
\newcommand{\nn}{\nonumber \\}

\newcommand{\order}[2]{\accentset{#2}{#1}}

\usepackage{amsmath}

\newcommand{\dd}{{\rm d}}
\newcommand{\DD}{{\rm D}}
\usepackage[utf8]{inputenc}
\allowdisplaybreaks
\usepackage{color}
\usepackage{hyperref}
\hypersetup{
    colorlinks=true,
    linkcolor=blue,
    filecolor=magenta,      
   citecolor=blue
}

\usepackage{enumitem}
\usepackage{comment}

\usepackage{tikz}
\usetikzlibrary{shapes.geometric}
\usetikzlibrary{arrows.meta,arrows}

\begin{document}
\begin{flushright} {\footnotesize YITP-23-135}  
\end{flushright}
\title{Cosmological Perturbation Theory in Metric-Affine Gravity}

\author{Katsuki Aoki}
\email{katsuki.aoki@yukawa.kyoto-u.ac.jp}
\affiliation{Center for Gravitational Physics and Quantum Information,
Yukawa Institute for Theoretical Physics, Kyoto University, Kyoto 606-8502, Japan}

\author{Sebastian Bahamonde}
\email{sbahamondebeltran@gmail.com, sebastian.bahamonde@ipmu.jp}
\affiliation{Department of Physics, Tokyo Institute of Technology
1-12-1 Ookayama, Meguro-ku, Tokyo 152-8551, Japan.}
\affiliation{Kavli Institute for the Physics and Mathematics of the Universe (WPI), The University of Tokyo Institutes
for Advanced Study (UTIAS), The University of Tokyo, Kashiwa, Chiba 277-8583, Japan.}
\author{Jorge Gigante Valcarcel}
\email{gigante.j.aa@m.titech.ac.jp}
\affiliation{Department of Physics, Tokyo Institute of Technology
1-12-1 Ookayama, Meguro-ku, Tokyo 152-8551, Japan.}

\author{Mohammad Ali Gorji}
\email{gorji@icc.ub.edu}
\affiliation{Departament de Física Quàntica i Astrofísica, Institut de Ciències del Cosmos,
Universitat de Barcelona, Martí i Franquès 1, 08028 Barcelona, Spain}

\begin{abstract}
We formulate cosmological perturbation theory around the spatially curved FLRW background in the context of metric-affine gauge theory of gravity which includes torsion and nonmetricity. Performing scalar-vector-tensor decomposition of the spatial perturbations, we find that the theory displays a rich perturbation spectrum with helicities 0, 1, 2 and 3, on top of the usual scalar, vector and tensor metric perturbations arising from Riemannian geometry. Accordingly, the theory provides a diverse phenomenology, e.g. the helicity-2 modes of the torsion and/or nonmetricity tensors source helicity-2 metric tensor perturbation at the linear level leading to the production of gravitational waves. As an immediate application, we study linear perturbation of the nonmetricity helicity-3 modes for a general parity-preserving action of metric-affine gravity which includes quadratic terms in curvature, torsion, and nonmetricity. We then find the conditions to avoid possible instabilities in the helicity-3 modes of the spin-3 field.

\end{abstract}

\maketitle
\tableofcontents
\section{Introduction}

Over the past few decades, numerous cosmological observations have provided compelling evidence that the universe is undergoing accelerated expansion. Additionally, data from sources like the Cosmic Microwave Background (CMB) and galactic rotation curves indicate the presence of no ordinary matter, which does not interact with light, may exist at cosmological scales. These observations have led to the introduction of two mysterious components in cosmology, known as dark energy and dark matter~\cite{SupernovaSearchTeam:1998fmf,SupernovaCosmologyProject:1998vns,Peebles:2002gy,Gaitskell:2004gd}. The simplest model capable of explaining these phenomena relies on General Relativity (GR) and a cosmological constant, which together constitute the $\Lambda$CDM model that incorporates both dark constituents.

Nonetheless, recent cosmological observations have raised concerns about the viability of the standard cosmological model, primarily due to various tensions~\cite{DiValentino:2021izs,DiValentino:2020zio,DiValentino:2020vvd,Schoneberg:2021qvd}. For instance, there is a growing tension, currently at a significance level of approximately $4.4\sigma$, in the measurement of the Hubble constant $H_0$ as determined by model-dependent observations on early-time cosmology  (such as Planck)~\cite{Planck:2018vyg,ACT:2023kun,Schoneberg:2022ggi} and direct measurements in the late universe (for example, using ladder measurements)~\cite{Riess:2019cxk,Wong:2019kwg,Anderson:2023aga}. Additionally, there are other, albeit less severe, tensions, such as the $\sigma_8$ tension; a parameter related to the clustering of matter, which also implies discrepancies between local and early-time observations~\cite{Zarrouk:2018vwy,BOSS:2016wmc}. The scientific community is engaged in a lively debate regarding whether these tensions stem from new physics or systematic errors~\cite{Riess:2019qba,Pesce:2020xfe,Perivolaropoulos:2021jda,deJaeger:2020zpb}. One potential approach to addressing or mitigating these tensions is to modify the $\Lambda$CDM model, and one avenue for doing so is indeed to modify GR. Furthermore, there are other theoretical motivations for modifying GR, including the cosmological constant problem, the issue of singularities, and the quest for a consistent framework of quantum gravity~\cite{Weinberg:1972kfs,Weinberg:1988cp,Hawking:1973uf}.

There exist various strategies to formulate alternative theories of gravity beyond GR (for comprehensive reviews, see~\cite{Nojiri:2017ncd,Clifton:2011jh,CANTATA:2021ktz,Koyama:2015vza,Bull:2015stt,Heisenberg:2018vsk, Arai:2022ilw,Odintsov:2023weg}). In this work, our focus is on a geometrical extension of GR, which introduces the torsion and nonmetricity tensors as post-Riemannian degrees of freedom (d.o.f.) into the geometrical structure of the space-time. From a theoretical point of view, the resulting geometry can be related to the existence of a new fundamental symmetry by applying the gauge principles, which leads to the appearance of the Metric-Affine Gauge (MAG) theory of gravity ~\cite{Hehl:1994ue}. In fact, within MAG there are several subclasses of theories, depending on the construction of the geometry and the choice of the gravitational action. For instance, one of the simplest versions is the Einstein-Cartan theory, where both curvature and torsion are present, the latter being nondynamical and tied to spinning sources~\cite{Kibble:1961ba,Sciama:1964wt,Hehl:1976kj}. Another subset is teleparallel gravity, in which the general curvature is absent, and gravity is solely represented by torsion and/or nonmetricity~\cite{Aldrovandi:2013wha,Bahamonde:2021gfp,BeltranJimenez:2019odq}. 

Numerous studies have been conducted on Friedmann-Lema\^{i}tre-Robertson-Walker (FLRW) cosmology at the background level within various MAG theories. Pioneering works can be traced back to~\cite{Kopczynski:1972fhu, Kopczynski:1973qim,Trautman:1973wy} within the Einstein-Cartan framework, where it was discovered that the cosmological singularity can be resolved by introducing a nonzero spin density of matter. Following these lines, studies involving couplings between Dirac fields and torsion found that the cosmological singularity can be indeed replaced by a cosmic bounce~\cite{Brechet:2008zz,Poplawski:2011jz,Magueijo:2012ug,Farnsworth:2017wzr,Bombacigno:2018tyw}. In addition, further investigations were also carried out within the general Poincar\'e and Weyl gauge theories of gravity, where the torsion tensor and the Weyl vector of the nonmetricity tensor constitute dynamical fields \cite{minkevich1980generalised,Blagojevic:1981wa,Gonner:1984rw,Obukhov:2022khx,Puetzfeld:2001hk,Puetzfeld:2001ur}.

There are also several studies in MAG concerning late-time cosmology. It has been found that particular models with dynamical torsion can provide the same description as $\Lambda$CDM, including an accelerating universe, with the torsion field accounting for the effects of the dark sector \cite{Yo:2006qs,Shie:2008ms}. These cosmological effects were originally driven by the scalar mode $0^{+}$ of torsion, but a further generalisation in the presence of the pseudoscalar mode $0^{-}$ was shortly achieved, finding that indeed the two modes decouple, which provides a suitable fitting of the supernova data \cite{Chen:2009at}. A natural extension with odd parity terms in the gravitational action found that the resulting parity-violating interactions might help to explain the imbalance between matter and antimatter at cosmological scales~\cite{Baekler:2010fr}.

Regarding inflationary models, several works have also shown interesting features in MAG. Following the original investigations performed on the cosmology of Einstein-Cartan theory, different inflationary and singularity-free solutions in the presence of torsion were soon found~\cite{Gasperini:1986mv,Demianski:1986zn}, while other natural extensions were also carried out by the introduction of scalar fields nonminimally coupled to the geometry of the space-time~\cite{DeRitis:1988kb}, including generalisations of Higgs-inflation with Holst and Nieh-Yan terms in the presence of torsion and nonmetricity~\cite{Langvik:2020nrs,Shaposhnikov:2020gts,Rigouzzo:2022yan,Gialamas:2022xtt}. As for the theories that endow these fields with dynamics by the presence of quadratic curvature invariants in the gravitational action, in~\cite{Aoki:2020zqm} it was found a model which can consistently describe the inflationary universe by a dynamical torsion field. In fact, linear tensor perturbations of torsion around a FLRW background were formerly performed and analysed in this model, revealing the general occurrence of a spontaneous parity-violation around such a background, which does not take place in the standard Starobinsky inflation. 

To sum up, the literature of cosmology in MAG has primarily focused on the role of torsion, exhibiting a large variety of interesting features in early and late cosmology, while nonmetricity has received relatively little attention. In particular, the physical implications at cosmological scales of the traceless part of the nonmetricity tensor have not been extensively investigated. Likewise, as stated in GR and other alternative theories of gravity, a thorough study on the cosmological perturbations arising from the theory turns out to be essential to carry out a more accurate assessment on its cosmological implications. Thereby, the main objective of this work is to develop a comprehensive theory of the cosmological perturbations in the generic MAG with curvature, torsion and nonmetricity. For this task, one must consider cosmological perturbations not only in the metric sector, but also in the torsion and nonmetricity sectors. In this sense, the physical role of the perturbations differs significantly based on whether they refer to dynamical fields, or on whether the geometry is particularly special. In the first case, for a nondynamical torsion tensor such as the one present in Einstein-Cartan theory, the cosmological perturbations encode all the dynamics purely in the metric sector. Furthermore, in the particular case of teleparallel gravity all the geometrical d.o.f. can be conveniently encoded in the tetrad field \cite{Hohmann:2020vcv}, while this is no longer the case in general metric-affine geometries.

Additionally, in order to conduct an exhaustive analysis in cosmology, it is also essential to incorporate the matter sector on top of the geometrical setup, which in the case of MAG can be described in terms of two fundamental quantities: the energy-momentum and hypermomentum tensors. Indeed, the latter plays a crucial role in the theory since it represents the source for both torsion and nonmetricity tensors. Therefore, the cosmological perturbations must include the corresponding d.o.f. in the geometry sector, associated with the metric, torsion and nonmetricity tensors, as well as d.o.f. encoded in the energy-momentum and hypermomentum tensors in the matter sector.

This paper is organised as follows. In Sec.~\ref{sec:metricaffine} we provide a brief introduction to MAG and define the most relevant mathematical quantities appearing in this framework. In Sec.~\ref{decomposition} we perform $3+1$ decompositions of arbitrary tensors up to rank-3, which can be systematically ascribed to the geometrical and material tensors of MAG in order to classify the corresponding perturbation modes by their spin and parity. Sec.~\ref{sec:background} is devoted to introducing the cosmological background, described by the FLRW metric along with the expressions for the torsion, nonmetricity, energy-momentum and hypermomentum tensors satisfying homogeneous and isotropic conditions. In Sec.~\ref{sec:perturbations} we provide Scalar-Vector-Torsion (SVT) decomposition of all spatial perturbations in the geometrical and matter sectors of MAG. As a pedagogical and relatively simple application, in Sec.~\ref{sec:spin3} we study linear perturbations of the helicity-3 modes of the traceless nonmetricity tensor for a generic quadratic action of MAG and we find conditions to avoid pathological instabilities. Finally, in Sec.~\ref{sec:conclusions} we conclude and discuss our main results. Some technical details are presented in the appendices.

We work in natural units $c=G=1$, and we consider the metric signature $(-,+,+,+)$. Tildes will be used to denote mathematical quantities that are defined from the general affine connection, whereas their unaccented counterparts will be only constructed from the Levi-Civita connection. Greek letters denote four-dimensional indices $\mu,\nu,\cdots=0,1,2,3$ while Latin letters denote the three-dimensional spatial indices $i,j,\cdots=1,2,3$.

\section{Metric-Affine Gravity with torsion and nonmetricity }\label{sec:metricaffine}

\subsection{Geometrical sector}
From a geometrical point of view, the standard framework of GR arises as a particular case of a more general class of metric-affine theories, where the geometry of the space-time is described by a metric tensor, a coframe field and an independent linear connection~\cite{Hehl:1994ue}. We assume that the theory enjoys the $GL(4,R)$ invariance by which we can fix the coframe field. Accordingly, the theory is described by the metric tensor and the connection, the latter including additional post-Riemannian d.o.f., which correspond to the torsion and nonmetricity tensors
\begin{equation}
    T^{\lambda}\,_{\mu \nu}=2\tilde{\Gamma}^{\lambda}\,_{[\mu \nu]}\,, \quad Q_{\lambda \mu \nu}=\tilde{\nabla}_{\lambda}g_{\mu \nu}\,.
\end{equation}
The components of the affine connection can then be split into a Levi-Civita part and two independent pieces
\begin{equation}
\tilde{\Gamma}^{\lambda}\,_{\mu \nu}=\Gamma^{\lambda}\,_{\mu \nu}+K^{\lambda}\,_{\mu \nu}+L^{\lambda}\,_{\mu \nu}\,,
\end{equation}
where $K^{\lambda}\,_{\mu \nu}$ is a metric-compatible contortion tensor containing torsion and $L^{\lambda}\,_{\mu \nu}$ a disformation tensor depending on nonmetricity:
\begin{align}
    K^{\lambda}\,_{\mu \nu}&=\frac{1}{2}\left(T^{\lambda}\,_{\mu \nu}-T_{\mu}\,^{\lambda}\,_{\nu}-T_{\nu}\,^{\lambda}\,_{\mu}\right)\,,\\
    L^{\lambda}\,_{\mu \nu}&=\frac{1}{2}\left(Q^{\lambda}\,_{\mu \nu}-Q_{\mu}\,^{\lambda}\,_{\nu}-Q_{\nu}\,^{\lambda}\,_{\mu}\right)\,.
\end{align}
One can further define the quantity
\begin{eqnarray}
    N^{\lambda}\,_{\mu \nu}=K^{\lambda}\,_{\mu \nu}+L^{\lambda}\,_{\mu \nu}\,,\label{distortion}
\end{eqnarray} which measures the deviation from Riemannian geometry as the distortion tensor. 
The resulting geometric structure is then characterised by a metric tensor and an asymmetric affine connection that in general does not preserve the lengths and angles of vectors under parallel transport. Hence, such a connection provides corrections in the covariant derivative operator, which further involves a generalisation of its commutation relations when these are applied on an arbitrary vector $v^{\lambda}$:
\begin{equation}
[\tilde{\nabla}_{\mu},\tilde{\nabla}_{\nu}]\,v^{\lambda}=\tilde{R}^{\lambda}\,_{\rho \mu \nu}\,v^{\rho}+T^{\rho}\,_{\mu \nu}\,\tilde{\nabla}_{\rho}v^{\lambda}\,,
\end{equation}
where the corresponding curvature tensor reads
\begin{equation}\label{totalcurvature}
\tilde{R}^{\lambda}\,_{\rho \mu \nu}=\partial_{\mu}\tilde{\Gamma}^{\lambda}\,_{\rho \nu}-\partial_{\nu}\tilde{\Gamma}^{\lambda}\,_{\rho \mu}+\tilde{\Gamma}^{\lambda}\,_{\sigma \mu}\tilde{\Gamma}^{\sigma}\,_{\rho \nu}-\tilde{\Gamma}^{\lambda}\,_{\sigma \nu}\tilde{\Gamma}^{\sigma}\,_{\rho \mu}\,.
\end{equation}
Note that the covariant derivatives do not commute in the presence of torsion even in Minkowski metric. From the expression of the curvature tensor, it is possible to define three independent traces, namely the Ricci and co-Ricci tensors:
\begin{eqnarray}\label{Riccitensor}
\tilde{R}_{\mu\nu}&=&\tilde{R}^{\lambda}\,_{\mu \lambda \nu}\,,\\
\label{co-Riccitensor}
\hat{R}_{\mu\nu}&=&\tilde{R}_{\mu}\,^{\lambda}\,_{\nu\lambda}\,,
\end{eqnarray}
and the homothetic curvature tensor, which depends on the trace part of nonmetricity:
\begin{equation}
    \tilde{R}^{\lambda}\,_{\lambda\mu\nu}=\nabla_{[\nu}Q_{\mu]\lambda}{}^{\lambda}\,.
\end{equation}
Furthermore, the trace of the Ricci and co-Ricci tensors provides a unique independent scalar curvature
\begin{equation}
    \tilde{R}=g^{\mu\nu}\tilde{R}_{\mu\nu}\,,
\end{equation}
whereas the contraction of the curvature tensor with the Levi-Civita tensor $\varepsilon_{\lambda\rho\mu\nu}=\epsilon_{\lambda\rho\mu\nu}\sqrt{-g}$ gives rise to the so-called Holst pseudoscalar
\begin{equation}
\ast\tilde{R}=\varepsilon^{\lambda\rho\mu\nu}\tilde{R}_{\lambda\rho\mu\nu}\,,
\end{equation}
where $\epsilon_{\lambda\rho\mu\nu}$ is the antisymmetric symbol with $\epsilon_{0123}=+\,1$.

Concerning the torsion and nonmetricity tensors, in four dimensions they carry twenty four and forty components, respectively, which represents a large number of independent geometrical d.o.f. included in the affine connection. In this sense, an irreducible decomposition under the four-dimensional pseudo-orthogonal group separates the different vector, pseudovector and tensor parts of these tensors.

Specifically, the torsion tensor can be decomposed in terms of a vector, a pseudovector and a traceless and pseudotraceless tensor
\begin{equation}\label{irreducibletorsion}
T^{\lambda}\,_{\mu \nu}=\frac{1}{3}\left(\delta^{\lambda}\,_{\nu}T_{\mu}-\delta^{\lambda}\,_{\mu}T_{\nu}\right)+\frac{1}{6}\,\varepsilon^{\lambda}\,_{\rho\mu\nu}S^{\rho}+t^{\lambda}\,_{\mu \nu}\,,
\end{equation}
where
\begin{align}\label{Tdec1}
T_{\mu}&=T^{\nu}\,_{\mu\nu}\,,\\
S_{\mu}&=\varepsilon_{\mu\lambda\rho\nu}T^{\lambda\rho\nu}\,,\\
t_{\lambda\mu\nu}&=T_{\lambda\mu\nu}-\frac{2}{3}g_{\lambda[\nu}T_{\mu]}-\frac{1}{6}\,\varepsilon_{\lambda\rho\mu\nu}S^{\rho}\,.\label{Tdec3}
\end{align}
On the other hand, the nonmetricity tensor can also be decomposed into a Weyl vector and a traceless tensor
\begin{equation}
Q_{\lambda\mu\nu}=g_{\mu\nu}W_{\lambda}+{\nearrow\!\!\!\!\!\!\!Q}_{\lambda\mu\nu}\,,\label{Qdecomposition}
\end{equation}
where the Weyl vector is related to the trace of the nonmetricity tensor
\begin{equation}
W_{\mu}=\frac{1}{4}\,Q_{\mu\nu}\,^{\nu}\,,
\end{equation}
and the traceless tensor is in turn decomposed into a vector, and two traceless and pseudotraceless tensors
\begin{equation}
\label{irreducibletracelessnonmetricity}
    {\nearrow\!\!\!\!\!\!\!Q}_{\lambda\mu\nu}=g_{\lambda(\mu}\Lambda_{\nu)}-\frac{1}{4}g_{\mu\nu}\Lambda_{\lambda}+\frac{1}{3}\varepsilon_{\lambda\rho\sigma(\mu}\Omega_{\nu)}\,^{\rho\sigma}+q_{\lambda\mu\nu}\,,
\end{equation}
with
\begin{align}
     \Lambda_{\mu}&=\frac{4}{9}\left(Q^{\nu}\,_{\mu\nu}-W_{\mu}\right)\,,\\
    \Omega_{\lambda}\,^{\mu\nu}&=-\,\left[\varepsilon^{\mu\nu\rho\sigma}Q_{\rho\sigma\lambda}+\varepsilon^{\mu\nu\rho}\,_{\lambda}\left(\frac{3}{4}\Lambda_{\rho}-W_{\rho}\right)\right]\,,\\
    q_{\lambda\mu\nu}&=Q_{(\lambda\mu\nu)}-g_{(\mu\nu}W_{\lambda)}-\frac{3}{4}g_{(\mu\nu}\Lambda_{\lambda)}\,.\label{qtensor}
\end{align}
As can be seen, the first tensor piece $\Omega_{\lambda}\,^{\mu\nu}$ is antisymmetric in the last pair of indices as the piece $t_{\lambda\mu\nu}$ of torsion, whereas the second tensor piece $q_{\lambda\mu\nu}$ constitutes a fully symmetric tensor.

\subsection{Matter sector}
As for the matter sources in the framework of MAG, not only an energy-momentum tensor of matter $\Theta_{\mu\nu}$ arises as source of curvature,
but also a hypermomentum density tensor $ \Delta_{\mu\nu\lambda}$, which operates as
source of torsion and nonmetricity~\cite{Hehl:1994ue}. Specifically, $\Theta_{\mu\nu}$ is defined from the variation of the matter action with respect to the metric, whereas $\Delta_{\mu\nu\lambda}$ arises from the corresponding variation with respect to the independent connection.

Following the same line as the previous subsection, the hypermomentum density tensor can then be decomposed into spin, dilation and shear currents as
\begin{eqnarray}\label{hypermomentum}
 \Delta_{\mu\nu\lambda}={}^{(\rm s)}\Delta_{[\mu\nu]\lambda}+\frac{1}{4}g_{\mu\nu}{}^{(\rm d)}\Delta_{\lambda}+{}^{(\rm sh)}{\nearrow\!\!\!\!\!\!\!\Delta}_{(\mu\nu)\lambda}\,,
\end{eqnarray}
where the trace part is given by
\begin{align}\label{dilation}
{}^{(\rm d)}\Delta_{\mu}=\Delta^{\nu}{}_{\nu\mu}\,,
\end{align} 
while the antisymmetric and symmetric traceless parts are reducible to
\begin{align}
    ^{(\rm s)}\Delta_{[\mu\nu]\lambda}&=\frac{1}{3}\left(g_{\lambda\nu}{}^{(\rm s)}\order{\Delta}{1}_{\mu}-g_{\mu\lambda}{}^{(\rm s)}\order{\Delta}{1}_{\nu}\right)+\frac{1}{6}\,\varepsilon_{\lambda\rho\mu\nu}{}^{(\rm s)}\order{\Delta}{2}^{\rho}+{}^{(\rm s)}\order{\Delta}{3}_{\mu\nu\lambda}\,,\label{irreduciblespin}\\
    ^{(\rm sh)}{\nearrow\!\!\!\!\!\!\!\Delta}_{(\mu\nu)\lambda}&=g_{\lambda(\mu}{}^{(\rm sh)}\order{{\nearrow\!\!\!\!\!\!\!\Delta}}{1}_{\nu)}-\frac{1}{4}g_{\mu\nu}{}^{(\rm sh)}\order{{\nearrow\!\!\!\!\!\!\!\Delta}}{1}_{\lambda}+\frac{1}{3}\varepsilon_{\lambda\rho\sigma(\mu}{}^{(\rm sh)}\order{{\nearrow\!\!\!\!\!\!\!\Delta}}{2}^{\rho\sigma}{}_{\nu)}+{}^{(\rm sh)}\order{{\nearrow\!\!\!\!\!\!\!\Delta}}{3}_{\mu\nu\lambda}\,\label{irreducibleshears}\,,
\end{align}
with
\begin{align}
    ^{(\rm s)}\order{\Delta}{1}_{\mu}&=\Delta_{[\mu\nu]}{}^{\nu}\,,\\
    ^{(\rm s)}\order{\Delta}{2}_{\mu}&=\varepsilon_{\mu\lambda\rho\nu}\Delta^{[\rho\nu]\lambda}\,,\\
    ^{(\rm s)}\order{\Delta}{3}_{\mu\nu\lambda}&=\Delta_{[\mu\nu]\lambda}-\frac{2}{3}g_{\lambda[\nu}{}^{(\rm s)}\order{\Delta}{1}_{\mu]}-\frac{1}{6}\,\varepsilon_{\lambda\rho\mu\nu}{}^{(\rm s)}\order{\Delta}{2}^{\rho}\,,\\
    {}^{(\rm sh)}\order{{\nearrow\!\!\!\!\!\!\!\Delta}}{1}_{\mu}&=\frac{4}{9}\left(\Delta_{(\nu\mu)}{}^{\nu}-\frac{1}{4}\Delta^{\nu}{}_{\nu\mu}\right)\,,\\
    {}^{(\rm sh)}\order{{\nearrow\!\!\!\!\!\!\!\Delta}}{2}^{\mu\nu}{}_{\lambda}&=-\,\left[\varepsilon^{\mu\nu\rho\sigma}\Delta_{(\sigma\lambda)\rho}+\varepsilon^{\mu\nu\rho}\,_{\lambda}\left(\frac{3}{4}{}^{(\rm sh)}\order{{\nearrow\!\!\!\!\!\!\!\Delta}}{1}_{\rho}-\frac{1}{4}\Delta^{\nu}{}_{\nu\rho}\right)\right]\,,\\
    {}^{(\rm sh)}\order{{\nearrow\!\!\!\!\!\!\!\Delta}}{3}_{\lambda\mu\nu}&=\Delta_{(\lambda\mu\nu)}-\frac{1}{4}g_{(\mu\nu|}\Delta^{\rho}{}_{\rho|\lambda)}-\frac{3}{4}g_{(\mu\nu}{}^{(\rm sh)}\order{{\nearrow\!\!\!\!\!\!\!\Delta}}{1}_{\lambda)}\,.
\end{align}

The aforementioned irreducible decompositions under the four-dimensional pseudo-orthogonal group will be useful in the following sections for constructing the corresponding cosmological perturbations in MAG. 

\section{$3+1$ decomposition}\label{decomposition}

In cosmology, time and space are clearly distinguished because of the expansion of the universe. Therefore, we inevitably decompose the space-time into the three-dimensional spatial parts and the temporal part to compute the cosmological perturbations. Here, we first explain a generic way of the $3+1$ decomposition, especially the decomposition of the torsion and nonmetricity tensors. Note that the decomposition in this section is generic: we do not assume any particular form of the metric and only assume the existence of a unit timelike vector $n_\mu$ satisfying
\begin{align}\label{n-def}
n_\mu n^\mu=-\,1 \,.
\end{align}
The corresponding projection tensor is defined by
\begin{align}
P_{\mu\nu} \equiv g_{\mu\nu}+n_{\mu}n_{\nu} \,,\label{Peq}
\end{align}
which is, by definition, orthogonal to $n_\mu$, i.e. $P_{\mu\nu}n^\mu=0$. 

Using the normal vector \eqref{n-def} and the projection tensor \eqref{Peq}, we can split any four-dimensional tensor into the parallel and orthogonal components with respect to $n_{\mu}$. This is easy to see for a vector: $X_\mu = \left(-\,n^\alpha{X}_\alpha\right) n_{\mu}+ \vec{X}_\mu$ with $\vec{X}_\mu\equiv{P}_\mu{}^\alpha{X}_\alpha$. Obviously, $\left(-\,n^\alpha{X}_\alpha\right)$ determines the temporal component of $X_\mu$ which is along $n_\mu$, while $\vec{X}_\mu$ characterises pure spatial components as $n^\mu\vec{X}_\mu=0$. Note that all four-dimensional tensors like the metric and energy-momentum tensors, and all four-dimensional irreducible pieces of the torsion, nonmetricity and hypermomentum tensors (e.g.~$T_{\mu}, S_{\mu}, t_{\lambda\mu\nu}$ for the torsion tensor) are indeed reducible to a further set of three-dimensional quantities. While the $3+1$ decomposition for a vector $X_{\mu}$ to two three-dimensional irreducible pieces $\left(-\,n^\alpha{X}_\alpha\right)$ and $\vec{X}_\mu$ is very simple, the decomposition of higher-rank tensors is not so straightforward at first glance, especially if it has a nontrivial symmetric property like $t_{\lambda\mu\nu}$. In this section, we directly present the form of the $3+1$ decomposition for generic rank-$n$ tensors with $n \leq 3$, while a systematic way of the decompositions is elaborated in the appendices. In Appendix~\ref{appendix1}, we first explain how to identify three-dimensional irreducible pieces embedded in a four-dimensional tensor by the use of Young tableaux. We then derive concrete tensorial expressions of the $3+1$ decomposition in Appendix~\ref{appendix2}.

For notational convenience, as we did above, we use an arrow on top to denote the corresponding spatial quantities, e.g. for an arbitrary rank-$n$ tensor $X_{\alpha_1\cdots \alpha_n}$, we extract the spatial part as
\begin{equation}
    \vec{X}_{\mu_1\cdots \mu_{n}} \equiv P_{\mu_{1}}{}^{\alpha_1}\cdots P_{\mu_{n}}{}^{\alpha_n}X_{\alpha_1\cdots \alpha_n}\,,\label{arrow1}
\end{equation}
which straightforwardly satisfies the following desired conditions:
\begin{equation}
  n^{\mu_1} \vec{X}_{\mu_1\cdots \mu_{n}}=\cdots = n^{\mu_n} \vec{X}_{\mu_1\cdots \mu_{n}}=0\,.\label{arrow3}
\end{equation}
Additionally, in three dimensions, two and three antisymmetric indices can be respectively dualised to one and zero indices by using the three-dimensional Levi-Civita tensor 
\begin{equation}
    \vec{\varepsilon}_{\mu\nu\rho} \equiv n^{\lambda}\varepsilon_{\lambda\mu\nu\rho}\,.
\end{equation} In general, all the three-dimensional irreducible tensors arising from the $3+1$ decomposition are symmetric and traceless, which means that the following expressions are also satisfied for irreducible tensors
\begin{align}
   &\vec{X}_{\mu_1\cdots \mu_{n}}=\vec{X}_{(\mu_1\cdots \mu_{n})}\,,\label{arrow4}\\
  &\vec{X}_{\mu_1\cdots\mu_i\cdots \mu_{n}}P^{\mu_1\mu_i}=\cdots=\vec{X}_{\mu_1\cdots\mu_i\cdots \mu_{n}}P^{\mu_1\mu_n}=\cdots=\vec{X}_{\mu_1\cdots\mu_i\cdots \mu_{n}}P^{\mu_i\mu_n}=0\,.\label{arrow2}
\end{align}
In short, all the tensors with an arrow on top are symmetric and traceless spatial tensors in the following. For simplicity in the notation, we will also put arrows on vectors and on the three-dimensional spatial Levi-Civita tensor.

For completeness, we present the decompositions of all generic tensors of rank-0,1,2, and 3.
For arbitrary rank-$0$ and rank-$1$ tensors, the decomposition is trivial:
\begin{align}
X_s&=X_s\,,\\
    X_\mu&= \vec{X}_\mu-n_\mu X_*\,,\label{Xvec}
\end{align}
where $X_*\equiv X_\mu n^\mu$. 

Next, for a rank-$2$ tensor, we can first separate it to the trace part and the traceless symmetric and antisymmetric parts
\begin{equation}
 X_{\mu\nu}= X_{(\mu\nu)}+X_{[\mu\nu]}=\frac{1}{4}g_{\mu\nu}X +  {\nearrow\!\!\!\!\!\!\!\!X}_{(\mu\nu)} + X_{[\mu\nu]}\,,
\end{equation}
where $X\equiv g^{\mu\nu}X_{\mu\nu}$ and ${\nearrow\!\!\!\!\!\!\!\!X}_{(\mu\nu)}\equiv X_{(\mu\nu)}-\frac{1}{4}g_{\mu\nu}X$, with $g^{\mu\nu} {\nearrow\!\!\!\!\!\!\!\!X}_{(\mu\nu)}=0$. The traceless symmetric part is further split into one tensor, one vector and one scalar\footnote{\label{footnote1}Note that one may more directly decompose the tensor; for instance, one can consider the $3+1$ decomposition without splitting into the symmetric and antisymmetric parts or without subtracting the trace piece from the symmetric part
	\begin{equation}
	X_{(\mu\nu)}= 
	\vec{X}_{\mu\nu}-2n_{(\mu}\vec{X}_{*\nu)} + n_{\mu}n_{\nu} X_{**}+\frac{1}{3}P_{\mu\nu} X_{***}\,,
	\end{equation}
	with $X_{**} \equiv n^\mu n^\nu X_{\mu\nu}$ and $X_{***} \equiv P^{\mu\nu}X_{\mu\nu}$, which are related to $\{X,  {\nearrow\!\!\!\!\!\!\!\!X}_{**}\}$ via
	\begin{align}
	X_{**}=-\,\frac{1}{4}X+{\nearrow\!\!\!\!\!\!\!\!X}_{**} \,, \quad X_{***}=\frac{3}{4}X+ {\nearrow\!\!\!\!\!\!\!\!X}_{**}
	\,.
	\end{align}}
\begin{align}
{\nearrow\!\!\!\!\!\!\!\!X}_{(\mu\nu)}= \vec{X}_{\mu\nu}-2n_{(\mu}\vec{X}_{*\nu)} + \left(n_{\mu}n_{\nu}+\frac{1}{3}P_{\mu\nu}\right){\nearrow\!\!\!\!\!\!\!\!X}_{**}\,,
\end{align}
where
\begin{align}
{\nearrow\!\!\!\!\!\!\!\!X}_{**} &\equiv n^{\mu}n^{\nu} {\nearrow\!\!\!\!\!\!\!\!X}_{\mu\nu} \,, \quad
\vec{X}_{*\mu}\equiv  n^{\alpha}P^{\beta}{}_{\mu}{\nearrow\!\!\!\!\!\!\!\!X}_{(\alpha\beta)} = n^{\alpha}P^{\beta}{}_{\mu}X_{(\alpha\beta)}
\,, \quad
\vec{X}_{\mu\nu} \equiv P^{\alpha}{}_{\mu} P^{\beta}{}_{\nu} {\nearrow\!\!\!\!\!\!\!\!X}_{(\alpha\beta)} - \frac{1}{3}P_{\mu\nu}{\nearrow\!\!\!\!\!\!\!\!X}_{**}
\,.
\end{align}
For the antisymmetric part we find
\begin{equation}\label{eee}
X_{[\mu\nu]}= -\,2\vec{X}^{\rm (E)}_{[\mu}n_{\nu]} + \frac{1}{2} \vec{\varepsilon}_{\mu\nu\lambda} \vec{X}^{\rm (B)}{}^{\lambda} \,,
\end{equation}
where
\begin{align}
\vec{X}^{\rm (E)}_{\mu} &\equiv P_{\mu}{}^{\alpha}n^{\beta}\vec{X}_{[\alpha\beta]}
\,, \quad
\vec{X}^{\rm (B)}{}^{\mu} \equiv \vec{\varepsilon}^{\mu\alpha\beta}\vec{X}_{[\alpha\beta]}
\,.\label{antisim_def}
\end{align}
The first term on the RHS of \eqref{eee} represents the electric temporal part and the second one the purely spatial magnetic part, each one carrying three d.o.f. of the antisymmetric part of the generic rank-$2$ tensor $X_{\mu\nu}$.

Finally, for an arbitrary rank-$3$ tensor, we have:
\begin{eqnarray}
X_{\lambda\mu\nu}= X_{\lambda[\mu\nu]}+X_{\lambda(\mu\nu)}\,.
\end{eqnarray}
As is shown, the antisymmetric part $X_{\lambda[\mu\nu]}$ corresponds to a tensor with the torsion-like symmetry, whereas the symmetric part $X_{\lambda(\mu\nu)}$ is analogous to nonmetricity. Hence, without any loss of generality, the $3+1$ decomposition of arbitrary rank-$3$ tensors is completely determined by the corresponding decompositions of the torsion and nonmetricity tensors. For this task, we shall use then the notation used in Sec.~\ref{sec:metricaffine} to denote the rank-3 tensor. Following the prescription explained above (see Appendix~\ref{appendix2} for the details of the decomposition of rank-2 and rank-3 tensors), the $3+1$ decomposition of the torsion tensor is provided by the expressions
\begin{align}
 T_{\mu}&=\vec{T}_{\mu}-n_{\mu}\phi \,,\label{torsion1} \\
 S_{\mu}&=\vec{\mathcal{S}}_{\mu}-n_{\mu}\varrho\,, \label{torsion2}  \\
 t_{\mu\nu\rho}&=2n_{[\nu}\vec{A}_{\rho]\mu}+2\Bigl(n_{\mu}n_{[\nu}+\frac{1}{2}P_{\mu[\nu} \Bigr) \vec{B}_{\rho]}
+\frac{1}{2}\varepsilon_{\nu\rho}{}^{\alpha\beta}\Bigl[-\,n_{\alpha}\vec{\mathcal{A}}_{\beta\mu}+\Bigl(n_{\beta}n_{\mu}+\frac{1}{2}P_{\beta\mu} \Bigr)\vec{\mathcal{B}}_{\alpha}\Bigr]\,,\label{torsion3}
\end{align}
where we recall that the quantities $\vec{A}_{\mu\nu}$ and $\vec{\mathcal{A}}_{\mu\nu}$ with an arrow on top are spatial, symmetric and traceless. For each tensor appearing in the decomposition, one can write the inverse relations as follows:
\begin{eqnarray}
    \phi&\equiv &T_\mu n^\mu\,,\quad 
    \vec{T}_\mu\equiv \,P^\nu{}_\mu T_\nu \,,\quad  
    \varrho\equiv S_\mu n^\mu\,,\quad \vec{\mathcal{S}}_\mu=\,P^\nu{}_\mu S_\nu \,,\label{inverse1}\\
    \vec{A}_{\mu\nu}&\equiv &\,t_{\alpha\beta\rho}P^\alpha{}_{(\mu}P^\beta{}_{\nu)}n^\rho\,,\quad 
    \vec{B}_\mu\equiv t_{\alpha\beta\rho}n^\alpha n^\beta P^\rho{}_\mu\,,\quad 
    \vec{\mathcal{A}}_{\mu\nu}\equiv \,t_{\alpha\beta\rho}P^\alpha{}_{(\mu}n^\lambda \varepsilon_{\nu)\lambda}{}^{\beta\rho}\,,\quad 
    \vec{\mathcal{B}}_\mu\equiv t_{\alpha\beta\rho}n^\alpha n^\lambda \varepsilon_\lambda{}^{\beta\rho}{}_\mu\,. \label{inverse2}
\end{eqnarray}
Similarly, the nonmetricity tensor turns out to be expressed under the $3+1$ decomposition as
\begin{align}
 W_{\mu}&=\vec{W}_{\mu}-n_{\mu} \theta\,,\label{Q1split}\\
\Lambda_{\mu}&=\vec{\Lambda}_{\mu}-n_{\mu} \sigma\,,\label{Q2split}\\
\Omega^{}_{\mu\nu\rho}&=
2n_{[\nu}\vec{\mathcal{Q}}_{\rho]\mu}+2\Bigl(n_{\mu}n_{[\nu}+\frac{1}{2}P_{\mu[\nu} \Bigr) \vec{\mathcal{Y}}_{\rho]}
+\frac{1}{2}\varepsilon_{\nu\rho}{}^{\alpha\beta}\Bigl[ -\,n_{\alpha}\vec{Q}_{\beta\mu}+\Bigl(n_{\beta}n_{\mu}+\frac{1}{2}P_{\beta\mu} \Bigr)\vec{Y}_{\alpha}\Bigr]
\,,\label{Q3split}\\ q_{\mu\nu\rho}&=\vec{C}_{\mu\nu\rho}-3n_{(\mu}\vec{\kappa}_{\nu\rho)}+\frac{3}{5}\bigl(5n_{(\mu}n_{\nu}+P_{(\mu\nu} \bigr)\vec{Z}_{\rho)}-\left(n_{(\mu}n_{\nu}+P_{(\mu\nu} \right) n_{\rho)} \xi\,,\label{Q4split}
\end{align}
where, once again, $\vec{C}_{\mu\nu\rho}$, $\vec{\kappa}_{\mu\nu}$, $\vec{\mathcal{Q}}_{\mu\nu}$ and $\vec{Q}_{\mu\nu}$ are spatial, symmetric and traceless tensors. Note that the inverse relations for the vectors $W_{\mu}$ and $\Lambda_\mu$ have the same form as the ones related to $T_\mu$ (see~\eqref{inverse1}), whereas the same holds for the inverse relations of $\Omega_{\mu\nu\rho}$ and $t_{\mu\nu\rho}$ (see~\eqref{inverse2}). Then, the remaining inverse relations are related to $q_{\mu\nu\rho}$:
\begin{eqnarray}
\xi&\equiv &n^{\alpha}n^{\beta}n^\rho q_{\alpha\beta\rho}\,,\quad 
\vec{Z}_{\mu}\equiv P^\alpha{}_\mu q_{\alpha\beta\rho}n^\beta n^\rho\,,\quad     
\vec{C}_{\mu\nu\lambda}\equiv P^\alpha{}_\mu  P^\beta{}_\nu P^\rho{}_\lambda q_{\alpha\beta\rho}-\frac{3}{5} P^{\alpha}{}_{(\mu}P_{\nu\lambda)}n^\beta n^\rho q_{\alpha\beta\rho}\,,\\
\vec{\kappa}_{\mu\nu}&\equiv &P^\alpha{}_\mu P^\beta{}_\nu n^\rho q_{\alpha\beta\rho}-\frac{1}{3}n^\alpha n^\beta n^\rho q_{\alpha\beta\rho}P_{\mu\nu}\,.
\end{eqnarray}
For later convenience, we use bold characters to denote the set of spatial scalars, vectors and tensors:
\begin{align}
{\bf X}&= \{\phi, \varrho, \theta, \sigma, \xi \}
\,, \\
\vec{\bf X}_{\mu}&=\{ \vec{T}_{\mu}, \vec{\mathcal{S}}_{\mu}, \vec{B}_{\mu}, \vec{\mathcal{B}}_{\mu}, \vec{W}_{\mu}, \vec{\Lambda}_{\mu}, \vec{Y}_{\mu}, \vec{\mathcal{Y}}_{\mu}, \vec{Z}_{\mu} \}
\,, \\
\vec{\bf X}_{\mu\nu}&=\{ \vec{A}_{\mu\nu}, \vec{\mathcal{A}}_{\mu\nu}, \vec{Q}_{\mu\nu}, \vec{\mathcal{Q}}_{\mu\nu}, \vec{\kappa}_{\mu\nu} \}
\,, \\
\vec{\bf X}_{\mu\nu\rho}&=\{\vec{C}_{\mu\nu\rho} \}
\,.
\end{align}
As we have explained in footnote~\ref{footnote1}, one may directly decompose the rank-3 tensor without splitting it into the four-dimensional irreducible pieces. However, using the four-dimensional irreducible pieces would be more systematic since any rank-3 tensor with an additional symmetric property (e.g. the torsion tensor $T^{\lambda}{}_{\mu\nu}$ and the traceless part of the nonmetricity tensor ${\nearrow\!\!\!\!\!\!\!Q}_{\lambda\mu\nu}$) is always given by a linear combination of them. In addition, there would be a practical advantage to consider the $3+1$ decomposition of the four-dimensional irreducible pieces in the cosmology. For instance, the tensors $t_{\mu\nu\rho}$ and $\Omega_{\mu\nu\rho}$ do not contain any scalar objects after performing the $3+1$ decomposition. Therefore, they vanish in the homogeneous and isotropic background and only contribute to the perturbations (see next section). One can thus discard higher-order terms of $t_{\mu\nu\rho}$ and $\Omega_{\mu\nu\rho}$ even before expanding the four-dimensional tensors into the $3+1$ objects.

As rank-$2$ and rank-$3$ tensors, it is straightforward to apply the above $3+1$ decomposition to the energy-momentum $\Theta_{\mu\nu}$ and hypermomentum $\Delta_{\mu\nu\lambda}$ tensors in the matter sector presented in Sec.~\ref{sec:metricaffine}. For the energy-momentum tensor, we find the well-known results in cosmological perturbation theory, which we will explain in Sec.~\ref{sec:perturbations}, while for the hypermomentum tensor, we find similar results as \eqref{torsion1}-\eqref{torsion3}, and \eqref{Q1split}-\eqref{Q4split} for the corresponding four-dimensional irreducible pieces that are defined in \eqref{dilation}-\eqref{irreducibleshears}. In the context of cosmology, $\Theta_{\mu\nu}$ and $\Delta_{\mu\nu\lambda}$ can be used to describe the form of perfect hyperfluids respecting spatial homogeneous and isotropic conditions, as well as more general relativistic fluids with heat flux and anisotropic stress.

It is also worth mentioning that the $3+1$ decomposition described above is generic for any metric, and then it can also be considered in Hamiltonian formalism. Thus, our setup can be straightforwardly implemented to perform nonlinear Hamiltonian analysis of any theory which includes torsion and nonmetricity, or any other rank-3 tensor.

\section{Cosmological background}\label{sec:background}
\subsection{Geometrical background}
Our current understanding of the evolution of the universe is based on the cosmological principle, which assumes that, on a sufficiently large scale, both the geometry of the space-time and the energy-momentum tensor of matter are spatially homogeneous and isotropic~\cite{Weinberg:2008zzc}. The first assumption directly leads to the FLRW space-time, whereas a spatially homogeneous and isotropic distribution of matter acquires the form of a perfect fluid, which is consistently realised in GR via the Einstein field equations in the FLRW space-time.
In the framework of MAG, the inclusion of the torsion, nonmetricity and hypermomentum tensors at the background level in the cosmological principle is theoretically and observationally possible. In this sense, in this section we assume that these tensors also satisfy the homogeneity and isotropy, and we provide the general ansatz for the cosmological background\footnote{In principle, field(s) can have a configuration without satisfying the homogeneity and isotropy but without contradicting the cosmological principle thanks to the internal symmetry of the field(s). Examples include the solid inflation~\cite{Dubovsky:2011sj,Endlich:2012pz,Aoki:2022ipw}, the two-form gauge field~\cite{Stein-Schabes:1986owe, Copeland:1994km, Elizalde:2018rmz, Aoki:2022ylc}, and the $SU(2)$ gauge fields~\cite{Maleknejad:2011jw}.}.

Following Weinberg~\cite{Weinberg:1972kfs}, we can think of the cosmological principle as the statement that the constant-time hypersurfaces are described by a maximally symmetric space; in the polar coordinates, the metric tensor is given by
\begin{align}
\gamma_{ij}\dd x^i \otimes\dd x^j=\frac{\dd r^2}{1-K r^2}+r^2 \dd\Omega^2\,,
\label{MSS}
\end{align}
where the constant $K=0, \pm 1$ is the spatial curvature and $\dd \Omega^2=\dd \vartheta^2+ \sin ^2\vartheta \,\dd \varphi^2$ is the line element of the unit two-sphere. More precisely, the maximally symmetric space means that the space admits the maximum number of Killing vectors $X_{\xi}$, six in the three dimensions, associated with translations and rotations. In the polar coordinates, the Killing vectors $X_{\xi}=\{R_i, X_i\}~(i=1,2,3)$ are given by
\begin{subequations}\label{eq:genvectrans}
\begin{align}
R_1&=\sin\varphi \,\partial_\vartheta+\frac{\cos\varphi}{\tan\vartheta}\,\partial_\varphi\,,\quad R_2=-\,\cos\varphi \,\partial_\vartheta+\frac{\sin\varphi}{\tan\vartheta}\,\partial_\varphi\,,\quad R_3=-\,\partial_\varphi\,, \\
X_1 &= \chi\sin\vartheta\cos\varphi\,\partial_r + \frac{\chi}{r}\cos\vartheta\cos\varphi\,\partial_{\vartheta} - \frac{\chi\sin\varphi}{r\sin\vartheta}\,\partial_{\varphi}\,,\\
X_2 &= \chi\sin\vartheta\sin\varphi\,\partial_r + \frac{\chi}{r}\cos\vartheta\sin\varphi\,\partial_{\vartheta} + \frac{\chi\cos\varphi}{r\sin\vartheta}\,\partial_{\varphi}\,,\\
X_3 &= \chi\cos\vartheta\,\partial_r - \frac{\chi}{r}\sin\vartheta\,\partial_{\vartheta}\,,
\end{align}
\end{subequations}
where \(\chi = \sqrt{1 - Kr^2}\).

Using the $3+1$ decomposition performed in Sec.~\ref{decomposition}, we can then fix the form of the FLRW background. The normal vector of the constant-time hypersurfaces is given by
\begin{equation}\label{n-BG}
    \bar{n}^{\mu}\partial_{\mu}=N^{-1}\partial_{t}\,,\quad \bar{n}_{\mu}\mathrm{d}x^{\mu}=-\,N\mathrm{d}t\,,
\end{equation}
where we added a bar to point out that the corresponding quantities are evaluated at the background.
The projector tensor $\bar{P}_{\mu\nu}$ is a three-dimensional quantity, so its tensorial structure has to be given by $\gamma_{ij}$, i.e.,
\begin{equation}\label{P-BG}
\bar{P}_{\mu\nu}\mathrm{d}x^{\mu}\otimes\mathrm{d}x^{\nu}=a^{2}\gamma_{ij}\mathrm{d}x^{i}\otimes\mathrm{d}x^{j}\,.
\end{equation}
Here, $N(t)$ and $a(t)$ are functions undetermined by the cosmological principle, called the lapse function and the scale factor, respectively. As a result, we obtain the well-known form of the FLRW metric
\begin{eqnarray}\label{metric-FLRW}
\bar{g}=\bar{g}_{\mu\nu}\mathrm{d}x^{\mu}\otimes\mathrm{d}x^{\nu}=-\,\bar{n}_{\mu}\bar{n}_{\nu}\mathrm{d}x^\mu \otimes \mathrm{d}x^\nu +\bar{P}_{\mu\nu}\mathrm{d}x^\mu \otimes \mathrm{d}x^\nu=-\,N^2\mathrm{d}t \otimes \mathrm{d}t +a^2\gamma_{ij}\mathrm{d}x^i \otimes \mathrm{d}x^j\,.
\end{eqnarray}

We use the same procedure to determine the most general form of the torsion and nonmetricity tensors under the symmetries of the FLRW space-time. Since the only nonzero (pseudo)-tensors in the maximally symmetric space are the metric $\gamma_{ij}$ and the Levi-Civita tensor $\epsilon_{ijk}$, we can immediately conclude
\begin{align}\label{spatial-BG}
\bar{\vec{\bf X}}_{\mu} &=\bar{\vec{\bf X}}_{\mu\nu}=\bar{\vec{\bf X}}_{\mu\nu\rho}=0
\,, \\\label{scalar-BG}
\bar{\bf{X}} &=  \bar{\bf{X}} (t)
\,,
\end{align}
at the background level. Note that the three-dimensional Levi-Civita tensor $\bar{\vec{\varepsilon}}_{\mu\nu\rho} \equiv \bar{n}^{\lambda}\bar{\varepsilon}_{\lambda\mu\nu\rho} $ introduced in the previous section and the Levi-Civita tensor of the maximally symmetric space $\varepsilon_{ijk}$ are related by
\begin{align}
\bar{\vec{\varepsilon}}_{\mu\nu\rho} \mathrm{d}x^{\mu} \otimes \mathrm{d}x^{\nu} \otimes \mathrm{d}x^{\rho}=
a^3 \varepsilon_{ijk} \mathrm{d}x^{i} \otimes \mathrm{d}x^{j} \otimes \mathrm{d}x^{k}
\,.
\end{align}
The factor $a^3$ arises because the projection tensor $\bar{P}_{\mu\nu}$ includes the factor $a^2$ in front of $\gamma_{ij}$. Hence, the irreducible pieces of the torsion and nonmetricity tensors are given by
\begin{align}
\bar{T}_\mu &=-\,\bar{n}_\mu \bar{\phi} \,,\quad \bar{S}_\mu=-\, \bar{n}_\mu \bar{\varrho}\,,\quad \bar{t}_{\lambda\mu\nu}=0\,,\label{background_torsion_modes}
\\
   \bar{W}_\mu&=-\,\bar{n}_\mu \bar{\theta} \,,\quad \bar{\Lambda}_\mu=-\,\bar{n}_\mu \bar{\sigma} \,,\quad \bar{\Omega}_{\lambda\mu\nu}=0\,, \quad
    \bar{q}_{\lambda\mu\nu}=-\left(\bar{n}_\lambda \bar{n}_\mu \bar{n}_\nu+\bar{n}_{(\lambda} \bar{P}_{\mu\nu)}\right)\bar{\xi} \,.\label{background_nonmetricity_modes}
\end{align}
The torsion and nonmetricity tensors then read
\begin{align}
    \bar{T}^\lambda{}_{\mu \nu}&=2T_1(t)\,\bar{n}_{[\mu}\bar{P}_{\nu]}{}^{\lambda}+2T_2(t)\,\bar{\varepsilon}^{\,\lambda}{}_{\mu\nu\rho}\bar{n}^{\rho} \,,\label{TorsionCos}\\
    \bar{Q}_{\lambda\mu\nu}&=2Q_{1}(t)\bar{n}_\lambda \bar{n}_\mu \bar{n}_\nu +2Q_2(t)\bar{n}_\lambda \bar{P}_{\mu\nu}+2Q_3(t)\bar{P}_{\lambda (\mu}\bar{n}_{\nu)}\,,\label{NonCos}
\end{align}
where $\{T_{i}\}_{i=1}^{2}$ and $\{Q_{i}\}_{i=1}^{3}$ are five arbitrary functions such as
\begin{align}
\bar{\phi}=-\,3T_1\,, \quad \bar{\varrho}=-\,12T_2\,, \quad
\bar{\theta}=\frac{1}{2}\left(Q_1-3Q_2\right)\,, \quad
\bar{\sigma}=\frac{2}{3}\left(Q_1+Q_2-2Q_3\right)
\,, \quad
\bar{\xi}=-\left(Q_1+Q_2+Q_3\right)
\,.
\end{align}
Note also that for simplicity in the notation, we omit bars on top for the five functions $\{T_i,Q_i\}$. By construction,~\eqref{TorsionCos} and~\eqref{NonCos} then solve the equations
\begin{align}
\mathcal{L}_{X_{\xi}} \bar{T}^\lambda{}_{\mu \nu} = \mathcal{L}_{X_{\xi}} \bar{Q}_{\lambda\mu\nu} =0
\,,
\end{align}
with $\mathcal{L}_{X_{\xi}} $ being the Lie derivative with respect to the Killing vector. Thus, these expressions constitute the most general ansatz for the torsion and nonmetricity tensors under the symmetries of the FLRW space-time. As previously mentioned, it is worthwhile to stress that the tensors $\bar{t}_{\lambda\mu\nu}$ and $\bar{\Omega}_{\lambda\mu\nu}$ vanish in FLRW background, since they cannot simultaneously satisfy spatially homogeneous and isotropic conditions.

\vskip\baselineskip
{\bf Component expressions.}
While Expressions~\eqref{TorsionCos} and~\eqref{NonCos} are coordinate independent, for practical purposes, it may be convenient to use particular coordinates $x^{\mu}=(t, x^i)$ (e.g. the ones used in~\eqref{MSS}) and summarise the component expressions in such a coordinate system.

Let us first not specify the spatial coordinates $x^i$ and keep the abstract indices for the spatial ones.
For the performance and mathematical treatment of the cosmological variables, it is convenient to use $\gamma_{ij}$, instead of $\bar{P}_{ij}=a^2 \gamma_{ij}$, in order to raise and lower the Latin indices, e.g. implying
\begin{align}
\vec{X}^{\mu}\equiv (0, a^{-1} X^i) \quad \implies \quad
\vec{X}_{\mu}=\bar{g}_{\mu\nu}\vec{X}^{\nu} = (0, a X_i)\,,\label{lower1}
\end{align}
with $X_i \equiv \gamma_{ij}X^j$. Furthermore, for any spatial rank-2 tensor we find
\begin{align}
\vec{X}^{\mu\nu}\equiv \begin{pmatrix} 
    0 & 0 \\
    0 & a^{-2} X^{ij} \\
\end{pmatrix} \quad \implies \quad \vec{X}_{\mu\nu}=\bar{g}_{\mu\alpha}\bar{g}_{\nu\beta}\vec{X}^{\alpha\beta}=\begin{pmatrix} 
    0 & 0 \\
    0 & a^2X_{ij} \\
\end{pmatrix}\,.\label{lower2}
\end{align}
Here, we have added a scalar factor and omitted the arrow in the components. The reason to include the scale factor in the way of \eqref{lower1} and \eqref{lower2} will be clarified shortly whereas the arrow is removed so as not to confuse the component of $\vec{X}^{\mu}$ with the rescaled quantity $X^i$: for instance, the spatial component of $\vec{X}^{\mu}$ reads $\vec{X}^i=a^{-1} X^i$.
Hereafter we will use this convention when raising and lowering Latin indices of general tensor quantities. In particular, the component expressions of the metric tensor are
\begin{align}\label{metric-BG-Matrix}
\bar{g}_{\mu\nu} = 
\begin{pmatrix}
-\,N^2 & 0 \\
0 & a^2 \gamma_{ij}
\end{pmatrix}
\,, \quad
\bar{g}^{\mu\nu} = 
\begin{pmatrix}
-\,N^{-2} & 0 \\
0 & a^{-2} \gamma^{ij}
\end{pmatrix}
\,.
\end{align}

From~\eqref{TorsionCos} and~\eqref{NonCos}, one can straightforwardly obtain the component expressions of the torsion and the nonmetricity as follows:
\begin{eqnarray}\label{uuu}
 \bar{T}^{\lambda}{}_{\mu\nu}&\longrightarrow& \arraycolsep=1.4pt\def\arraystretch{2.2}\left\{\begin{array}{l}
 \bar{T}^{0}{}_{0i} =0\\
 \bar{T}^{0}{}_{ij} =0\\
 \bar{T}^{i}{}_{0j} =
- \, N(t) T_1(t)\delta^{i}{}_{j} \\
     \bar{T}^{i}{}_{jk} =\displaystyle -\,2a (t)T_{2}(t) \varepsilon^i{}_{jk}  
     \displaystyle
   \end{array}\right.\,,\quad 
  \bar{Q}_{\lambda\mu\nu}\longrightarrow \arraycolsep=1.4pt\def\arraystretch{2.2}\left\{\begin{array}{l}
  \bar{Q}_{000}=-\,2N^{3}(t)Q_1(t)  \\
       \bar{Q}_{00i}=0\\
    \bar{Q}_{0ij} =  
  - \,2N(t)\,a^{2}(t) Q_2(t)\gamma_{ij}\\
     \bar{Q}_{i00} =0  \\
      \bar{Q}_{i0j} = 
     -\,N(t)\,a^{2}(t)Q_3(t)\gamma_{ij}  \\
       \bar{Q}_{ijk} = 0
   \end{array}\right.\,.
\end{eqnarray}
One may notice that the lapse and the scale factor appear in a systematic way: one lower (upper) temporal and spatial indices yield factors $N$ and $a$ ($N^{-1}$ and $a^{-1}$), respectively. This is indeed what we wrote in \eqref{lower1} and \eqref{lower2}.

When one further specifies the spatial coordinates, one obtains all the components explicitly; in the polar coordinates $x^{\mu}=(x^0, x^1, x^2, x^3) = (t, r, \vartheta, \varphi)$, we have
\begin{gather}\label{T1}
\bar{T}^1{}_{01}=\bar{T}^2{}_{02}=\bar{T}^3{}_{03}=-\,N(t)T_1(t)\,,\quad \bar{T}^2{}_{31}=\frac{\bar{T}^1{}_{23}}{r^2\chi^2}=\bar{T}^3{}_{12}\sin^{2}\vartheta =-\,\frac{2a(t)T_2(t)\sin\vartheta }{\chi}\,,\\
\bar{Q}_{000}=-\,2N^{3}(t)Q_1(t)\,,\quad \bar{Q}_{022}=r^{2}\chi^2\bar{Q}_{011}=\frac{\bar{Q}_{033}}{\sin^2\vartheta}=-\,2r^2a^{2}(t)N(t)Q_2(t)\,,\\
\bar{Q}_{202}=\frac{\bar{Q}_{303}}{\sin^2\vartheta}=r^2\chi^2\bar{Q}_{101}=-\,r^2a^{2}(t)N(t)Q_{3}(t)\,.\label{Q1}
\end{gather}
Then, the affine connection is composed of the Levi-Civita part determined by the metric tensor, and a distortion tensor containing torsion and nonmetricity, which depends on the five functions $\{T_{i}\}_{i=1}^{2}$ and $\{Q_{i}\}_{i=1}^{3}$ (c.f.~\cite{Hohmann:2019fvf}):
\begin{eqnarray}
\bar{\tilde{\Gamma}}^0{}_{00} &=&\frac{\dot{N}(t)}{N(t)}- N(t)Q_1(t)\,, \quad \bar{\tilde{\Gamma}}^1{}_{01} = \bar{\tilde{\Gamma}}^{2}{}_{02} = \bar{\tilde{\Gamma}}^{3}{}_{03} = \frac{\dot{a}(t)}{a(t)}+N(t)(Q_2(t)-T_1(t))\,, \\
\bar{\tilde{\Gamma}}^1{}_{10} &=& \bar{\tilde{\Gamma}}^{2}{}_{20} = \bar{\tilde{\Gamma}}^{3}{}_{30} = \frac{\dot{a}(t)}{a(t)}+N(t)Q_2(t)\,,\\
\bar{\tilde{\Gamma}}^0{}_{33}&=&\bar{\tilde{\Gamma}}^0{}_{22}\sin^2\vartheta=r^{2}\chi^2\bar{\tilde{\Gamma}}^0{}_{11}\sin^2\vartheta=\left(\frac{a(t)}{N(t)}\right)^2 \left[\frac{\dot{a}(t)}{a(t)}+N(t)\left(Q_2(t)-Q_3(t)-T_1(t)\right)\right]r^2\sin^2\vartheta\,,\\
\bar{\tilde{\Gamma}}^1{}_{32}&=&-\,\bar{\tilde{\Gamma}}^1{}_{23}= r^2\chi^2\bar{\tilde{\Gamma}}^{2}{}_{13}= -\,r^2\chi^2 \bar{\tilde{\Gamma}}^{2}{}_{3 1}=-\,r^2\chi^2\sin^2\vartheta \, \bar{\tilde{\Gamma}}^{3}{}_{12}=r^2\chi^2\sin^2\vartheta \, \bar{\tilde{\Gamma}}^{3}{}_{21} = r^2\chi a(t)T_2(t)\sin\vartheta\,,\\
\bar{\tilde{\Gamma}}^1{}_{11} &=& \frac{Kr}{\chi^2}\,, \quad \bar{\tilde{\Gamma}}^{2}{}_{12} = \bar{\tilde{\Gamma}}^{2}{}_{21} = \bar{\tilde{\Gamma}}^{3}{}_{13} = \bar{\tilde{\Gamma}}^{3}{}_{31} = \frac{1}{r}\,, \quad \bar{\tilde{\Gamma}}^{3}{}_{23} = \bar{\tilde{\Gamma}}^{3}{}_{32} = \cot\vartheta\,, \quad \bar{\tilde{\Gamma}}^{2}{}_{33} = -\,\sin\vartheta\cos\vartheta\,,\\
\bar{\tilde{\Gamma}}^1{}_{33}&=&\sin^{2}\vartheta\,\bar{\tilde{\Gamma}}^1{}_{22}=-\,r\chi^2\sin^2\vartheta \,,
\end{eqnarray}
where dots denote differentiation with respect to the time coordinate $t$.

\subsection{Matter background}\label{secmatter1}

Let us now introduce the matter content at the background level. First, by assuming the cosmological principle, the corresponding metric energy-momentum tensor fulfilling the symmetries of FLRW space-times acquires the form of a perfect fluid
\begin{align}
    \bar{\Theta}_{\mu\nu}&=\left(\bar{\rho}(t)+\bar{p}(t)\right)\bar{n}_\mu \bar{n}_\nu+\bar{p}(t)\bar{g}_{\mu\nu}=\bar{\rho}(t)\bar{n}_\mu \bar{n}_\nu+\bar{p}(t) \bar{P}_{\mu\nu}\,,\label{energymomentum}
\end{align}
with $\bar{\rho}(t)$ and $\bar{p}(t)$ being the energy density and the pressure of the fluid, respectively. Note that the normal vector $n_\mu$ coincides with the four-velocity of the perfect fluid $u_{\mu}$ at the background level ${\bar u}_\mu={\bar n}_\mu = (-N(t),0,0,0)$. On the other hand, the spatially homogeneous and isotropic form of the hypermormentum tensor reads~\cite{Iosifidis:2020gth}:
\begin{equation}
    	\bar{\Delta}_{\lambda\mu\nu}=\frac{1}{3}\bar{\Delta}_1(t) \bar{P}_{\lambda\mu}\bar{n}_{\nu}+\bar{\Delta}_2(t) \bar{P}_{\lambda\nu}\bar{n}_{\mu}+\bar{\Delta}_3(t) \bar{n}_{\lambda}\bar{P}_{\mu\nu}+\frac{1}{4}\bar{\Delta}_4(t) \bar{n}_{\lambda}\bar{n}_{\mu} \bar{n}_{\nu}+\bar{\Delta}_5(t)\bar{\varepsilon}_{\lambda\mu\nu\rho}\bar{n}^{\rho} \label{hyper}\,,
\end{equation}
which in line with the background values of torsion and nonmetricity tensors depends on five arbitrary functions $\{\bar{\Delta}_{i}\}_{i=1}^{5}$; in this case, representing the intrinsic spin, dilation and shear currents of matter. Indeed, we can straightforwardly separate these characteristics by a redefinition of the aforementioned functions:
\begin{eqnarray}
\bar{ \Delta}_1={}^{(\rm d)}\bar{\Delta}_1+{}^{(\rm sh)}\bar{\Delta}_1\,,\quad  \bar{\Delta}_2={}^{(\rm s)}\bar{\Delta}_2+{}^{(\rm sh)}\bar{\Delta}_2\,,\quad \bar{\Delta}_3={}^{(\rm s)}\bar{\Delta}_3+{}^{(\rm sh)}\bar{\Delta}_3\,,\quad \bar{\Delta}_4={}^{(\rm d)}\bar{\Delta}_4+{}^{(\rm sh)}\bar{\Delta}_4\,,\quad \bar{\Delta}_5={}^{(\rm s)}\bar{\Delta}_5\,,
\end{eqnarray}
with  
\begin{eqnarray}
 {}^{(\rm s)}\bar{\Delta}_2=-{}^{(\rm s)}\bar{\Delta}_3\,,\quad  {}^{(\rm sh)}\bar{\Delta}_3={}^{(\rm sh)}\bar{\Delta}_2\,,\quad   {}^{(\rm d)}\bar{\Delta}_1=-\,\frac{3}{4}{}^{(\rm d)}\bar{\Delta}_4\,,\quad  {}^{(\rm sh)}\bar{\Delta}_4=4\,{}^{(\rm sh)}\bar{\Delta}_1\,, 
\end{eqnarray}
in such a way that the intrinsic spin, dilation and shear currents of matter acquire the following form:
\begin{align}
    \bar{\Delta}_{[\lambda\mu]\nu}&=2\,{}^{(\rm s)}\bar{\Delta}_3\bar{n}_{[\lambda}\bar{P}_{\mu]\nu}+{}^{(\rm s)}\bar{\Delta}_5\bar{\varepsilon}_{\lambda\mu\nu\rho}\bar{n}^\rho\,,\label{hyyper1}\\
    \bar{\Delta}^{\lambda}{}_{\lambda\mu}&=-\,{}^{(\rm d)}\bar{\Delta}_4 \bar{n}_\mu\,,\label{hyyper2}\\
    \bar{\nearrow\!\!\!\!\!\!\!\Delta\,}_{(\lambda\mu)\nu}&=\,{}^{(\rm sh)}\bar{\Delta}_1\,\bar{n}_\lambda \bar{n}_\mu \bar{n}_\nu+\frac{1}{3}{}^{(\rm sh)}\bar{\Delta}_1 \bar{P}_{\lambda\mu}\bar{n}_\nu+2\,{}^{(\rm sh)}\bar{\Delta}_2\,\bar{n}_{(\lambda}\bar{P}_{\mu)\nu}\,.\label{hyyper3}
\end{align}
Since the three main parts of the hypermomentum tensor fulfil the same algebraic symmetries as the torsion and nonmetricity tensors, the intrinsic spin current of matter also receive both scalar and pseudoscalar contributions:
\begin{equation}
    ^{(\rm s)}\order{\bar{\Delta}}{1}_{\mu}=3{}^{(\rm s)}\bar{\Delta}_{3}\bar{n}_{\mu}\,, \quad ^{(\rm s)}\order{\bar{\Delta}}{2}_{\mu}=6{}^{(\rm s)}\bar{\Delta}_{5}\bar{n}_{\mu}\,, \quad ^{(\rm s)}\order{\bar{\Delta}}{3}_{\mu\nu\lambda}=0\,,\label{hyper1}
\end{equation}
whereas the intrinsic dilation and shear currents of matter are given by scalar parts:
\begin{align}
    ^{(\rm d)}\bar{\Delta}_{\mu}&=-\,{}^{(\rm d)}\bar{\Delta}_4 \bar{n}_\mu\,,\label{hyper2}\\
    {}^{(\rm sh)}\order{\bar{{\nearrow\!\!\!\!\!\!\!\Delta}}}{1}_{\mu}&=\frac{4}{9}\bigl(3{}^{(\rm sh)}\bar{\Delta}_2-{}^{(\rm sh)}\bar{\Delta}_1\bigr)\bar{n}_{\mu}\,, \quad
    {}^{(\rm sh)}\order{\bar{{\nearrow\!\!\!\!\!\!\!\Delta}}}{2}^{\mu\nu}{}_{\lambda}=0\,, \quad
    {}^{(\rm sh)}\order{\bar{{\nearrow\!\!\!\!\!\!\!\Delta}}}{3}_{\lambda\mu\nu}=\left(\frac{2}{3}{}^{(\rm sh)}\bar{\Delta}_1+{}^{(\rm sh)}\bar{\Delta}_2\right)\left(\bar{n}_{\lambda}\bar{n}_{\mu}\bar{n}_{\nu}+\bar{n}_{(\lambda}\bar{P}_{\mu\nu)}\right)\,.\label{hyper3}
\end{align}

\vskip\baselineskip
{\bf Component expressions.}
Overall, the background configuration in the matter sector described by the metric energy-momentum tensor~\eqref{energymomentum} and the hypermomentum tensor~\eqref{hyper} can be expressed in component form as
\begin{eqnarray}
 \bar{\Theta}_{\mu\nu}&\longrightarrow& \arraycolsep=1.4pt\def\arraystretch{2.2}\left\{\begin{array}{l}
\bar{ \Theta}_{0i} =0\\
  \bar{ \Theta}_{00} =N^{2}(t)\bar{\rho}(t)\\
\bar{ \Theta}_{ij}=\displaystyle \,a^{2}(t)\bar{p}(t)\,\gamma_{ij}   
   \end{array}\right.\,,\quad 
  \bar{\Delta}_{\alpha\mu\nu}\longrightarrow \arraycolsep=1.4pt\def\arraystretch{2.2}\left\{\begin{array}{l}
  \bar{\Delta}_{000}= -\,\displaystyle\frac{1}{4}N^{3}(t)\bar{\Delta}_4(t)  \\
       \bar{\Delta}_{00i}=0\\
    \bar{\Delta}_{0ij}
    =-\,N(t)\, a^{2}(t)\bar{\Delta}_3(t)\gamma_{ij}
   \\
     \bar{\Delta}_{i00} =0  \\
      \bar{\Delta}_{i0j}
      =-\,N(t)\,a^{2}(t)\bar{\Delta}_2(t)\gamma_{ij}\\
       \bar{\Delta}_{ij0} 
      =\displaystyle -\,\frac{1}{3}\,N(t)\,a^{2}(t)\bar{\Delta}_{1}(t)\gamma_{ij}\\
       \bar{\Delta}_{ijk} =\displaystyle -\,\bar{\Delta}_5(t)a^{3}(t)\varepsilon_{ijk}
   \end{array}\right.\,.
\end{eqnarray}
Being all the background quantities already displayed, the next sections will be devoted to performing their SVT decompositions, which directly provides all the possible cosmological linear perturbations of MAG.

\section{Cosmological perturbations}\label{sec:perturbations}

In this section, we perform perturbations around the cosmological background configuration characterised by the FLRW metric \eqref{metric-FLRW}, background torsion \eqref{TorsionCos} and nonmetricity \eqref{NonCos} in the geometry sector, as well as by the background perfect fluid energy-momentum tensor \eqref{energymomentum} and hypermomentum \eqref{hyper} in  the matter sector.

\subsection{Geometrical perturbations}

In the geometry sector, we consider perturbations of the metric tensor and the affine connection around their cosmological background quantities:
\begin{eqnarray}
    g_{\mu\nu}=\bar{g}_{\mu\nu}+\delta g_{\mu\nu}\,,\quad    \tilde{\Gamma}^{\lambda}{}_{\mu\nu}= \bar{\tilde{\Gamma}}^{\lambda}{}_{\mu\nu} + \delta \tilde{\Gamma}^{\lambda}{}_{\mu\nu} \,,
\end{eqnarray}
or equivalently in terms of the torsion and nonmetricity tensors:
\begin{eqnarray}\label{geometry-pert}
    g_{\mu\nu}=\bar{g}_{\mu\nu}+\delta g_{\mu\nu}\,,\quad    T^\lambda{}_{\mu\nu}=\bar{T}^\lambda{}_{\mu\nu}+\delta T^\lambda{}_{\mu\nu}\,,\quad Q_{\lambda\mu\nu}=\bar{Q}_{\lambda\mu\nu}+\delta Q_{\lambda\mu\nu}\,,
\end{eqnarray}
where quantities with bar refer to the background and quantities with ``$\delta$" denote the corresponding perturbations.

We have performed the $3+1$ decomposition of all building blocks of the theory in terms of the corresponding irreducible pieces in Sec.~\ref{decomposition}. Having presented the cosmological background configuration in Sec.~\ref{sec:background}, it is then straightforward to find the $3+1$ expressions for the perturbations. For instance, the $3+1$ decomposition of the metric perturbation $\delta g_{\mu\nu}$ is
\begin{align}\label{deltag}
\delta g_{\mu\nu}= -\,2\bar{n}_{\mu} \bar{n}_{\nu} \alpha + 2 \bar{P}_{\mu\nu} \psi + \vec{h}_{\mu\nu} - 2\bar{n}_{(\mu} \vec{\beta}_{\nu)}
\,,
\end{align}
where $\bar{n}_{\mu}$ and $\bar{P}_{\mu\nu}$ are given by \eqref{n-def} and \eqref{P-BG} and the perturbations with arrow on top, $\delta \vec{X}^{\mu\nu \cdots} = \bar{P}^{\mu}{}_{\alpha} \bar{P}^{\nu}{}_{\beta} \cdots \delta X^{\alpha \beta \cdots}$, are pure spatial, symmetric and traceless tensors, i.e.~$\bar{n}^{\mu}\vec{\beta}_{\mu}=0,~ \vec{h}_{\mu\nu}=\vec{h}_{(\mu\nu)},~\bar{n}^{\mu} \vec{h}_{\mu\nu}=0$ and $\bar{P}^{\mu\nu}\vec{h}_{\mu\nu}=0$.

The perturbations are regarded as tensors on the background space-time so their indices are raised and lowered by the FLRW background metric $\bar{g}_{\mu\nu}$ given by \eqref{metric-FLRW}. Thus, in the $3+1$ decomposition, we deal with the background quantities
\begin{align}
\bar{n}^{\mu}&=(N^{-1}, \vec{0})
\,, \qquad
\bar{P}^{\mu}{}_{\nu}=
\begin{pmatrix}
0         &  \vec{0} \\
\vec{0} & \delta^i{}_j
\end{pmatrix}
\,.
\end{align}
Using the above results in \eqref{metric-BG-Matrix} and \eqref{deltag}, we find the following form for the full metric tensor:
\begin{align}\label{metric-matrix-form}
g_{\mu\nu}=
\begin{pmatrix}
-\,N^2\left(1+2\alpha\right) & aN \vec{\beta}_i \\
a N \vec{\beta}_j & a^2\bigl[\left(1+2\psi\right) \gamma_{ij} + \vec{h}_{ij} \bigr]\,
\end{pmatrix}
\,.
\end{align}
This component $3+1$ form of the metric tensor in terms of the Latin indices is very useful in practice. In order to do so for the other quantities like torsion and nonmetricity, similar to Sec.~\ref{sec:background}, e.g. \eqref{lower1} and \eqref{lower2}, we introduce the scale factors:
\begin{align}
\delta \vec{{\bf X}}_{\mu} \dd x^{\mu} &= a \delta \vec{{\bf X}}_i \dd x^i \,, \\
\delta \vec{{\bf X}}_{\mu\nu} \dd x^{\mu} \otimes \dd x^{\nu} &= a^2 \delta \vec{{\bf X}}_i \dd x^i \otimes \dd x^j\,, \\
\delta \vec{{\bf X}}_{\mu\nu\rho} \dd x^{\mu} \otimes \dd x^{\nu} \otimes \dd x^{\rho}&= a^3 \delta \vec{{\bf X}}_{ijk} \dd x^i \otimes \dd x^j \otimes \dd x^k \,,
\end{align}
where the bold letters will be used to denote the set of all the perturbation quantities of the metric, torsion and nonmetricity tensors with the same number of indices. Using then similar expressions to \eqref{torsion1}-\eqref{torsion3} and \eqref{Q1split}-\eqref{Q4split} for the perturbations $\delta T^{\lambda}{}_{\mu\nu}$ and $\delta Q_{\lambda\mu\nu}$, together with the background quantities \eqref{background_torsion_modes} and \eqref{background_nonmetricity_modes}, we obtain:
\begin{eqnarray}
 T^{\lambda}{}_{\mu\nu}&\longrightarrow& \arraycolsep=1.4pt\def\arraystretch{2.2}\left\{\begin{array}{l}
 T^{0}{}_{0i}
 =\displaystyle -\,\frac{a}{3}\bigl(\vec{T}_{i}+3\vec{B}_{i}\bigr) \\
      T^{0}{}_{ij}=\displaystyle -\,\frac{a^2}{6N}\varepsilon_{ijk}\big(\vec{\mathcal{S}}^{k}+3\vec{\mathcal{B}}^{k}\big)\\
    T^{i}{}_{0j} = 
  \displaystyle -\,N\Bigl[\vec{A}\,^{i}{}_{j}-\frac{1}{3}\delta^{i}\,_{j} \left(\bar{\phi}+\phi\right) +\frac{1}{6}\varepsilon^i{}_{jk}\Bigl(\vec{\mathcal{S}}^{k}-\frac{3}{2}\vec{\mathcal{B}}^{k}\Bigr)\Bigr]\\
    T^{i}{}_{jk}=\displaystyle -\,\frac{a}{6}\Bigl\{4\delta^{i}\,_{[j} \vec{T}_{k]}-6\delta^i{}_{[j} \vec{B}_{k]}+\Bigl[3\varepsilon_{jkl}\vec{\mathcal{A}}^{il}-\varepsilon^{i}\,_{jk}\left(\bar{\varrho}+\varrho\right)\Bigr] \Bigr\}
   \end{array}\right.\,,
\end{eqnarray}
and
\begin{eqnarray}
 Q_{\lambda\mu\nu}&\longrightarrow& \arraycolsep=1.4pt\def\arraystretch{2.2}\left\{\begin{array}{l}
 Q_{000}=-\,\displaystyle N^3\Bigl[\bar{\theta}+\theta+\frac{3}{4}\left(\bar{\sigma}+\sigma\right)-\left(\bar{\xi}+\xi\right)\Bigr]  \\
 Q_{00i}=\displaystyle -\,\frac{N^{2}a}{2}\Bigl(\vec{\Lambda}_{i}-\frac{1}{3}\vec{Y}_{i}-2\vec{Z}_{i}\Bigr)\\
 Q_{0ij} = \displaystyle -\,\frac{Na^2}{3}\Bigl\{\vec{Q}_{ij}-3\vec{\kappa}_{ij}-3\Big[\bar{\theta}+\theta-\frac{1}{4}\left(\bar{\sigma}+\sigma\right)+\frac{1}{3}\left(\bar{\xi}+\xi\right)\Bigr]\gamma_{ij}\Bigl\}
   \\
 Q_{i00} = \displaystyle -\,N^2a\Bigl(\vec{W}_{i}-\frac{1}{4}\vec{\Lambda}_{i}+\frac{1}{3}\vec{Y}_{i}-\vec{Z}_{i}\Bigr) \\
 Q_{i0j} = \displaystyle \frac{Na^2}{6}\Big\{\bigl[\vec{Q}_{ij}+6\vec{\kappa}_{ij}+\bigl(3\left(\bar{\sigma}+\sigma\right)+2\left(\bar{\xi}+\xi\right)\bigr)\gamma_{ij}\bigl]-3\varepsilon_{ijk}\vec{\mathcal{Y}}^{k}\Big\} \\
 Q_{ijk} =  \displaystyle a^3\Big\{\vec{C}_{ijk}+\frac{1}{4}\bigl(4\vec{W}_{i}-\vec{\Lambda}_{i}\bigr)\gamma_{jk}+\gamma_{i(j}\vec{\Lambda}_{k)}-\frac{1}{6}\bigl(\gamma_{jk}\vec{Y}_{i}-\gamma_{i(j}\vec{Y}_{k)}\bigr)+\frac{3}{5}\gamma_{(ij}\vec{Z}_{k)}+\frac{2}{3}\varepsilon_{li(j}\vec{\mathcal{Q}}^{\,l}{}_{k)}\Big\}
   \end{array}\right.\,,
\end{eqnarray}
where quantities denoted by bar refer to the background. Note that for notational convenience in the calculations, we do not add ``$\delta$'' to the spatial perturbations in the geometry sector. This will not bring any confusion since: i) the spatial quantities with spin higher than zero do not have any background values as shown in \eqref{spatial-BG}, ii) we have decomposed the spin zero quantities, that may have nonvanishing background values \eqref{scalar-BG}, to the background and perturbation parts as $\phi_{\rm tot}=\bar{\phi}+\phi$ and so on. In summary, we have the following set of perturbation variables in the generic MAG:
\begin{align}
\delta \vec{{\bf X}}&= \{\alpha, \psi, \phi, \varrho, \theta, \sigma, \xi \}
\,, \\
\delta \vec{{\bf X}}_i&=\{ \vec{\beta}_i , \vec{T}_{i}, \vec{\mathcal{S}}_{i}, \vec{B}_{i}, \vec{\mathcal{B}}_{i}, \vec{W}_{i}, \vec{\Lambda}_{i}, \vec{Y}_{i}, \vec{\mathcal{Y}}_{i}, \vec{Z}_{i} \}
\,, \\
\delta \vec{{\bf X}}_{ij}&=\{\vec{h}_{ij}, \vec{A}_{ij}, \vec{\mathcal{A}}_{ij}, \vec{Q}_{ij}, \vec{\mathcal{Q}}_{ij}, \vec{\kappa}_{ij} \}
\,, \\
\delta \vec{{\bf X}}_{ijk}&=\{\vec{C}_{ijk} \}
\,.
\end{align}
Each element of $\delta \vec{{\bf X}}_i, \delta \vec{{\bf X}}_{ij}$ and $\delta \vec{{\bf X}}_{ijk}$ have 3, 5 and 7 independent components, respectively. The 24 d.o.f. of the torsion perturbations are distributed as $3+3+9+9$ d.o.f. among the above components, whereas the 40 d.o.f. of the nonmetricity perturbations are split as $1+3+6+3+9+18$ d.o.f..
Thereby, under the $3+1$ decomposition of the space-time, these tensors are split into a large number of d.o.f., which correspond to representations of the three-dimensional orthogonal group and thus can be further classified around FLRW geometries according to their spin and parity, whose leading values are displayed in Table~\ref{Table:spinparity}.

\begin{table}[H]
    \centering
    \begin{tabular}{|l|c|c|c|c|c|c|c|}
        \hline
        \textbf{Spin and Parity} & $\boldsymbol{0^+}$ & $\boldsymbol{0^-}$ & $\boldsymbol{1^+}$ & $\boldsymbol{1^-}$ & $\boldsymbol{2^+}$ & $\boldsymbol{2^-}$ & $\boldsymbol{3^-}$ \\ \hline
        \textbf{Metric sector} $g_{\mu\nu}$ & $\alpha, \psi$ & - & - & $\vec{\beta}_i$ & $\vec{h}_{ij}$ & - & - \\ \hline
        \textbf{Torsion sector} $T^\lambda{}_{\mu\nu}$ & $\phi$ & $\varrho$ & $\vec{\mathcal{S}}_{i},\vec{\mathcal{B}}_{i}$ & $\vec{T}_{i},\vec{B}_{i}$ & $\vec{A}_{ij}$ & $\vec{\mathcal{A}}_{ij}$ & - \\ \hline
        \textbf{Nonmetricity sector} $Q_{\lambda\mu\nu}$ & $\theta,\sigma,\xi$ & - & $\vec{\mathcal{Y}}_{i}$ & $\vec{W}_{i},\vec{\Lambda}_{i},\vec{Y}_{i},\vec{Z}_{i}$ & $\vec{Q}_{ij}, \vec{\kappa}_{ij}$ & $\vec{\mathcal{Q}}_{ij}$ & $\vec{C}_{ijk}$ \\ \hline
    \end{tabular}
    \caption{Species in MAG.}
    \label{Table:spinparity}
\end{table}

\subsubsection{SVT decomposition}\label{SVT}

After the $3+1$ decomposition, we only need to deal with the spatial scalar and tensors with Latin indices belonging to the spatial maximally symmetric space with metric $\gamma_{ij}$ (and with the ``external" parameter $t$). We shall use $\DD_i$ to denote the spatial Riemannian covariant derivative compatible with $\gamma_{ij}$.

While the tensors $\delta \vec{{\bf X}}_i, \delta \vec{{\bf X}}_{ij}$ and $\delta \vec{{\bf X}}_{ijk}$ have a large number of components, the symmetries of the FLRW space-time allow us to significantly simplify the problem: spatial objects with different helicities evolve completely independently at the linear level of perturbations \cite{Kodama:1984ziu}. In this regard, let us decompose perturbations into different helicity sectors:
\begin{align}
\delta \vec{{\bf X}}_i& = \DD_i \mathbf{S} + \mathbf{V}^{(\rm T)}_i \,, \label{vectorSVT}\\
\delta \vec{{\bf X}}_{ij} &=\Big(\DD_{(i}\DD_{j)} -\frac{1}{3}\gamma_{ij}\DD^2 \Big)\mathbf{S}+\DD_{(i} \mathbf{V}^{(\rm T)}_{j)}+ \mathbf{T}^{(\rm TT)}_{ij} \,, \label{tensorSVT}\\
\delta \vec{{\bf X}}_{ijk} &= \Bigl[ \DD_{(i}\DD_j \DD_{k)} -\frac{1}{5}\gamma_{(ij}\DD_{k)} \bigl(3\DD^2+4K\bigr) \Bigr] \mathbf{S} 
+ \Bigl[ \DD_{(i} \DD_{j} -\frac{1}{5}\gamma_{(ij} \bigl(\DD^2+2K\bigr)\Bigr] \mathbf{V}^{(\rm T)}_{k)}
+\DD_{(i} \mathbf{T}^{(\rm TT)}_{jk)}+\mathbf{TT}^{(\rm TT)}_{ijk}
\,,\label{spin3SVT}
\end{align}
where $\mathbf{S}$, $\mathbf{V}^{(\rm T)}_i$, $\mathbf{T}^{(\rm TT)}_{ij}$, and $\mathbf{TT}^{(\rm TT)}_{ijk}$ denote a scalar (rank-0 spatial tensor), a vector (rank-1 spatial tensor), a rank-2 spatial tensor, and a rank-3 spatial tensor respectively. Here, the superscript ``$(\rm T)$" means that the quantity is transverse, while the tensors denoted by the superscript ``$(\rm TT)$" are transverse-traceless and symmetric:
\begin{align}
\DD^i \mathbf{V}^{(\rm T)}_{i}&=0\,, \\
\DD^i \mathbf{T}^{(\rm TT)}_{ij}&=\mathbf{T}^{(\rm TT)}_{ij}\gamma^{ij}=0\,, \quad \mathbf{T}^{(\rm TT)}_{ij}=\mathbf{T}^{(\rm TT)}_{(ij)}\,, \\
\DD^i  \mathbf{TT}^{(\rm TT)}_{ijk}&=\mathbf{TT}^{(\rm TT)}_{ijk}\gamma^{ij}=0\,, \quad
\mathbf{TT}^{(\rm TT)}_{ijk}=\mathbf{TT}^{(\rm TT)}_{(ijk)}\,.
\end{align}
Given the fact that the spatial Riemannian covariant derivatives do not commute in the presence of spatial curvature, i.e. $[\DD_i,\DD_j]X^k=  {}^{(3)}\!R^k{}_{pij}X^p$ with  ${}^{(3)}\!R_{ijkl}=2K\gamma_{k[i}\gamma_{j]l}$, as is shown, the SVT decomposition of a spatially rank-3 symmetric and traceless tensor requires contributions from the Riemannian Ricci tensor and the Riemannian Ricci scalar, in order to maintain the traceless property of such a tensor. The modes $\mathbf{S}, \mathbf{V}^{(\rm T)}_i, \mathbf{T}^{(\rm TT)}_{ij},$ and $\mathbf{TT}^{(\rm TT)}_{ijk}$ describe the helicity $0, \pm1, \pm2$ and $\pm3$ modes, and contain 1, 2, 2 and 2 d.o.f., respectively. The uniqueness of the helicity decomposition, also called the SVT decomposition being short for the Scalar-Vector-Tensor decomposition\footnote{In the present case, we have the rank-3 tensor in addition to the conventional scalar, vector, and (rank-2) tensor modes.}, on the maximally symmetric space is studied in Appendix~\ref{appendix3} (see~\cite{Kodama:1984ziu} for the proof up to the rank-2 tensor). In particular, the uniqueness of this decomposition guarantees that the different helicity sectors are decoupled at the linearised equations of motion. One can also confirm that the different sectors are certainly decoupled at the quadratic order action: any scalar quantities constructed by a pair of different helicity sectors, the metric $\gamma^{ij}$, the spatial Levi-Civita tensor $\varepsilon^{ijk}$, and the covariant derivative $\DD_i$ vanish up to the freedom of integration by parts (recall that the maximally symmetric space only allows the metric and the Levi-Civita tensor as the background tensors). For instance, a scalar $(\DD^i \mathbf{S}) \mathbf{V}^{(\rm T)}_i = \DD^i (\mathbf{S} \mathbf{V}^{(\rm T)}_i)- \mathbf{S} \DD^i \mathbf{V}^{(\rm T)}_i $ is apparently nonzero but is just a boundary term thanks to the transverse condition; thus, it does not contribute to the equations of motion. All in all, if one is interested in the linearised equations of motion (or, equivalently, the quadratic action) around the FLRW space-time, one can separately discuss the different helicity sections\footnote{In other words, a coupling between different helicity modes arises only when the background breaks the symmetries of the FLRW space-time and/or when nonlinear interactions are taken into account.}.

The perturbations of each helicity mode of the metric, torsion and nonmetricity tensors are then given as follows:\begin{itemize}
	\item Helicity-3 modes:
	\begin{align}
	\delta g_{\mu\nu}&= \delta T^{\lambda}{}_{\mu\nu}=0
	\,, \\
	\delta Q_{\lambda\mu\nu} & \longrightarrow
	\begin{dcases}
	\delta Q_{000}
	&\!\!\!\! =\delta  Q_{00i}=\delta Q_{0ij}=
	\delta Q_{i00}= \delta Q_{i0j}=0\,,\\
	\delta Q_{ijk} &\!\!\!\! =a^3 C^{(\rm TT)}_{ijk}
	\,.
	\end{dcases}
	\end{align}
	\item Helicity-2 modes:
	\begin{eqnarray}\label{metric-pert-helicity2}
	\delta g_{\mu\nu}&\longrightarrow& \arraycolsep=1.4pt\def\arraystretch{2.2}\left\{\begin{array}{l}
	\delta g_{00}=\delta g_{0i}=0 \\
	\delta g_{ij}=a^2 h^{(\rm TT)}_{ij}
	\end{array}\right.\,, \\
	\delta T^{\lambda}{}_{\mu\nu}&\longrightarrow& \arraycolsep=1.4pt\def\arraystretch{2.2}\left\{\begin{array}{l}
	\delta T^{0}{}_{0i}=
	\delta  T^{0}{}_{ij}=0 
	\\
	\delta T^{i}{}_{0j} = 
	\displaystyle -\,NA^{(\rm TT)}{}^{i}{}_{j}
	\\
	\delta T^{i}{}_{jk}=\displaystyle -\,\frac{a}{2}\varepsilon_{jkl}\mathcal{A}^{(\rm TT)}{}^{il}
	\end{array}\right.\,, \\
	\delta Q_{\lambda\mu\nu}&\longrightarrow& \arraycolsep=1.4pt\def\arraystretch{2.2}\left\{\begin{array}{l}
	\delta Q_{000}=
	\delta  Q_{00i}=
	\delta Q_{i00}=0
	\\
	\delta Q_{0ij} = \displaystyle -\,\frac{Na^2}{3}\bigl(Q^{(\rm TT)}_{ij}-3\kappa^{(\rm TT)}_{ij}\bigr) \\
	\delta Q_{i0j} =\displaystyle \frac{Na^2}{6}\bigl(Q^{(\rm TT)}_{ij}+6\kappa^{(\rm TT)}_{ij}\bigr)\\
	\delta Q_{ijk} =\displaystyle a^3\Bigl(\DD_{(i} C^{(\rm TT)}_{jk)}+ \frac{2}{3}\varepsilon_{li(j}\mathcal{Q}^{(\rm TT)}_{k)}{}^{l}\Bigr) \displaystyle
	\end{array}\right.\,.
	\end{eqnarray}
	\item Helicity-1 modes:
	\begin{eqnarray}\label{metric-pert-helicity1}
	\delta g_{\mu\nu}&\longrightarrow& \arraycolsep=1.4pt\def\arraystretch{2.2}\left\{\begin{array}{l}
	\delta g_{00}=0 \\
	\delta g_{0i}=Na \beta^{(\rm T)}_i \\
	\delta g_{ij}=a^2 \DD_{(i} h^{(\rm T)}_{j)}
	\end{array}\right.\,, \\
	\delta T^{\lambda}{}_{\mu\nu}&\longrightarrow& \arraycolsep=1.4pt\def\arraystretch{2.2}\left\{\begin{array}{l}
	\delta T^{0}{}_{0i}
	=\displaystyle 
	-\,\frac{a}{3}\bigl(T_i^{(\rm T)}+3B_i^{(\rm T)}\bigr)\\
	\delta  T^{0}{}_{ij}=\displaystyle -\,\frac{a^2}{6N}\varepsilon_{ijk}\big(\mathcal{S}^{(\rm T)}{}^k+3\mathcal{B}^{(\rm T)}{}^k\big) \\
	\delta T^{i}{}_{0j} = 
	\displaystyle - \,
	N\Big[\gamma^{ki}\DD_{(k} A^{(\rm T)}_{j)}+\frac{1}{6}\varepsilon^i{}_{jk}\Bigl(\mathcal{S}^{(\rm T)}{}^k-\frac{3}{2}\mathcal{B}^{(\rm T)}{}^k\Bigr)\Big] \\
	\delta T^{i}{}_{jk}=\displaystyle -\,\frac{a}{6}\bigl(4\delta^{i}\,_{[j} T_{k]}^{(\rm T)}-6\delta^i{}_{[j} B_{k]}^{(\rm T)}+3\varepsilon_{jkl}\DD^{(i} \mathcal{A}^{(\rm T)}{}^{l)} \bigr)
	\end{array}\right.\,, \\
	\delta Q_{\lambda\mu\nu}&\longrightarrow& \arraycolsep=1.4pt\def\arraystretch{2.2}\left\{\begin{array}{l}
	\delta Q_{000}=0  \\
	\delta  Q_{00i}=\displaystyle -\,\frac{N^{2}a}{2}\Bigl(\Lambda^{(\rm T)}_i-\frac{1}{3} Y^{(\rm T)}_i-2Z^{(\rm T)}_i\Bigr)
	\\
	\delta Q_{0ij} =\displaystyle -\,\frac{Na^2}{3}\bigl(\DD_{(i}Q^{(\rm T)}_{j)}-3\DD_{(i}\kappa^{(\rm T)}_{j)}\bigr) \\
	\delta Q_{i00} =  \displaystyle -\,N^2a\Bigl(W^{(\rm T)}_i-\frac{1}{4}  \Lambda^{(\rm T)}_i+\frac{1}{3}Y^{(\rm T)}_i - Z^{(\rm T)}_i \Bigr) \\
	\delta Q_{i0j} = \displaystyle \frac{Na^2}{6}\Bigl(\DD_{(i} Q^{(\rm T)}_{j)}+6\DD_{(i} \kappa^{(\rm T)}_{j)}-3\varepsilon_{ijk} \mathcal{Y}^{(\rm T)}{}^k \Bigr) \\
	\delta Q_{ijk} = \displaystyle a^3\Bigl\{
	\Bigl[\DD_{(i} \DD_{j} -\frac{1}{5}\gamma_{(ij} \bigl(\DD^2+2K\bigr)\Bigr] C^{(\rm T)}_{k)}+\frac{1}{4}\bigl(4W^{(\rm T)}_i  - \Lambda ^{(\rm T)}_i \bigr)\gamma_{jk}+\gamma_{i(j}\Lambda^{(\rm T)}_{k)}\bigr.\\
	\displaystyle\,\,\,\,\,\,\,\,\,\,\,\,\,\,\,\,\,-\,\frac{1}{6}\bigl(\gamma_{jk}Y^{(\rm T)}_i-\gamma_{i(j} Y^{(\rm T)}_{k)}\bigr)+\frac{3}{5}\gamma_{(ij}Z^{(\rm T)}_{k)}+\frac{1}{3}\gamma^{ml}\bigl(\varepsilon_{lij}\DD_{(k} \mathcal{Q}^{(\rm T)}_{m)}+\varepsilon_{lik}\DD_{(j}\mathcal{Q}^{(\rm T)}_{m)}\bigr)\bigl.\Bigr\}
	\end{array}\right.\,.
	\end{eqnarray}
	\item Helicity-0 modes: \begin{eqnarray}\label{metric-pert-helicity0}
	\delta g_{\mu\nu}&\longrightarrow& \arraycolsep=1.4pt\def\arraystretch{2.2}\left\{\begin{array}{l}
	\delta g_{00}=-2\,N^2 \alpha \\
	\delta g_{0i}=Na \DD_i \beta \\
	\delta g_{ij}=\displaystyle a^2\Bigl[ 2\psi \gamma_{ij} + \Bigl(\DD_i \DD_j -\frac{1}{3}\gamma_{ij} \DD^2 \Bigr)h \Bigr]
	\end{array}\right.\,, \\
	\delta T^{\lambda}{}_{\mu\nu}&\longrightarrow& \arraycolsep=1.4pt\def\arraystretch{2.2}\left\{\begin{array}{l}
	\delta T^{0}{}_{0i}
	=\displaystyle  -\,\frac{a}{3}\bigl(\DD_{i}T+3\DD_{i}B\bigr)  \\
	\delta  T^{0}{}_{ij}=\displaystyle -\,\frac{a^2}{6N}\varepsilon_{ijk}\big(\DD^{k}\mathcal{S}+3\DD^{k}\mathcal{B}\big) \\
	\delta T^{i}{}_{0j} = \displaystyle - \,
	N\Big[\Big(\DD^{i}\DD_{j} -\frac{1}{3}\delta^i{}_j\DD^2 \Big)A-\frac{1}{3}\delta^{i}\,_{j} \phi +\frac{1}{6}\varepsilon^i{}_{jk}\Bigl(\DD^{k}\mathcal{S}-\frac{3}{2}\DD^{k}\mathcal{B}\Bigr)\Big] 
	\\
	\delta T^{i}{}_{jk}=\displaystyle -\,\frac{a}{6} \Bigl[4\delta^{i}\,_{[j} \DD_{k]}T-6\delta^i{}_{[j} \DD_{k]}B+3\varepsilon_{jkl}\Big(\DD^{i}\DD^{l} -\frac{1}{3}\gamma^{il}\DD^2 \Big)\mathcal{A}-\varepsilon^{i}\,_{jk}\, \varrho \Bigr]
	\end{array}\right.\,, \\
	\delta Q_{\lambda\mu\nu}&\longrightarrow& \arraycolsep=1.4pt\def\arraystretch{2.2}\left\{\begin{array}{l}
	\delta Q_{000}=-\,\displaystyle N^3\Bigl(\theta+\frac{3}{4}\sigma-\xi\Bigr)
	\\
	\delta  Q_{00i}=\displaystyle -\,\frac{N^{2}a}{2}\Bigl(\DD_i \Lambda-\frac{1}{3}\DD_i Y -2\DD_i Z \Bigr)
	\\
	\delta Q_{0ij} = \displaystyle -\,\frac{Na^2}{3}\Bigl[\Bigl(\DD_{i}\DD_{j} -\frac{1}{3}\gamma_{ij}\DD^2 \Bigr)Q-3\Bigl(\DD_{i}\DD_{j} -\frac{1}{3}\gamma_{ij}\DD^2 \Bigr)\kappa-3\Bigl(\theta-\frac{1}{4}\sigma+\frac{1}{3}\xi\Bigr)\gamma_{ij}\Bigl]  
	\\
	\delta Q_{i00} = \displaystyle -\,N^2a\Bigl(\DD_i W-\frac{1}{4}\DD_i \Lambda+\frac{1}{3}\DD_i Y -\DD_i Z \Bigr) \\
	\delta Q_{i0j} = \displaystyle \frac{Na^2}{6}\Bigl[\Bigl(\DD_{i}\DD_{j} -\frac{1}{3}\gamma_{ij}\DD^2 \Bigr)Q+6\Bigl(\DD_{i}\DD_{j} -\frac{1}{3}\gamma_{ij}\DD^2 \Bigr)\kappa+\bigl(3\sigma+2\xi\bigr)\gamma_{ij}-3\varepsilon_{ijk}\DD^k \mathcal{Y}\Bigr] \\
	\delta Q_{ijk} =  \displaystyle a^3\Big\{\Bigl[\DD_{(i}\DD_j \DD_{k)} -\frac{1}{5}\gamma_{(ij}\DD_{k)} \bigl(3\DD^2+4K\bigr)\Bigr]C
	+\frac{1}{4}\bigl(4\DD_i W -\DD_i \Lambda\bigr)\gamma_{jk}+\gamma_{i(j}\DD_{k)} \Lambda  \\
	\displaystyle\,\,\,\,\,\,\,\,\,\,\,\,\,\,\,\,\,\,\,-\,\frac{1}{6}\bigl(\gamma_{jk}\DD_i Y-\gamma_{i(j}\DD_{k)} Y\bigr)+\frac{3}{5}\gamma_{(ij}\DD_{k)} Z+\frac{2}{3}\varepsilon_{li(j}\Bigl(\gamma^{ml}\DD_{k)}\DD_{m} -\frac{1}{3}\delta^l{}_{k)}\DD^2 \Bigr)\mathcal{Q}\Big\}\\
	\end{array}\right.\,.
	\end{eqnarray}
\end{itemize}
Thereby, the 10 d.o.f. described by the metric perturbations are split in terms of four scalars $\{\alpha,\beta,\psi,h\}$ (1 d.o.f. each), two transverse vectors $\{\beta^{(\rm T)}_i,h^{(\rm T)}_i\}$ (2 d.o.f. each), and one symmetric and transverse-traceless tensor $h^{(\rm TT)}_{ij}$ (2 d.o.f.); in Table~\ref{Table:Torsion} and Table~\ref{Table:Q}, we list all the helicity modes appearing in the torsion and nonmetricity perturbations and their number of d.o.f..

\begin{table}[H]
	\centering
	\begin{tabular}{|l|c|c|c|}\hline
		\textbf{\hspace{1.2cm}SVT} & \textbf{Quantities} & \textbf{d.o.f.}& \textbf{Total d.o.f.}\\ \hline
		\textrm{4 scalars } &$\{T,B,\phi,A\}$  & \textrm{1 d.o.f. each} & \textrm{4}\\ \hline
		\textrm{4 pseudoscalars } &$\{\mathcal{S},\mathcal{B},\varrho,\mathcal{A}\}$  & \textrm{1 d.o.f. each} & \textrm{4}\\ \hline
		\textrm{3 vectors } &$\{T_i^{(\rm T)},B_i^{(\rm T)},A_i^{(\rm T)}\}$  & \textrm{2 d.o.f. each} & \textrm{6}\\ \hline
		\textrm{3 pseudovectors } &$\{\mathcal{S}_i^{(\rm 
			T)},\mathcal{B}_i^{(\rm T)},\mathcal{A}_i^{(\rm T)}\}$  & \textrm{2 d.o.f. each} & \textrm{6}\\ \hline
		\textrm{1 rank-2 tensor } &$\{A_{ij}^{(\rm TT)}\}$  & \textrm{2 d.o.f. each} & \textrm{2}\\ \hline
		\textrm{1 rank-2 pseudotensor } &$\{\mathcal{A}_{ij}^{(\rm TT)}\}$  & \textrm{2 d.o.f. each} & \textrm{2}\\ \hline
	\end{tabular} 
	\caption{Helicity decomposition of the perturbations for the torsion tensor.}
	\label{Table:Torsion}
\end{table}

\begin{table}[H]
	\centering
	\begin{tabular}{|l|c|c|c|}\hline
		\textbf{\hspace{1.2cm}SVT} & \textbf{Quantities} & \textbf{d.o.f.}& \textbf{Total d.o.f.}\\ \hline
		\textrm{10 scalars } &$\{\theta,\sigma,\xi,\Lambda,Y,Z,\kappa,Q,W,C\}$  & \textrm{1 d.o.f. each} & \textrm{10}\\ \hline
		\textrm{2 pseudoscalars } &$\{\mathcal{Y},\mathcal{Q}\}$  & \textrm{1 d.o.f. each} & \textrm{2}\\ \hline
		\textrm{7 vectors } &$\{\Lambda^{(\rm T)}_i,Y^{(\rm T)}_i,Z^{(\rm T)}_i,\kappa^{(\rm T)}_i,Q^{(\rm T)}_i,W^{(\rm T)}_i,C^{(\rm T)}_i\}$  & \textrm{2 d.o.f. each} & \textrm{14}\\ \hline
		\textrm{2 pseudovectors } &$\{\mathcal{Y}^{(\rm T)}_i,\mathcal{Q}^{(\rm T)}_i\}$  & \textrm{2 d.o.f. each} & \textrm{4}\\ \hline
		\textrm{3 rank-2 tensor } &$\{\kappa^{(\rm TT)}_{ij},Q^{(\rm TT)}_{ij},C^{(\rm TT)}_{ij}\}$  & \textrm{2 d.o.f. each} & \textrm{6}\\ \hline
		\textrm{1 rank-2 pseudotensor } &$\{\mathcal{Q}^{(\rm TT)}_{ij}\}$  & \textrm{2 d.o.f. each} & \textrm{2}\\ \hline
		\textrm{1 rank-3 tensor } &$\{C^{(\rm TT)}_{ijk}\}$  & \textrm{2 d.o.f. each} & \textrm{2}\\ \hline
	\end{tabular} 
	\caption{Helicity decomposition of the perturbations for the nonmetricity tensor.}
	\label{Table:Q}
\end{table}

\subsubsection{Fourier space}

After performing the SVT decomposition, it is convenient to move to the momentum space and work with Fourier amplitudes of the perturbations. For scalars, we have\footnote{For the closed universe, the momentum is discrete and the integral is replaced with the summation.}
\begin{align}
\mathbf{S}(t, \vec{x}) = \frac{1}{(2\pi)^3} \int \dd^3k \, \frac{1}{\sqrt{2}} \mathbf{S}(t, \vec{k}) E(\vec{x} ; \vec{k}) + {\rm c.c.}\,,
\end{align}
with real functions $\mathbf{S}(t, \vec{k})$ and where $E$ are the orthogonal eigenstates of $\gamma_{ij}$:
\begin{align}
\DD^2 E = -\,k^2 E \,, \quad \int \dd x^3 \sqrt{\gamma} \, E^*(\vec{x} ; \vec{k}) E(\vec{x} ; \vec{k}') = (2\pi)^3 \delta^{(3)}(\vec{k}-\vec{k}')
\,,
\end{align}
being $k$ the norm of the momentum. We have chosen the normalisation so that
\begin{align}
\int \dd^3 x \sqrt{\gamma} \, [\mathbf{S}(t, \vec{x})]^2 = \int \frac{\dd^3 k}{(2\pi)^3} [\mathbf{S}(t, \vec{k})]^2
\,.
\label{quadraticS}
\end{align}
The helicity-1, 2, and 3 modes can be similarly mapped to the momentum space but each helicity still contains two different modes. In MAG, it would be particularly convenient to characterise them by using the circular polarisation bases $ E^{(A)}_i, E^{(A)}_{ij},$ and $ E^{(A)}_{ijk}~({A}=L, R)$~\cite{Aoki:2019snr, Aoki:2020zqm}; they are defined by the eigenstates of the operators
\begin{align}
\DD^2 E^{(A)}_i &= -\,k^2 E^{(A)}_i\,, \quad \varepsilon_{i}{}^{pq} \DD_p E^{(A)}_{q} = \lambda_A k E^{(A)}_i \,, \\
\DD^2 E^{(A)}_{ij} &= -\,k^2 E^{(A)}_{ij}\,, \quad \varepsilon_{(i}{}^{pq} \DD_p E^{(A)}_{q|j)} = \lambda_A k E^{(A)}_{ij} \,, \\
\DD^2 E^{(A)}_{ijk} &= -\,k^2 E^{(A)}_{ijk}\,, \quad \varepsilon_{(i}{}^{pq} \DD_p E^{(A)}_{q|jk)} = \lambda_A k E^{(A)}_{ijk} \,,
\end{align}
with $\lambda_L=-\,1, \lambda_R=+\,1$, and satisfy the orthogonal conditions
\begin{align}
\int \dd^3x \sqrt{\gamma} [E^{(A)}_{i}(\vec{x};\vec{k})]^* E^{(A')}_{i' }(\vec{x};\vec{k}') \gamma^{ii'}  &=(2\pi)^3 \delta_{AA'}\delta^{(3)}(\vec{k}-\vec{k}') 
\,, \\
\int \dd^3x \sqrt{\gamma} [E^{(A)}_{ij}(\vec{x};\vec{k})]^* E^{(A')}_{i'j' }(\vec{x};\vec{k}') \gamma^{ii'}\gamma^{jj'}  &=(2\pi)^3 \delta_{AA'}\delta^{(3)}(\vec{k}-\vec{k}') 
\,, \\
\int \dd^3x \sqrt{\gamma} [E^{(A)}_{ijk}(\vec{x};\vec{k})]^* E^{(A')}_{i'j'k' }(\vec{x};\vec{k}') \gamma^{ii'}\gamma^{jj'}\gamma^{kk'}  &=(2\pi)^3 \delta_{AA'}\delta^{(3)}(\vec{k}-\vec{k}') 
\,.
\end{align}
We omit the superscripts ``(T)'' and ``(TT)'' for notational simplicity, but note that $ E^{(A)}_i$ is transverse, and $\{ E^{(A)}_{ij}, E^{(A)}_{ijk} \}$ are transverse-traceless and symmetric in all the indices. Also, one should not confuse the spatial index $k$ with the momentum $k$; the difference between them is clear from the context. Note also that thanks to the isotropy of the FLRW space-time, the final results depend on the norm $k$ but do not depend on the specific direction $\vec{k}/k$.

While we have not classified the quantities with respect to the parity so far, we now take the parity into account. We collectively write them as
\begin{align}
\bm{S}&=\{ \alpha, \beta, \psi, h, 
\phi, T, B, A,
\theta,  \sigma,  \xi, W, \Lambda, Y, Z, Q, \kappa, C\}
\,, \\
\bm{\mathcal{S}}&= \{ \varrho, \mathcal{S}, \mathcal{B}, \mathcal{A}, \mathcal{Y}, \mathcal{Q} \}
\,, \\
\bm{V}^{(\rm T)}_i&=\{ \beta^{(\rm T)}_i, h^{(\rm T)}_i, T^{(\rm T)}_i, B^{(\rm T)}_i, A^{(\rm T)}_i, W^{(\rm T)}_i, \Lambda^{(\rm T)}_i, Y^{(\rm T)}_i, Z^{(\rm T)}_i, Q^{(\rm T)}_i, \kappa^{(\rm T)}_i, C^{(\rm T)}_i 
\}
\,, \\
\bm{\mathcal{V}}^{(\rm T)}_i &= \{ \mathcal{S}^{(\rm T)}_i, \mathcal{B}^{(\rm T)}_i, \mathcal{A}^{(\rm T)}_i, \mathcal{Y}^{(\rm T)}_i, \mathcal{Q}^{(\rm T)}_i
\}
\,, \\
\bm{T}^{(\rm TT)}_{ij} &= \{ h^{(\rm TT)}_{ij} , A^{(\rm TT)}_{ij} , Q^{(\rm TT)}_{ij} , \kappa^{(\rm TT)}_{ij}  
\}
\,, \\
\bm{\mathcal{T}}^{(\rm TT)}_{ij}  &= \{
\mathcal{A}^{(\rm TT)}_{ij} , \mathcal{Q}^{(\rm TT)}_{ij} 
\}
\,, \\
\bm{TT}^{(\rm TT)}_{ijk} &= \{ C^{(\rm TT)}_{ijk} \}
\,.
\end{align}
Then, the corresponding Fourier transformations are
\begin{align}
\bm{S}(t, \vec{x}) &= \frac{1}{(2\pi)^3} \int \dd^3k \, \frac{1}{\sqrt{2}} \bm{S}(t, \vec{k}) E(\vec{x} ; \vec{k}) + {\rm c.c.}\,, 
\label{FourierS} \\
\bm{\mathcal{S}}(t, \vec{x}) &= \frac{1}{(2\pi)^3} \int \dd^3k \, \frac{1}{\sqrt{2}} \bm{\mathcal{S}}(t, \vec{k}) E(\vec{x} ; \vec{k}) + {\rm c.c.}\,, 
\label{FourierSc} \\
\bm{V}^{(\rm T)}_i (t, \vec{x}) &= \frac{1}{(2\pi)^3} \int \dd^3k \, \frac{1}{\sqrt{2}} [ \bm{V}^{(L)} (t, \vec{k}) E^{(L)}_i (\vec{x} ; \vec{k}) +\bm{V}^{(R)} (t, \vec{k}) E^{(R)}_i (\vec{x} ; \vec{k}) ]+ {\rm c.c.}\,, \\
\bm{\mathcal{V}}^{(\rm T)}_i (t, \vec{x}) &= \frac{1}{(2\pi)^3} \int \dd^3k \, \frac{1}{\sqrt{2}} [ \bm{\mathcal{V}}^{(L)} (t, \vec{k}) E^{(L)}_i (\vec{x} ; \vec{k}) - \bm{\mathcal{V}}^{(R)} (t, \vec{k}) E^{(R)}_i (\vec{x} ; \vec{k}) ] + {\rm c.c.}\,,
\end{align}
and the transformations of the rank-2 and 3 tensors are performed similarly to the vectors. The quadratic form of the (pseudo)-scalar is explained in \eqref{quadraticS} whereas that of the (pseudo)-vector is
\begin{align}
\int \dd^3 x \sqrt{\gamma}\, [\bm{V}^{(\rm T)}_i (t, \vec{x}) ]^2
=\int \frac{\dd^3k}{(2\pi)^3}
\left\{ [\bm{V}^{(L)} (t, \vec{k}) ]^2+[\bm{V}^{(R)} (t, \vec{k})]^2\right\}
\,,
\end{align}
where a contraction of the indices is understood.
Note that we have added the minus sign in front of the right-handed mode of the parity odd sector to systematically deal with couplings between the parity even and odd sectors (see \cite{Aoki:2019snr, Aoki:2020zqm}).

For illustrative purposes, let us explicitly display the circular polarisation bases in the flat FLRW universe $K=0$ with the Cartesian metric $\gamma_{ij}=\delta_{ij}$. The scalar harmonics are nothing but the exponential function
\begin{align}
E=e^{i \vec{k}\cdot \vec{x}}
\,.
\end{align}
By the use of the isotropy of the FLRW space-time, we can choose $\vec{k}=(0,0,k)$ without loss of generality. We then find
\begin{align}
E^{(A)}_i= e^{(A)}_i e^{i \vec{k}\cdot \vec{x}}\,,
\end{align}
with
\begin{align}
e^{(L)}_i =\bigl(1/\sqrt{2}, -\, i/\sqrt{2}, 0\bigr) 
\,, \quad
e^{(R)}_i =\bigl(1/\sqrt{2},  i/\sqrt{2}, 0\bigr) 
\,.
\end{align}
As $e^{(A)}_i$ satisfies $e^{(A)}_i e^{(A)}_j \delta^{ij}=0$, the bases of the helicity-2 and 3 modes are straightfowardly found to be
\begin{align}
E^{(A)}_{ij} &= e^{(A)}_i e^{(A)}_j e^{i \vec{k}\cdot \vec{x}} \,, \\
E^{(A)}_{ijk} &= e^{(A)}_i e^{(A)}_j e^{(A)}_k e^{i \vec{k}\cdot \vec{x}} 
\,.
\end{align}
One can explicitly confirm that $ E^{(A)}_i$, $E^{(A)}_{ij}$ and $E^{(A)}_{ijk}$ satisfy all the desired properties.

\subsection{Matter perturbations}\label{subsec:matter-perturbations}
For the matter perturbations, we have to consider perturbations around the background matter configuration that is described in Sec.~\ref{secmatter1}.  

The matter sector is characterised by the energy-momentum and hypermomentum tensors. Considering general perturbations around them
\begin{eqnarray}
\Theta_{\mu\nu}=\bar{\Theta}_{\mu\nu}+  \delta \Theta_{\mu\nu}\,,\quad  \Delta_{\mu\nu\lambda}=\bar{\Delta}_{\mu\nu\lambda}+ \delta \Delta_{\mu\nu\lambda}\,,
\end{eqnarray}
the aim is to find SVT decomposition of $\delta \Theta_{\mu\nu}$ and $\delta \Delta_{\mu\nu\lambda}$. 

Let us first look at the perturbations around the energy-momentum tensor which is quite well-known in the context of cosmological perturbation theory. In $3+1$ decomposition, the components are:
\begin{eqnarray}
\delta \Theta_{\mu\nu}\longrightarrow\arraycolsep=1.4pt\def\arraystretch{2.2}\left\{\begin{array}{l}
\delta \Theta_{00}= N^2 \delta\rho - \bar{\rho} \delta g_{00} \\
\delta \Theta_{0i}= - \,N(\bar{\rho}+\bar{p})\delta \vec{u}_i+\bar{p}\delta g_{0i}\\
\delta \Theta_{ij}= \displaystyle\bar{p}\delta g_{ij} + a^2 \left( \gamma_{ij}\delta p + \vec{\pi}_{ij} \right) \\
\end{array}\right.\,,\label{deltaTmunu}
\end{eqnarray}
where $\delta\rho$ and $\delta{p}$ are the perturbations in the energy density and pressure respectively, $\delta\vec{ u}_i$ is the spatial velocity vector, which is defined in the four-velocity of the fluid as $u_{\mu}={\bar u}_\mu+\delta{u}_\mu$ with ${\bar u}_\mu={\bar n}_\mu$ and $\delta{u}_\mu=({\delta u}_0,\delta \vec{u}_i)$, and $\vec{\pi}_{ij}$ is the spatial stress tensor. Note that in obtaining the above result we have used $\delta{u}_0=\delta{g}_{00}/2N$ that can be deduced from $g^{\mu\nu}u_\mu {u}_\nu = -\,1$. 

The SVT decomposition of the energy-momentum tensor is now straightforward. The SVT decomposition of the metric perturbations are already presented in \eqref{metric-pert-helicity2}, \eqref{metric-pert-helicity1}, and \eqref{metric-pert-helicity0} and the SVT decomposition of the spatial velocity vector and the spatial stress tensor can be read from \eqref{vectorSVT} and \eqref{tensorSVT}:
\begin{align}
\delta\vec{u}_i& = \DD_i \delta{u} + \delta{u}^{(\rm T)}_i \,, \label{du-SVT}\\
\vec{\pi}_{ij} &=\Big(\DD_{(i}\DD_{j)} -\frac{1}{3}\gamma_{ij}\DD^2 \Big)\pi + \DD_{(i} \pi^{(\rm T)}_{j)}+ \pi^{(\rm TT)}_{ij} \,. \label{pi-SVT}
\end{align}

On the other hand, the perturbation of the hypermomentum $\delta\Delta_{\mu\nu\lambda}$ can introduce up to 64 d.o.f. with the SVT decomposition behaving in the same way as the sum of the perturbation of torsion (24 d.o.f.) and nonmetricity (40 d.o.f.) as we described in the previous subsections. We thus do not explicitly present them here. As we noticed above, the matter perturbation for the energy-momentum tensor has a clear physical interpretation as the perturbation of the 4-velocity of the fluid and the anisotropic pressure. However, for the hypermomentum, still, it is unclear how to physically interpret all its perturbation d.o.f.. This could be interesting to analyse further in the future.

\section{Cosmological perturbations of the nonmetricity helicity-3 modes}\label{sec:spin3}
 
Based on the SVT decomposition theorem, the helicity-3 perturbation decouples from all of the modes with different helicities at the linear order of perturbations. Moreover, we only have two helicity-3 modes that are characterised by $C_{ijk}^{(\rm TT)}$. Thus, it is easy to perform linear perturbation analysis for the helicity-3 modes. For this task, we consider the most general parity-preserving gravitational action in MAG, constructed from four-dimensional invariant quantities of the curvature, torsion and nonmetricity tensors under the general linear group $GL(4,R)$ up to their quadratic order\footnote{Note that the action~\eqref{MAGLag} in general develops a ghost instability~\cite{Percacci:2020ddy}. Here, however, our aim is to study linear perturbations of the helicity-3 modes, keeping in mind that the absence of ghosts implies certain relations between the coefficients in the action.}
\begin{eqnarray}
S_{\rm g}&=&\frac{1}{16\pi}\int d^{4}x \sqrt{-g}\Bigl(\tilde{R}+a_{1}\tilde{R}^{2}+a_{2}\tilde{R}_{\lambda\rho\mu\nu}\tilde{R}^{\lambda\rho\mu\nu}+a_{3}\tilde{R}_{\lambda\rho\mu\nu}\tilde{R}^{\rho\lambda\mu\nu}+\,a_{4}\tilde{R}_{\lambda\rho\mu\nu}\tilde{R}^{\mu\nu\lambda\rho}+a_{5}\tilde{R}_{\lambda\rho\mu\nu}\tilde{R}^{\lambda\mu\rho\nu}
\Bigr.
\nonumber\\
&&
\left.
\Bigl.
\,\,\,\,\,\,\,\,\,\,\,\,\,\,\,\,\,\,\,\,\,\,\,\,\,\,\,\,\,\,\,\,\,\,\,\,\,\,\,\,\,\,\,\,+\,a_{6}\tilde{R}_{\lambda\rho\mu\nu}\tilde{R}^{\mu\lambda\rho\nu}+a_{7}\tilde{R}_{\rho\lambda\mu\nu}\tilde{R}^{\mu\lambda\rho\nu}+\,a_{8}\tilde{R}_{\mu\nu}\tilde{R}^{\mu\nu}+a_{9}\tilde{R}_{\mu\nu}\tilde{R}^{\nu\mu}+a_{10}\hat{R}_{\mu\nu}\hat{R}^{\mu\nu}+a_{11}\hat{R}_{\mu\nu}\hat{R}^{\nu\mu}
\Bigr.
\right.
\nonumber\\
&&
\left.
\Bigl.
\,\,\,\,\,\,\,\,\,\,\,\,\,\,\,\,\,\,\,\,\,\,\,\,\,\,\,\,\,\,\,\,\,\,\,\,\,\,\,\,\,\,\,\,+\,a_{12}\tilde{R}_{\mu\nu}\hat{R}^{\mu\nu}+\,a_{13}\tilde{R}_{\mu\nu}\hat{R}^{\nu\mu}+a_{14}\tilde{R}^{\lambda}\,_{\lambda\mu\nu}\tilde{R}^{\rho}\,_{\rho}\,^{\mu\nu}+a_{15}\tilde{R}^{\lambda}\,_{\lambda\mu\nu}\tilde{R}^{\mu\nu}+a_{16}\tilde{R}^{\lambda}\,_{\lambda\mu\nu}\hat{R}^{\mu\nu}
\Bigr.
\right.
\nonumber\\
&&
\left.
\Bigl.
\,\,\,\,\,\,\,\,\,\,\,\,\,\,\,\,\,\,\,\,\,\,\,\,\,\,\,\,\,\,\,\,\,\,\,\,\,\,\,\,\,\,\,\,+\,b_{1}T_{\lambda\mu\nu}T^{\lambda\mu\nu}+b_{2}T_{\lambda\mu\nu}T^{\mu\lambda\nu}+b_{3}T^{\lambda}\,_{\lambda\nu}T^{\mu}\,_{\mu}\,^{\nu}+c_{1}T_{\lambda\mu\nu}Q^{\mu\lambda\nu}+\,c_{2}T^{\lambda}\,_{\lambda\nu}Q^{\nu\mu}\,_{\mu}+c_{3}T^{\lambda}\,_{\lambda\nu}Q^{\mu\nu}\,_{\mu}
\Bigr.
\right.
\nonumber\\
&&\,\,\,\,\,\,\,\,\,\,\,\,\,\,\,\,\,\,\,\,\,\,\,\,\,\,\,\,\,\,\,\,\,\,\,\,\,\,\,\,\,\,\,\,\,
+\,d_{1}Q_{\lambda\mu\nu}Q^{\lambda\mu\nu}+d_{2}Q_{\lambda\mu\nu}Q^{\mu\lambda\nu}\Bigl.
+\,d_{3}Q^{\lambda}\,_{\lambda\nu}Q^{\mu}\,_{\mu}\,^{\nu}+d_{4}Q_{\nu}\,^{\lambda}\,_{\lambda}Q^{\nu\mu}\,_{\mu}+d_{5}Q^{\lambda}\,_{\lambda\nu}Q^{\nu\mu}\,_{\mu}\Bigr)\,,\label{MAGLag}
\end{eqnarray}
where all coefficients $a_i, b_i, c_i$, and $d_i$ are constant.

The helicity-3 modes are encoded in the nonmetricity tensor $Q_{\lambda\mu\nu}$, or more specifically, in the irreducible piece $q_{\lambda\mu\nu}$ defined in~\eqref{qtensor}. In Appendix~\ref{appendix4}, we have classified the action \eqref{MAGLag} in terms of $q_{\lambda\mu\nu}$, where the interaction of the spin-3 field with all the other fields can be explicitly seen. Here, before performing explicit calculations of perturbations, some comments are in order. First, the derivatives of the nonmetricity tensor appear linearly in the definition of the curvature tensor \eqref{totalcurvature} and, that is why we have considered quadratic terms in curvature in action \eqref{MAGLag}. Otherwise, nonmetricity will not be a dynamical quantity. Second, all terms with coefficients $b_i$ and $c_i$ in \eqref{MAGLag} do not contribute to the quadratic action of perturbations for the helicity-3 modes. Third, in principle, the terms with coefficients $d_i$ can generate mass for the helicity-3 modes in any background.

For the sake of simplicity, we will concentrate on the spatially flat $K=0$ FLRW background and we also work with the cosmic time $N=1$. Note that, as we have explained in Appendix~\ref{app-gauge}, the helicity-3 modes are gauge-invariant since they do not have any background value and the transformation parameters (encoded in the four-dimensional vector $\xi^\mu$ that is defined in Appendix~\ref{app-gauge}) involve only helicity-0 and 1 modes. We thus do not need to fix any gauge for them. Substituting SVT decomposition of all quantities, it is cumbersome but straightforward to show that the quadratic action of perturbations for the helicity-3 modes takes the following form:
\begin{eqnarray}
S_{\rm C}^{(2)}&=&\frac{1}{16\pi}\int \dd t \,\dd^{3}x \,a^3 \mathcal{G}_{T}^{(C)} \Biggl[ \left(\dot{C}^{(\rm TT)}_{ijk}\right)^2 -\frac{1}{a^2} \left(\DD_l C^{(\rm TT)}_{ijk}\right)^2 +\frac{4}{a}T_2 \varepsilon_{ijk}C^{(\rm TT)}{}^i{}_{lm}\DD^j C^{(\rm TT)}{}^{klm} - m_{C,{\rm eff}}^2 \left(C^{(\rm TT)}_{ijk}\right)^2 \Biggl]
\nonumber \\
&=& \frac{1}{16\pi}\sum_{A=L,R} \int \dd t \, \frac{\dd^3 k}{(2\pi)^3} \, a^3 \mathcal{G}_{T}^{(C)}
\Biggl[ [\dot{C}^{(A)}]^2 - [\omega^{(A)}_{C,k}]^2 [ C^{(A)}]^2
\Biggl] \,,
\label{spin3pert}
\end{eqnarray}
where a dot denotes derivative with respect to the cosmic time $t$ and we have defined the frequency as
\begin{align}\label{omega}
   &[\omega^{(A)}_{C,k}]^2 \equiv \frac{k^2}{a^2} - 4\lambda_A T_2 \frac{k}{a} + m_{C,{\rm eff}}^2 \,,
   \end{align}
   and also the constant kinetic coefficient and the time-dependent effective mass as
   \begin{align}\label{GT}
   \mathcal{G}_{T}^{(C)} \equiv-\,\frac{1}{4} (2 a_2+2 a_3+a_5+a_6+a_7)\,,\quad
   m_{C,{\rm eff}}^2 \equiv \frac{\mathcal{H}_{T}^{(C)}}{\mathcal{G}_{T}^{(C)}} \,,
   \end{align}
   with
   \begin{eqnarray}
   \mathcal{H}_{T}^{(C)}&\equiv&\frac{1}{4}-d_1-d_2+\frac{1}{2} \left[2\alpha_1-7\left(\alpha_2+\alpha_3\right)\right] \dot{H}+\left(2 \alpha_1-6\alpha_2-\alpha_3\right)H^2
   \nonumber\\
   &&+\left(7\alpha_2-3\alpha_1\right) T_{1} H
   +\frac{1}{2}\left[2\alpha_4 Q_1+\left(3\alpha_1-7\alpha_2\right)\left(2Q_2-Q_3\right)+\alpha_4 \left(2Q_2+Q_3\right)\right]H
   \nonumber\\
   && +\left(3\alpha_2+4\alpha_3-\alpha_1\right)\dot{T}_1+\left[\alpha_1+2\left(\alpha_3-\alpha_2\right)\right]T_1^2-\alpha_1 T_2^2
   \nonumber\\
   &&-\, \alpha_4 T_1 Q_1+\left(4\alpha_2-2\alpha_1-4\alpha_3-\alpha_4\right)T_{1}Q_{2} + \left[\alpha_1+2\left(\alpha_3-\alpha_2\right)\right]T_1 Q_3 
   \nonumber\\
   &&+\,\frac{1}{2}\bigl[\left(\alpha_1-3\alpha_2-4\alpha_3\right)\bigl(2\dot{Q}_2-\dot{Q}_3\bigr)+\alpha_4\dot{Q}_3\bigr]+\left(\alpha_1+2\alpha_3+\alpha_4-2\alpha_2\right)Q_2^2
   \nonumber\\
   &&+\,\frac{1}{2}\left[\left(\alpha_1-3\alpha_2-4 \alpha_3-\alpha_4\right)Q_1+\left(\alpha_2-\alpha_1-8 \alpha_3-\alpha_4\right)Q_2\right]Q_3+\alpha_4Q_1 Q_2 \,. \label{HT}
\end{eqnarray}
In the above, $H=\dot{a}/a$ denotes the Hubble parameter, whereas we have introduced the following constants:
\begin{eqnarray}
 \alpha_1&=&3a_{1}+3a_{2}+a_{3}+a_{4}+\frac{3a_{5}}{2}+\frac{a_{6}}{2}+\frac{3a_{7}}{2}+a_{8}+a_{9}+a_{10}+a_{11}+a_{12}+a_{13}\,,\\
 \alpha_2&=& \frac{1}{14}\left(14a_{2}+10a_{3}+2a_{4}+7 a_{5}+5a_{6}+7a_{7}+a_{8}+a_{9}+a_{10}+a_{11}+a_{12}+a_{13}\right)\,,\\
 \alpha_3&=&-\,\frac{1}{14}\left(4a_{3}-2a_{4}+2a_{6}-a_{8}-a_{9}-a_{10}-a_{11}-a_{12}-a_{13}\right)\,,\\
 \alpha_4&=&-\,\frac{1}{2}\left(a_8+a_9-a_{10}-a_{11}\right)\,,\quad \alpha_5=-\,\frac{1}{2}\left(2 a_{10}+2a_{11}+a_{12}+a_{13}\right)\,,\quad \alpha_6=4 \left(a_{10}+a_{11}\right)\,.
\end{eqnarray}

The action \eqref{MAGLag} determines the gravitational sector for the spin-3 field. In general, there can be spin-3 fields in the matter sector. Taking into account this point, variation of the quadratic action~\eqref{spin3pert} gives the following equations for the linear perturbations of the helicity-3 modes:
\begin{eqnarray}\label{EoMs-helicity3}
\ddot{C}^{(A)}+ 3H \dot{C}^{(A)} + [\omega^{(A)}_{C,k}]^2 {C}^{(A)} = \, {J}^{(A)} \,,
\end{eqnarray}
where ${J}^{(A)}$ describe the Fourier amplitudes of the transverse-traceless source provided by the matter sector. More precisely, at the level of equation of motion, ${J}^{(A)}$ are encoded in the helicity-3 modes of the intrinsic hypermomentum, such that ${J}^{(A)}\propto{}^{(\rm sh)}\order{{\nearrow\!\!\!\!\!\!\!\Delta}}{3}^{(\rm TT)(A) }$.

Having the quadratic action \eqref{spin3pert} in hand, we can look for the stability conditions. Avoiding ghost instability implies
\begin{align}\label{ghost-condition}
\mathcal{G}_{T}^{(C)} > 0 \,,
\end{align} 
which, as is clear from \eqref{GT}, sets constraints on some coefficients $a_i$ in the action. To avoid gradient and tachyonic instabilities, one may assume $[\omega^{(A)}_{C,k}]^2>0$, while this condition is too strong as we will discuss it in detail in the following.

The parity-violating term, that is linear in $k$, is present in the quadratic action \eqref{spin3pert} as far as torsion has a nonvanishing background $T_2\neq0$. Indeed, it originates from the covariant interaction $ \varepsilon_{\alpha \nu \lambda \mu } q^{\rho \tau \nu } S^{\alpha } \nabla^{\mu }q_{\rho \tau }{}^{\lambda }$ since $S^0 \propto T_2 \neq 0$ (see Appendix~\ref{appendix4}). While our original gravitational action \eqref{MAGLag} is parity-preserving, the background evolution of the universe breaks the time reflection symmetry and then $S^0 \propto T_2 \neq 0$ spontaneously breaks the parity invariance of the perturbations: the left-handed $\lambda_L=-\,1$ and right-handed $\lambda_R=+\,1$ modes of $C^{(\rm TT)}_{ijk}$ obey different equations of motion, as can be seen from \eqref{EoMs-helicity3}. In this sense, it is important to stress that the coupling constant of the parity-violating term is the same as the coefficient of the kinetic term. Hence, the parity invariance of the helicity-3 modes, if it is dynamical, is inevitably broken in a cosmological background with $T_2 \neq 0$ in the quadratic MAG. A similar observation can be found in the helicity-2 sector in Poincar\'{e} gauge theory~\cite{Aoki:2020zqm}. Moreover, the parity-violating term can only develop instability for one of the modes (the right-handed $\lambda_R=+\,1$ for $T_2>0$ and the left-handed $\lambda_R=-\,1$ for $T_2<0$) for low momenta that satisfy $[\omega^{(A)}_{C,k}]^2<0$. In fact, this type of infrared instability is very interesting in cosmology as it provides particle production (see e.g. \cite{Banks:1993en} and \cite{Sorbo:2011rz}). 

For the mass term, due to the Hubble friction, we do not need to assume $m_{C,{\rm eff}}^2>0$; even if the helicity-3 modes develop instabilities, the growth rate can be slow enough as long as the size is comparable to the Hubble rate ($|\mathcal{H}_{T}^{(C)}|\sim H^2$), similarly to the Jeans' instability \cite{Gumrukcuoglu:2016jbh}. It is also important to elaborate more on the different contributions to the effective mass of the spin-3 modes $m_{C,{\rm eff}}^2$ presented in \eqref{HT}. The first three terms in the RHS of \eqref{HT}, $1/4-d_1-d_2$, are present at any background while the other terms vanish in the absence of background curvature $H=0$, and background torsion and nonmetricity $T_i=Q_i=0$. Therefore, if $1/4-d_1-d_2=0$, the spin-3 field $q_{\mu\nu\lambda}$ is massless when $H=T_i=Q_i=0$ (the Minkowski vacuum) while it acquires nonvanishing mass if $H\neq0$, or $T_i\neq0$ or $Q_i\neq0$. As is well known, the massless spin-3 field presents the issue that it cannot have a consistent interaction preserving the gauge symmetry. The nonvanishing mass around nontrivial backgrounds might be indeed a sign of such an inconsistency.

Let us now analyse a particular MAG model, recently explored in~\cite{Bahamonde:2022kwg}, which interestingly displays the broadest family of static and spherically symmetric black hole solutions
with spin, dilation and shear charges in MAG. This model is characterised by a gravitational action with dynamical torsion, Weyl vector and traceless nonmetricity tensor\footnote{In~\cite{Bahamonde:2022kwg}, the signature $(+---)$ was used, so that the action in our signature $(-+++)$ shows an opposite sign in $R$ and also in $T_{[\lambda\mu\nu]}T^{[\lambda\mu\nu]}$.}
\begin{align}
S = &\,\frac{1}{64\pi}\int
\Bigl[
4R-6\tilde{d}_{1}\tilde{R}_{\lambda\left[\rho\mu\nu\right]}\tilde{R}^{\lambda\left[\rho\mu\nu\right]}-9\tilde{d}_{1}\tilde{R}_{\lambda\left[\rho\mu\nu\right]}\tilde{R}^{\mu\left[\lambda\nu\rho\right]}+2\tilde{d}_{1}\left(\tilde{R}_{[\mu\nu]}+\hat{R}_{[\mu\nu]}\right)\left(\tilde{R}^{[\mu\nu]}+\hat{R}^{[\mu\nu]}\right)
\Bigr.
\nonumber\\
\Bigl.
&+18\tilde{d}_{1}\tilde{R}_{\lambda\left[\rho\mu\nu\right]}\tilde{R}^{(\lambda\rho)\mu\nu}-3\tilde{d}_{1}\tilde{R}_{(\lambda\rho)\mu\nu}\tilde{R}^{(\lambda\rho)\mu\nu}+6\tilde{d}_{1}\tilde{R}_{(\lambda\rho)\mu\nu}\tilde{R}^{(\lambda\mu)\rho\nu}+2\left(2\tilde{e}_{1}-\tilde{f}_{1}\right)\tilde{R}^{\lambda}\,_{\lambda\mu\nu}\tilde{R}^{\rho}\,_{\rho}\,^{\mu\nu}
\Bigr.
\nonumber\\
\Bigl.
&+8\tilde{f}_{1}\tilde{R}_{(\lambda\rho)\mu\nu}\tilde{R}^{(\lambda\rho)\mu\nu}-2\tilde{f}_{1}\left(\tilde{R}_{(\mu\nu)}-\hat{R}_{(\mu\nu)}\right)\left(\tilde{R}^{(\mu\nu)}-\hat{R}^{(\mu\nu)}\right)-3\left(1-2\tilde{a}_{2}\right)T_{[\lambda\mu\nu]}T^{[\lambda\mu\nu]}
\Bigr]d^4x\sqrt{-g}\,,\label{full_action_model}
\end{align}
which indeed contains contributions from the spin-3 field. In particular, Expression~\eqref{full_action_model} can be directly obtained from the general quadratic action of MAG, up to boundary terms, by setting the following combination of the Lagrangian coefficients:
\begin{align}
    a_1&=0\,,\quad a_2=\tilde{f}_1-\frac{\tilde{d}_1}{8}\,,\quad a_3=\frac{\tilde{d}_1}{8}+\tilde{f}_1\,,\quad a_4=\frac{\tilde{d}_1}{2}\,,\quad a_5=a_7=-\,\frac{\tilde{d}_1}{8}\,,\quad a_6=\frac{\tilde{d}_1}{4}\,,\quad a_8=a_{10}=\frac{\tilde{d}_1-\tilde{f}_1}{4}\,,\\
     a_9&=a_{11}=-\,\frac{ \tilde{d}_1+\tilde{f}_1}{4}\,,\quad a_{12}=\frac{\tilde{d}_1+\tilde{f}_1}{2}\,,\quad a_{13}=\frac{\tilde{f}_1-\tilde{d}_1}{2}\,,\quad  a_{14}=\tilde{e}_1-\frac{\tilde{f}_1}{2}\,,\quad  a_{15}=a_{16}=0\,,\\
  b_1&=-\,\frac{1-\tilde{a}_2}{2}\,,\quad b_2=-\,\tilde{a}_2\,,\quad b_3=c_1=c_2=-\,c_3=1\,,\\
  d_1&=-\,d_4=-\,\frac{1}{4}\,,\quad d_2=-\,d_5=\frac{1}{2}\,,\quad d_3=0\,.
\end{align}
Thus, for this model, the functions~\eqref{GT} and~\eqref{HT} become
\begin{eqnarray}
     \mathcal{G}_{T}^{(C)}=-\,\tilde{f}_1\,,\quad m_{C,{\rm eff}}^2 =\,\dot{H}+2 H^2+4 T_2^2\,.
\end{eqnarray}

The condition to avoid ghost instability \eqref{ghost-condition} implies
\begin{eqnarray}
    \tilde{f}_1 <  0\,.
\end{eqnarray}
This result is consistent with the ghost-free condition for the vector modes of the traceless nonmetricity tensor of the model~\cite{Bahamonde:2022kwg}. 
The tachyonic instability can also be avoided if $\dot{H}+2H^2>0$, since $m_{C,{\rm eff}}^2$ is always positive in this case. In the other case $\dot{H}+2H^2<0$, $m_{C,{\rm eff}}^2<0$ may happen which, as we explained above, is not necessarily a pathology of the helicity-3 perturbations.

Before concluding this section, we stress that a massive spin-3 field should have 7 dynamical d.o.f., while the present analysis only guarantees the stability of the 2 polarisation modes thereof (the helicity $\pm 3$ modes). One should make sure that other helicity sectors are stable as well. We also note that $1/4-d_1-d_2=0$ for the action \eqref{full_action_model}. As we have explained above, theories with $1/4-d_1-d_2=0$ can be problematic in the Minkowski limit due to the no-go theorems of massless higher-spin fields. Nonetheless, one might also regard \eqref{full_action_model} as an effective theory around a nontrivial background if the other 5 polarisation modes of the spin-3 (and other dynamical modes) are all well-behaved there. In this sense, it is important to systematically analyse the perturbations of all sectors, which we leave for future study.

\section{Conclusions}\label{sec:conclusions}

In this work, we laid the foundations to obtain and analyse the cosmological perturbations in the framework of MAG, which includes torsion and nonmetricity on top of the curvature tensor. Indeed, cosmological perturbation theory in Riemannian geometry plays a fundamental role in the description of the inhomogeneities and anisotropies of the universe, such as the structure formation at large scales, the primordial fluctuations in the CMB and the propagation of gravitational waves at cosmological distances, but its formulation in MAG has remained largely unaddressed in the literature.

Specifically, the geometrical perturbations in MAG are associated with the metric, torsion and nonmetricity tensors, whereas the matter perturbations are encoded in the energy-momentum and hypermomentum tensors. Thus, we performed $3+1$ and SVT decompositions of these tensors around spatially curved FLRW background and presented all the possible spatial perturbations arising in MAG. As shown in Tables~\ref{Table:Torsion} and~\ref{Table:Q}, the theory displays a rich perturbation spectrum for the torsion and nonmetricity tensors, which includes a large number of d.o.f. with helicity states $0$, $1$, $2$, and $3$, on top of the well-known metric perturbations of Riemannian geometry. Therefore, MAG provides a diverse phenomenology at cosmological scales. For instance, the additional helicity-$2$ modes can source the helicity-2 modes of metric perturbations at the linear level of perturbations and lead to a significant production of gravitational waves \cite{Gorji:2023ziy,Gorji:2023sil}. Moreover, the extra d.o.f. with different helicities may provide a geometrical source for dark matter. In particular, recently, a lot of attention has been paid to higher spin dark matter models \cite{Alexander:2020gmv,Falkowski:2020fsu,Kolb:2023dzp,Maeda:2013bha,Aoki:2014cla,Aoki:2016zgp,Babichev:2016bxi,Aoki:2017cnz,Aoki:2017ffl,Marzola:2017lbt,Chu:2017msm,GonzalezAlbornoz:2017gbh,Criado:2020jkp,Jain:2021pnk,Manita:2022tkl,Wu:2023dnp,Gorji:2023cmz,Manita:2023mnc}. 

It is worth mentioning that in order to fully determine the number of physical d.o.f. in MAG, one would need to apply Hamiltonian analysis since perturbations around FLRW backgrounds might not exhibit all the possible d.o.f. appearing in a particular theory. Several studies regarding the Hamiltonian analysis of theories with torsion and nonmetricity have already been considered in the literature \cite{Yo:1999ex, Yo:2001sy, Blagojevic:2018dpz, Lin:2018awc, Glavan:2023cuy}.

As an immediate application of our setup, we studied linear perturbation of the nonmetricity helicity-3 modes around the spatially flat FLRW background for a general parity-preserving MAG action that is quadratic in curvature, torsion and nonmetricity. As guaranteed by the SVT decomposition theorem, these modes decouple from the rest of the modes at the linear level of perturbations, which makes their perturbation analysis relatively simple. We thus found the general stability conditions of the helicity-3 modes and applied them to a particular MAG model with dynamical torsion, Weyl vector and traceless nonmetricity tensor, recently studied in the literature in the context of black hole physics~\cite{Bahamonde:2022kwg}. We showed that the stability conditions can be satisfied around the spatially flat FLRW background. However, we found that the spin-3 field is massless in Minkowski background, while it acquires a nonvanishing mass in FLRW background when curvature and/or torsion have a nonvanishing background. This result shows that five modes of the spin-3 field may become strongly coupled around the Minkowski background. It is then interesting to look for possible solutions to this issue. One simple possibility is to add an explicit quadratic term of the field in the gravitational action to have a well-behaved Minkowski limit. We remit the research following these lines for future works.

\noindent
\section*{Acknowledgements}
The work of K.A. was supported in part by Grants-in-Aid from the Scientific Research Fund of the Japan Society for the Promotion of Science, No.~20K14468 and No.~24K17046. S.B. and J.G.V. were supported by JSPS Postdoctoral Fellowships for Research in Japan and KAKENHI Grant-in-Aid for Scientific Research (Grants No. JP21F21789 and No. JP22F22044, respectively). At the end of finishing this manuscript, S.B. was also supported by ``Agencia Nacional de Investigación y Desarrollo" (ANID),  Grant ``Becas Chile postdoctorado al extranjero" No. 74220006. The work of M.A.G. was supported by Mar\'{i}a Zambrano fellowship. M.A.G. thanks the Institute for Basic Science (IBS), Tokyo Institute of Technology, and Yukawa Institute for Theoretical Physics (YITP) for the support and hospitality when this work was in progress.

\appendix

\section{Irreducible decomposition}\label{appendix1}

In this appendix, we present how to systematically find all the irreducible pieces of general tensors of up to rank 3 in the $3+1$ decomposition. A scalar is already irreducible, so we shall start with a four-dimensional vector $X^{\mu}$. The vector is irreducible with respect to the four-dimensional pseudo-orthogonal group. On the other hand, applying the $3+1$ decomposition, the four-dimensional vector is decomposed into a scalar and vector with respect to the three-dimensional orthogonal group, which are, loosely speaking, temporal and spatial components of $X^{\mu}$, respectively. For later convenience, we write this decomposition by using Young tableaux as follows:
\begin{align}
\ydiagram[]{1} &= \ytableaushort{0} \oplus \ytableaushort{i}
\,.
\label{Young1}
\end{align}
We use the Young diagrams (boxes without characters) and the Young tableaux (boxes filled with either $0$ or Latin characters) to denote four-dimensional and three-dimensional tensors, respectively. The boxes filled by $0$ correspond to temporal components (precisely speaking, components multiplied by $n_{\mu}$) while the $i,j,k, \cdots$ boxes are spatial components (components multiplied by $P^{\mu}{}_{\nu}$). The number and shape of boxes filled with Latin characters determine the rank and symmetry of the spatial tensors. For instance, in \eqref{Young1}, the first term of RHS has no Latin character, so it corresponds to the scalar part, whereas the second term represents the spatial vector.

We then consider a generic rank-$2$ tensor which is described by a tensor product
\begin{align}
\ydiagram[]{1} \otimes \ydiagram[]{1}
\,.
\end{align}
As is well-known, this tensor can be decomposed into symmetric and anti-symmetric parts. However, it should be noted that indices can be contracted. Hence, the irreducible decomposition of the rank-$2$ tensor is
\begin{align}
\ydiagram[]{1} \otimes \ydiagram[]{1} = \ydiagram[]{2} \oplus \ydiagram[*(gray)]{2} \oplus \ydiagram[]{1,1}
\,,
\end{align}
where the grey boxes are understood as contracted indices and the white boxes are traceless; that is, the first, second, and third terms on RHS respectively describe a symmetric-traceless tensor, a scalar, and an anti-symmetric tensor in four dimensions. The symmetric and antisymmetric parts are further decomposed into
\begin{align}
\ydiagram[]{2} &= \ytableaushort{00} \oplus \ytableaushort{i0} \oplus \ytableaushort{ij}
\,, \label{young2}\\
\ydiagram[]{1,1} &= \ytableaushort{i,0} \oplus \ytableaushort{i,j}
\,.\label{young_anti}
\end{align}
The boxes filled with Latin characters should be top-left aligned in order that Young sub-tableaux with Latin characters describe the rank and symmetry of spatial tensors. According to this rule, one may avoid double-counting of spatial pieces: for instance, the tableau $\scalebox{0.7}{\ytableaushort{0i}}$ does not appear in \eqref{young2}.
Note that $0$ does not appear twice (or more than twice) in different rows within the same column because the indices are antisymmetrised.
We now use the speciality of three dimensions: two antisymmetric indices can be reduced to one index by taking the Hodge dual,
\begin{align}
\ytableaushort{i,j} = \ytableaushort{{\normalsize \hat{i}} }
\,.
\end{align}
Here we added a hat to record that the indices are dualised. In summary, we find
\begin{align}
\ydiagram[]{1} \otimes \ydiagram[]{1}
=\ytableaushort{00} \oplus \ytableaushort{i0} \oplus \ytableaushort{ij}
\oplus \ydiagram[*(gray)]{2} 
\oplus \ytableaushort{i,0} \oplus \ytableaushort{{\hat{i}} }
\,,
\end{align}
showing that 16 components of rank-$2$ tensor are decomposed into two scalars $(2\times 1)$, three vectors $(3\times 3)$, and one tensor $(1\times 5)$ of three dimensions.

We proceed to consider a rank-$3$ tensor
\begin{align}
\ydiagram[]{1} \otimes \ydiagram[]{1} \otimes \ydiagram[]{1}
\,.
\end{align}
All the essential ingredients have been already explained above. We first perform the irreducible decompositions with respect to the four-dimensional orthogonal group:
\begin{align}
\ydiagram[]{1} \otimes \ydiagram[]{1} \otimes \ydiagram[]{1}
&=\left(\, \ydiagram[]{1} \otimes \ydiagram[]{1} \,\right) \otimes \ydiagram[]{1}
\nn
&=\underbrace{\left[\left(\,  \ydiagram[]{2} \oplus \ydiagram[*(gray)]{2} \, \right)  \otimes \ydiagram[]{1} \,\right]}_{\text{nonmetricity-type}}
\oplus \underbrace{\left[ \, \ydiagram[]{1,1}  \otimes \ydiagram[]{1} \, \right]}_{\text{torsion-type}}
\,,
\end{align}
and
\begin{align}
\left(\,  \ydiagram[]{2} \oplus \ydiagram[*(gray)]{2} \, \right)  \otimes \ydiagram[]{1} 
&=  \ydiagram[]{3} \oplus \ydiagram[*(gray)]{1+2}*[*(white)]{3} \oplus \ydiagram[]{2,1} \oplus \ydiagram[*(gray)]{1+1,1}*[*(white)]{2,1}
\,,
\\
\ydiagram[]{1,1}  \otimes \ydiagram[]{1}
&= \ydiagram[]{1,1,1} \oplus \ydiagram[]{2,1} \oplus \ydiagram[*(gray)]{1+1,1}*[*(white)]{2,1}
\nn
&=  \ydiagram[]{1} \oplus  \ydiagram[]{2,1} \oplus  \ydiagram[*(gray)]{1+1,1}*[*(white)]{2,1}
\,.
\end{align}
where the Young diagram of the shape $(1,1,1)$ can be dualised to the shape $(1)$ in four dimensions.
Here, the white boxes, which are supposed to be filled with either $0$ or Latin characters in the $3+1$ decomposition, are top-left aligned.
The $3+1$ decomposition of vectors (the terms with only one white box) has been already presented. The $3+1$ decompositions of other blocks are
\begin{align}
\ydiagram[]{3}
&=\ytableaushort{000} \oplus \ytableaushort{i0} \oplus \ytableaushort{ij0} \oplus \ytableaushort{ijk}
\,, \label{Young3}
\\
\ydiagram[]{2,1}
&=\ytableaushort{i0, 0} \oplus \ytableaushort{ij,0} \oplus
\ytableaushort{i0,j} \oplus \ytableaushort{ij,k}
\nn
&=
\ytableaushort{i0, 0} \oplus \ytableaushort{ij,0} \oplus
\ytableaushort{{\hat{i}}0} \oplus \ytableaushort{{\hat{i}}j }
\,, \label{Young21}
\\
\ydiagram[]{1,1,1}
&=\ydiagram[]{1}
\nn
&=\ytableaushort{{\hat{0}}} \oplus \ytableaushort{{\hat{i}}}
\,.
\end{align}
One may notice that there are two different ways of embedding rank-2 spatial tensors in $\scalebox{0.7}{\ydiagram[]{2,1}}$.
In this way, the generic four-dimensional rank-$3$ tensor is decomposed into three-dimensional symmetric traceless tensors. Let us count the number of independent components. 64 independent components of a rank-$3$ tensor are first decomposed into 24 torsion-type components and 40 nonmetricity-type components. They are further decomposed into four-dimensional irreducible pieces as follows (see e.g.~\cite{frolov2011introduction}):
\begin{align}
\ydiagram[]{3} &: 16~{\rm components}
\,, \\
\ydiagram[*(gray)]{1+2}*[*(white)]{3}  &: 4~{\rm components}
\,, \\
\ydiagram[]{2,1} &: 16~{\rm components}
\,, \\
\ydiagram[*(gray)]{1+1,1}*[*(white)]{2,1} &: 4~{\rm components}
\end{align}
and
\begin{align}
\ydiagram{1,1,1}=\ydiagram{1} &: 4~{\rm components}
\,, \\
\ydiagram[]{2,1} &: 16~{\rm components}
\,, \\
\ydiagram[*(gray)]{1+1,1}*[*(white)]{2,1} &: 4~{\rm components}
\,.
\end{align}
Note that our Young diagrams are traceless. The nontrivial parts are $\scalebox{0.7}{\ydiagram[]{3}}$ and $\scalebox{0.7}{\ydiagram[]{2,1}}$. As shown in \eqref{Young3} and \eqref{Young21}, they are decomposed into a scalar, vector, rank-$2$ tensor, and rank-$3$ tensor for $\scalebox{0.7}{\ydiagram[]{3}}$, and two vectors and two rank-$2$ tensors for $\scalebox{0.7}{\ydiagram[]{2,1}}$, respectively. Hence, we find
\begin{align}
\ydiagram[]{3} : 16&=1+3+5+7\,, \\
\ydiagram[]{2,1}: 16&=2\times 3 + 2\times 5 \,,
\end{align}
which indeed agree with each other.

\section{$3+1$ decompositions of rank-2 and rank-3 tensors}\label{appendix2}

In this section, we will demonstrate the form of the splitting from a rank-2 tensor to rank-3 tensor following the $3+1$ decomposition with a unit timelike vector $n_\mu$ satisfying $n_\mu n^\mu=-\,1$ and a projector as~\eqref{Peq} satisfying $P_{\mu\nu}n^\mu=0$. Then, our tensors will be constructed from all the possible linear combinations of their three-dimensional irreducible pieces and expressed in terms of $n_\mu$ and $P_{\mu\nu}$. 

Following the Young diagram displayed in Eq.~\eqref{young2}, a symmetric-traceless rank-2 tensor ${\nearrow\!\!\!\!\!\!\!\!X}_{(\mu\nu)}$ is decomposed into a symmetric traceless tensor, a vector and a scalar. Thereby, the most general linear combination constructed from all these quantities can be written as
\begin{equation}
    {\nearrow\!\!\!\!\!\!\!\!X}_{(\mu\nu)}= c_{1}n_{\mu}n_{\nu}+c_{2} P_{\mu\nu}+c_{3}\vec{X}_{(\mu\nu)}-2c_{4}\vec{X}_{*(\mu}n_{\nu)}\,,
\end{equation}
but note that constants $c_3$ and $c_4$ can be normalised as $c_3=c_4=1$, whereas $c_1$ and $c_2$ contribute to the trace unless $-c_1+3c_2=0$. Hence, we find
\begin{equation}
  {\nearrow\!\!\!\!\!\!\!\!X}_{(\mu\nu)}= \left(n_{\mu}n_{\nu}+\frac{1}{3}P_{\mu\nu}\right){\nearrow\!\!\!\!\!\!\!\!X}_{**}+ \vec{X}_{(\mu\nu)}-2 \vec{X}_{*(\mu}n_{\nu)}\,,
\end{equation}
with
\begin{equation}
    {\nearrow\!\!\!\!\!\!\!\!X}_{**} =c_1 = n^{\mu}n^{\nu} {\nearrow\!\!\!\!\!\!\!\!X}_{\mu\nu}\,.
\end{equation}
In addition, by taking into account the Young diagram~\eqref{young_anti} and the dualisation of two antisymmetric indices to one index, the antisymmetric rank-2 tensor $X_{[\mu\nu]}$ turns out to be determined by a vector and a pseudovector
\begin{equation}
    X_{[\mu\nu]}= -\,2\vec{X}^{\rm (E)}_{[\mu}n_{\nu]} + \frac{1}{2} \vec{\varepsilon}_{\mu\nu\lambda} \vec{X}^{\rm (B)}{}^{\lambda} \,,
\end{equation}
where the respective constants in the linear combination have also been normalised to match our conventions~\eqref{antisim_def}.

For an arbitrary rank-3 tensor $X_{\mu\nu\rho}$, the fact that, in terms of algebraic symmetries, the antisymmetric part $X_{\mu[\nu\rho]}$ corresponds to a torsion-like tensor and the symmetric part $X_{\mu(\nu\rho)}$ to a nonmetricity-like tensor, means that without any loss of generality its $3+1$ decomposition is completely determined by the corresponding decompositions of the torsion and nonmetricity tensors. A straightforward analysis is then to perform the $3+1$ decomposition of all their four-dimensional irreducible pieces, which trivially reduces to the problem of finding the decompositions of the rank-3 tensor modes $t_{\mu\nu\rho}$, $\Omega_{\mu\nu\rho}$ and $q_{\mu\nu\rho}$, since the respective decompositions of the other rank-1 vectors are well-known.

Following these lines, let us now obtain the $3+1$ decomposition of the tensor $t_{\mu\nu\rho}$ expressed in Expression~\eqref{torsion3}. First, we notice from the Young diagram~\eqref{Young21} that such a tensor is decomposed into a traceless tensor $\vec{A}_{\mu\nu}$, a vector $\vec{B}_\mu$, a traceless pseudotensor $\vec{\mathcal{A}}_{\mu\nu}$ and a pseudovector $\vec{\mathcal{B}}_\mu$. Thereby, by taking into account the skew symmetry $t_{\mu\nu\rho}=t_{\mu[\nu\rho]}$ of this tensor, the most general linear combination constructed from these four quantities can be written then in terms of $n_\mu$ and $P_{\mu\nu}$ as
\begin{eqnarray}\label{form1}
    t_{\mu\nu\rho}&=& c_1n_{[\nu}\vec{A}_{\rho]\mu}+c_2n_{\mu}n_{[\nu}\vec{B}_{\rho]}+c_3P_{\mu[\nu}\vec{B}_{\rho]}+\varepsilon_{\nu\rho}{}^{\alpha\beta}\bigl(c_4 n_\alpha \vec{\mathcal{A}}_{\beta\mu}+c_5n_{\mu}n_\beta\vec{\mathcal{B}}_\alpha+c_{6} P_{\mu\beta}\vec{\mathcal{B}}_{\alpha}\bigr)\nonumber\\
    &&+\,c_{7}\varepsilon_{\mu[\rho}{}^{\alpha\beta} \vec{\mathcal{A}}_{\nu]\beta}n_\alpha +c_{8}\varepsilon_{\mu[\rho}{}^{\alpha\beta} P_{\nu]\beta}\vec{\mathcal{B}}_\alpha \,.
\end{eqnarray}
However, by virtue of the identities
\begin{align}
    \varepsilon_{\nu\rho}{}^{\alpha\beta}n_\alpha \vec{\mathcal{A}}_{\beta\mu}&=2\varepsilon_{\mu[\rho}{}^{\alpha\beta} \vec{\mathcal{A}}_{\nu]\beta}n_\alpha\,,\\
    \varepsilon_{\mu[\rho}{}^{\alpha\beta} P_{\nu]\beta}\vec{\mathcal{B}}_{\alpha}&=\frac{1}{2}\varepsilon_{\nu\rho}{}^{\alpha\beta}\vec{\mathcal{B}}_{\alpha}\bigl(2n_{\mu}n_\beta- P_{\mu\beta}\bigr)\,,
\end{align}
the term related to the constant $c_8$ is linearly dependent on the terms related to $c_5$ and $c_6$, whereas the same holds for the terms given by the constants $c_7$ and $c_4$. Therefore, without any loss of generality, we can directly set $c_7=c_8=0$. Next, in order for the tensor mode $t_{\mu\nu\rho}$ to be totally traceless and pseudotraceless, we must demand
\begin{equation}\label{con1t}
    c_3=\frac{1}{2}c_2\,,\quad c_6=\frac{1}{2}c_5\,,
\end{equation}
which automatically satisfies the condition $t_{[\alpha\beta\mu]}=0$, since
\begin{equation}
  3 \vec{\mathcal{B}}^\lambda n_{[\mu}\varepsilon_{\nu\rho]\lambda\xi}n^\xi -\vec{\mathcal{B}}^\lambda \varepsilon_{\mu\nu\rho\lambda}=n^\lambda \vec{\mathcal{A}}^\xi{}_{[\mu}\varepsilon_{\nu\rho]\lambda\xi}=0\,.
\end{equation}
Then, by replacing the relations~\eqref{con1t} into Expression~\eqref{form1}, and choosing $c_1=c_2=2,c_4=-\,c_5=-\,1/2$ for normalisation purposes, we find that such a tensor can be finally written as follows:
\begin{equation}
t_{\mu\nu\rho}=\displaystyle 2n_{[\nu}\vec{A}_{\rho]\mu}+2\Bigl(n_{\mu}n_{[\nu}+\frac{1}{2}P_{\mu[\nu} \Bigr) \vec{B}_{\rho]}
+\frac{1}{2}\varepsilon_{\nu\rho}{}^{\alpha\beta}\Bigl[ -\,n_{\alpha}\vec{\mathcal{A}}_{\beta\mu}+\Bigl(n_{\beta}n_{\mu}+\frac{1}{2}P_{\beta\mu} \Bigr)\vec{\mathcal{B}}_{\alpha}\Bigr]\,.
\end{equation}

Similarly, the fact that $\Omega_{\mu\nu\rho}$ presents the same algebraic symmetries as $t_{\mu\nu\rho}$ leads to an analogous result:
\begin{equation}
    \Omega^{}_{\mu\nu\rho}=2n_{[\nu}\vec{\mathcal{Q}}_{\rho]\mu}+2\Bigl(n_{\mu}n_{[\nu}+\frac{1}{2}P_{\mu[\nu} \Bigr) \vec{\mathcal{Y}}_{\rho]}+\frac{1}{2}\varepsilon_{\nu\rho}{}^{\alpha\beta}\Bigl[ -\,n_{\alpha}\vec{Q}_{\beta\mu}+\Bigl(n_{\beta}n_{\mu}+\frac{1}{2}P_{\beta\mu} \Bigr)\vec{Y}_{\alpha}\Bigr]\,.
\end{equation}

Finally, let us consider the Young diagram~\eqref{Young3}, which represents the fully symmetric and traceless tensor $q_{\lambda\mu\nu}$. In this case, the building blocks of the $3+1$ decomposition are $\vec{C}_{\lambda\mu\nu},\vec{\kappa}_{\mu\nu},\vec{Z}_{\rho}$ and $\xi$. Thus, the most general linearly independent combination formed by these building blocks can be written in terms of $n_\mu$ and $P_{\mu\nu}$ as
\begin{equation}
q_{\mu\nu\rho}=\displaystyle\vec{C}_{\mu\nu\rho}+c_1 n_{(\mu}\vec{\kappa}_{\nu\rho)}+\left(c_2n_{(\mu}n_{\nu}+c_3P_{(\mu\nu} \right)\vec{Z}_{\rho)}+\left(c_4n_{(\mu}n_{\nu}+c_5 P_{(\mu\nu} \right) n_{\rho)} \xi\,,
\end{equation}
where the following relations must be satisfied in order to fulfil the traceless condition:
\begin{equation}
    c_3=\frac{1}{5}c_2\,,\quad c_5=c_4\,.
\end{equation}
In that case, by normalising the values of the constants as $c_1=-\,c_2=-\,3, c_4=-\,1$, we obtain:
\begin{equation}
q_{\mu\nu\rho}=\displaystyle\vec{C}_{\mu\nu\rho}-3n_{(\mu}\vec{\kappa}_{\nu\rho)}+\frac{3}{5}\left(5n_{(\mu}n_{\nu}+P_{(\mu\nu} \right)\vec{Z}_{\rho)}-\left(n_{(\mu}n_{\nu}+P_{(\mu\nu} \right) n_{\rho)} \xi\,,
\end{equation}
which concludes the derivation of the $3+1$ decomposition of arbitrary rank-2 and rank-3 tensors.

\section{Uniqueness of the SVT decomposition}\label{appendix3}

In this appendix, we will show that the SVT decomposition performed in Sec.~\ref{SVT} for rank-1, rank-2 and rank-3 tensors is uniquely determined. For this task, we will follow the same reasoning as the one considered in~\cite{Kodama:1984ziu, Nakamura:2004rm,Uggla:2011jn} for rank-1 and rank-2 tensors, thus extending these results to the case of rank-3 tensors.  First, it is useful to recall the following equations:
\begin{eqnarray}
    {}^{(3)}\!R_{ijkl}=2K\gamma_{k[i}\gamma_{j]l}\,,\quad  {}^{(3)}\!R_{ij}=2K\gamma_{ij}\,,\quad  {}^{(3)}\!R=6K\,,
\end{eqnarray}
with $K$ being the constant spatial curvature of the FLRW space-time. Then, one can find the following commutator rules for an arbitrary vector $U_{i}$ and an arbitrary tensor $U_{ij}$:
\begin{eqnarray}\label{comm1}
  \DD^{i}\DD^2 U_{i} - \DD^2 \DD^{i} U_{i} 
  &=& \DD^{i}\bigl({}^{(3)}\!R_{ij} U^{j}\bigr)
  = 2K \DD^{i} U_{i}
  , \\
  \DD^{i}\DD^2 U_{ij} - \DD^2 D^{i} U_{ij} 
  &=& \DD^{i}\bigl(
    {}^{(3)}\!R_{i}^{\;\;l} U_{lj}
    +
    {}^{(3)}\!R_{\;\;ij}^{l\;\;\;\;k}U_{lk}
  \bigr)
  + 
  {}^{(3)}\!R_{ikjl} \DD^{k}U^{il}
  = 4K \Bigl( \DD^{i} U_{(ij)}
    -
    \frac{1}{2}
    \DD_{j} U_{i}^{\;\;i}
  \Bigr)\,.\label{comm2}
\end{eqnarray}

For the vector decomposition~\eqref{vectorSVT}, let us apply $\DD_i$ which gives us
\begin{equation}
    \DD_i\delta X^i=\DD^2 \mathbf{S} + \DD_i\mathbf{V}^{(\rm T)}{}^i=\DD^2 \mathbf{S}\,.
\end{equation}
Thus, one can use the inverse of the operator $\DD^2$ to determine $\mathbf{S} $ and then $\mathbf{V}^{(\rm T)}{}^i$ is uniquely determined by~\eqref{vectorSVT}. Thereby, if the inverse operators exist, Eq.~\eqref{vectorSVT} determines $\mathbf{S}$ and $\mathbf{V}^{(\rm T)}{}^i$ uniquely in terms of $\delta X^i$.

Similarly, for the rank-2 tensor~\eqref{tensorSVT}, let us first apply the operator $\DD_i$, yielding
\begin{equation}
    \DD_i \delta\vec{X}^{ij}=\displaystyle \frac{2}{3}\DD^j\bigl(\DD^2+3K\bigr)\mathbf{S}+\bigl(\DD^2 +2K\bigr)\mathbf{V}^{(\rm T)}{}^{j}\,,
\end{equation}
where we have used~\eqref{comm1}. Next, by applying the operator $\DD_{(i}\DD_{j)} -\frac{1}{3}\gamma_{ij}\DD^l\DD_l$ to the rank-2 tensor, we get
\begin{equation}
   \Bigl(\DD_{(i}\DD_{j)} -\frac{1}{3}\gamma_{ij}\DD^2\Bigr)\delta\vec{X}^{ij}
   =\frac{2}{3}\DD^2\bigl(\DD^2 +3K\bigr)\mathbf{S}\,,
\end{equation}
where we have used~\eqref{comm2}. Then, if the inverses of $\DD^2, \DD^2+2K$ and $\DD^2+3K$ exist, Eq.~\eqref{tensorSVT} determines $\mathbf{S},\mathbf{V}^{(\rm T)}{}^{i}$ and $\mathbf{T}^{(\rm TT)}{}^{ij}$ uniquely in terms of $\delta\vec{X}^{ij}$.

Finally, let us consider the rank-3 tensor given by Eq.~\eqref{spin3SVT}. By applying the operator $\DD_i$, we find
\begin{eqnarray}
    \DD_i \delta \vec{X}^{ijk}&=&\frac{1}{3}\bigl(\DD^2 +6 K \bigr)\mathbf{T}^{(\rm TT)}{}^{jk} +\frac{4}{15} \DD^{k}\bigl( \DD^2 + 7K \bigr)\mathbf{V}^{(\rm T)}{}^{j}+\frac{4}{15} \DD^{j}\bigl(\DD^2+7K\bigr)\mathbf{V}^{(\rm T)}{}^{k} \nonumber\\
    &&+\frac{3}{5}\Bigl( \DD^{k}\DD^{j}\DD^2-\frac{1}{3}\gamma^{jk}\DD^4+8K\,\DD^k\DD^j-\frac{8}{3}K\,\gamma^{jk}\DD^2\Bigr)\mathbf{S}\,,
\end{eqnarray}
while applying two covariant derivatives provides
\begin{eqnarray}
    \DD_i\DD_j\delta\vec{X}^{ijk}&=&\frac{4}{15}\bigl( \DD^4+9K\,\DD^2 + 14\, K^2 \bigr)\mathbf{V}^{(\rm T)}{}^{k} +\frac{2}{5}\DD^{k}\bigl(\DD^4+ 11 K\, \DD^2 + 24 K^2\,   \bigr)\mathbf{S}
    \nn
    &=&\frac{4}{15}(\DD^2+7K)(\DD^2+2K)\mathbf{V}^{(\rm T)}{}^{k} +\frac{2}{5}\DD^{k}(\DD^2+3K)(\DD^2+8K)\mathbf{S}\,,
\end{eqnarray}
where $\DD^4\equiv\DD^2\DD^2$. Furthermore, we can also compute the following operator from three covariant derivatives:
\begin{eqnarray}
 \Bigl[ \DD_{i}\DD_{j}\DD_{k} -  \frac{1}{5} \gamma_{ij} 
\bigl(\DD_{k}\DD^2 + \DD^{l}\DD_{k}\DD_{l} + \DD^2 \DD_{k}\bigr)\Bigr]\delta\vec{X}^{ijk}
    &=&\frac{2}{5}\DD^2\bigl(\DD^4+11 K \DD^2+24K^2\bigr)\mathbf{S}
    \nn
    &=&\frac{2}{5}\DD^2(\DD^2+3K)(\DD^2+8K)\mathbf{S}\,.
\end{eqnarray}
Then, if the inverses of the operators
\begin{eqnarray}
   \DD^2\,, \quad \DD^2+2K\,, \quad \DD^2+3K\,, \quad \DD^2 +6 K \,,\quad \DD^2+7K\,,\quad \DD^2+8K \,,
\end{eqnarray}
exist, the SVT decomposition~\eqref{spin3SVT} of the rank-3 tensor is also uniquely determined.

\section{Gauge transformation and gauge invariance}
\label{app-gauge}

In this appendix, we briefly review the gauge transformation and introduce some common gauges that are used in the context of cosmological perturbation theory. Although this is a quite well-known subject, we present it for the benefit of the readers and completeness of the paper.

The MAG theory is invariant under the four-dimensional spacetime diffeomorphisms. On the other hand, all of the perturbations are defined in a component form and, therefore, they change if we perform a coordinate transformation. For example, one may change spacetime coordinate in the background metric \eqref{metric-FLRW} and generate some fake perturbations. There are two ways to overcome this difficulty and deal with the real perturbations: i) working with gauge-invariant perturbations, ii) fixing the gauge. In order to understand these approaches in more detail, let us consider a general coordinate transformation
\begin{align}\label{coordinate-trans}
x^\mu \to x^\mu + \xi^\mu \,.
\end{align}
Note that the symbol $\xi$ in this appendix always refers to the vector $\xi^\mu$ defined above. Under the above transformation, perturbations of a spacetime tensor $T_I=\{s,v_\mu,t_{\mu\nu}, f_{\mu\nu\lambda}, \cdots\}$ change as
\begin{align}\label{Lie-pert}
\delta{T}_I \to \delta{T}_I + \pounds_{\xi} \bar{T}_I \,,
\end{align} 
where perturbations are shown by a $\delta$, background values are labeled by a bar, and $\pounds_{\xi}$ denotes the Lie derivative along $\xi^\mu$. Performing SVT decomposition of $\xi^\mu$, following the method presented in subsection \ref{SVT}, we find
\begin{align}\label{xi-SVT}
\xi^\mu = \left( \xi^0, \DD^i\xi + \xi^{{(\rm T)}i} \right) \,,
\end{align}
where $\DD_i \xi^{{(\rm T)}i}=0$. 

Let us apply \eqref{Lie-pert} to the metric perturbations. Substituting \eqref{metric-pert-helicity2}, \eqref{metric-pert-helicity1}, \eqref{metric-pert-helicity0} in \eqref{metric-matrix-form}, we find the following form for the metric in terms of the helicity modes
\begin{align}\label{metric-matrix-form-helicity}
g_{\mu\nu}=
\begin{pmatrix}
-N^2\left(1+2\alpha\right) & Na \left(\DD_i \beta + \beta^{(\rm T)}_i \right) \\
Na \left( \DD_j \beta + \beta^{(\rm T)}_j \right)  & a^2\left[\left(1+2\psi\right) \gamma_{ij} + \Bigl(\DD_i \DD_j -\frac{1}{3}\gamma_{ij} \DD^2 \Bigr)h + \DD_{(i} h^{(\rm T)}_{j)} + h^{(\rm TT)}_{ij} \right]\,
\end{pmatrix}
\,.
\end{align}
Working with conformal time $\tau$ with $N=a$ and substituting $T_I=\{g_{\mu\nu}\}$ in \eqref{Lie-pert} and using \eqref{xi-SVT}, it is straightforward to show that the different helicities of the metric perturbations change as follows:
\begin{itemize}
\item Helicity-2 modes: \begin{align}\label{tensor-GT}
&h^{(\rm TT)}_{ij} \to h^{(\rm TT)}_{ij} \,,
\end{align}
\item Helicity-1 modes:
\begin{align}\label{vector-GT}
&\beta^{(\rm T)}_i \to \beta^{(\rm T)}_i - \left(\xi^{(\rm T)}_i\right)' \,, 
&&h^{(\rm T)}_i \to h^{(\rm T)}_i +2 \xi^{(\rm T)}_i  \,,
\end{align}
\item Helicity-0 modes:
\begin{align}\label{scalar-GT}
&\alpha \to \alpha + {\cal H} \xi^0 + \left(\xi^0\right)' \,, 
&&\beta \to \beta - \xi^0 + \xi' \,,
&& \psi \to \psi + {\cal H} \xi^0 + \frac{1}{6} \DD^2\xi \,,
&&h \to h + 2 \xi \,,
\end{align}
\end{itemize}
where a prime denotes derivative with respect to the conformal time $\tau$ and ${\cal H}=a'/a$ is the conformal Hubble parameter. 

\subsection{Helicity-2 tensor modes}
As it can be clearly seen from \eqref{tensor-GT}, the helicity-2 tensor modes $h^{(\rm TT)}_{ij}$ are gauge-invariant. Thus, as far as we are interested in linear perturbations of the helicity-2 modes in MAG, we can safely consider the following metric
\begin{align}\label{metric-helicity2}
g_{\mu\nu}=
\begin{pmatrix}
-N^2 & 0 \\
0  & a^2\left( \gamma_{ij} + h^{(\rm TT)}_{ij} \right)\,
\end{pmatrix}
\,.
\end{align}

\subsection{Helicity-1 vector modes}
On the other hand, the helicity-1 vector modes are not gauge-invariant as shown in  \eqref{vector-GT}. We can then either work with a gauge-invariant variable or fix the gauge. From \eqref{vector-GT}, it is clear that the combination
\begin{align}\label{helicity1-GI}
\beta^{(\rm T)}_i+\left(h^{(\rm T)}_i\right)'/2 \,,
\end{align}
is gauge-invariant. So, for the linear perturbations, one can work with metric \eqref{metric-matrix-form-helicity} keeping both $\beta^{(\rm T)}_i$ and $h^{(\rm T)}_i$ while ignoring helicity-2 and helicity-0 modes. At the end, it can be shown that working with \eqref{helicity1-GI}, there are only two real d.o.f. for the system. Alternatively, we can fix the gauge by choosing $\xi^{(\rm T)}_i$ such that $\beta^{(\rm T)}_i=0$ or $h^{(\rm T)}_i=0$ (while the former does not completely fix the gauge). For the latter choice, we find
\begin{align}\label{metric-helicity1}
g_{\mu\nu}=
\begin{pmatrix}
-N^2 & Na \beta^{(\rm T)}_i \\
Na \beta^{(\rm T)}_j & a^2 \gamma_{ij} \,
\end{pmatrix}
\,.
\end{align}
The above metric can be used for the linear perturbation analysis of the helicity-1 modes in MAG.

Note that working with either gauge-invariant variable \eqref{helicity1-GI} or a gauge-fixed metric for the helicity-1 modes, e.g. \eqref{metric-helicity1}, there are only two real helicity-1 d.o.f. in the metric sector.  Thus, in any case we deal with two real vector d.o.f. which is the direct consequence of the freedom in choosing $\xi^{(\rm T)}_{i}$.

\subsection{Helicity-0 scalar modes}
The helicity-0 scalar modes are also not gauge-invariant as shown in \eqref{scalar-GT}. Again, one can either fix the gauge or work with gauge-invariant variables. In practice, it is easier to fix the gauge and work with the real d.o.f. and we follow this approach here. We have freedom in choosing $\xi^0$ and $\xi$. Thus, we can work in the following gauges without any ambiguities:
\begin{itemize}
	\item {\it Longitudinal (Newtonian) gauge:} $\beta=0$ and $h=0$
	\begin{align}\label{metric-helicity0-LG}
	g_{\mu\nu}=
	\begin{pmatrix}
	-N^2\left(1+2\alpha\right) & 0 \\
	0 & a^2\left[\left(1+2\psi\right) \gamma_{ij}  \right]\,
	\end{pmatrix}
	\,.
	\end{align}
	As it can be seen the metric takes a diagonal form in this case.
	\item {\it Spatially uniform gauge:} $\psi=0$ and $h=0$
	\begin{align}\label{metric-helicity0-SUG}
	g_{\mu\nu}=
	\begin{pmatrix}
	-N^2\left(1+2\alpha\right) & Na \DD_i \beta \\
	Na \DD_j \beta & a^2 \gamma_{ij}\,
	\end{pmatrix}
	\,.
	\end{align}
	Note that, in this gauge, the spatial sector of metric remains the same as the background spatial metric.	
\end{itemize}

One can safely work with either \eqref{metric-helicity0-LG} and \eqref{metric-helicity0-SUG} as far as linear perturbations of the helicity-0 modes are concerned in MAG. In both cases, we deal with two real d.o.f. in the metric sector which is the direct consequences of the freedom in choosing $\xi^0$ and $\xi$. 

It is also worth mentioning that fixing the gauge in the metric sector, one can safely deal with all other perturbations with different helicities in both geometry (torsion and nonmetricity) and matter sectors at any level of perturbations. However, one may be interested in different gauges such that fixing the gauge in the torsion and/or nonmetricity sector. Although cumbersome, this is very straightforward: one needs to find how the helicity modes of torsion and nonmetricity change under a general coordinate transformation \eqref{coordinate-trans} by applying the formula \eqref{Lie-pert}. The same also holds if one is interested in fixing the gauge in the matter sector: one needs to look how helicity modes of the energy-momentum and hypemomentum tensors change under \eqref{coordinate-trans}. For example, the so-called comoving gauge fixes the gauge such that the four-velocity of the comoving observer in the energy-momentum tensor remains similar to its background value all the time. There might be some interesting gauges with clear physical meaning in the torsion and nonmetricity or hypemomentum sectors as well. However, this is beyond the scope of this paper and we leave it for future works.

\section{Interactions of spin-3 field $q_{\lambda\mu\nu}$}\label{appendix4}

In Sec.~\eqref{sec:spin3}, we have worked out the resulting action for the helicity-3 modes arising from the general quadratic action~\eqref{MAGLag} of MAG. As can be seen from \eqref{qtensor} and \eqref{Q4split}, the helicity-3 modes are encoded in the irreducible part $q_{\mu\nu\lambda}$ of the traceless nonmetricity tensor. In this appendix, we thus classify the Lagrangian in the action~\eqref{MAGLag} based on the kinetics and the interactions of $q_{\mu\nu\lambda}$ as follows:
\begin{eqnarray}
\mathcal{L}&=& \mathcal{L}_{\rm g}^{(q-R)}+\mathcal{L}_{\rm g}^{(q-q)}+\mathcal{L}_{\rm g}^{(q-W)}+\mathcal{L}_{\rm g}^{(q-\Lambda)}+\mathcal{L}_{\rm g}^{(q-T)}+\mathcal{L}_{\rm g}^{(q-S)}+ \mathcal{L}_{\rm g}^{(q-RTSW\Lambda)}\nonumber\\
&&+\,\mathcal{L}_{\rm g}^{(q-t)}+\mathcal{L}_{\rm g}^{(q-\Omega)}+\mathcal{L}_{\rm g}^{(q-RTStW\Lambda)}+\mathcal{L}_{\rm g}^{(q-RTSW\Lambda\Omega)}+\mathcal{L}_{\rm g}^{(q-TStW\Lambda\Omega)}\,,\label{Lagterm}
\end{eqnarray}
where
\small{
\begin{eqnarray}
16\pi\mathcal{L}_{\rm g}^{(q-R)}&=&   \frac{1}{6} (- \alpha_{1}{} + 3 \alpha_{2}{} + 4 \alpha_{3}{} -  \alpha_{4}{} -  \alpha_{5}{}) q_{\alpha \rho \tau } q^{\alpha \rho \tau } R+ (\alpha_{2}{} + 6 \alpha_{3}{} + \alpha_{4}{} + \alpha_{5}{}) q_{\alpha }{}^{\lambda \mu } q^{\alpha \rho \tau } R_{\rho \lambda \tau \mu } \nonumber\\
&&+2 \alpha_{4}{} R^{\alpha \rho } \nabla_{\tau }q_{\alpha \rho }{}^{\tau }+(\alpha_{4}{} + \alpha_{5}{}) q_{\alpha }{}^{\tau \lambda } q_{\rho \tau \lambda } R^{\alpha \rho } \,,\\
  16\pi \mathcal{L}_{\rm g}^{(q-q)}&=&\Big(-\frac{1}{4} + d_{2}{} + d_{1}{}\Big) q_{\lambda \mu \alpha } q^{\lambda \mu \alpha }+ \frac{1}{8} (\alpha_{2}{} + 6 \alpha_{3}{}) q_{\alpha }{}^{\tau \nu } q_{\rho \tau \nu } q_{\lambda \mu }{}^{\rho } q^{\lambda \mu \alpha } + \frac{1}{4} (-2 \alpha_{4}{} + 2 \alpha_{5}{} + \alpha_{6}{}) \nabla_{\rho }q_{\mu \alpha }{}^{\rho } \nabla_{\lambda }q^{\lambda \mu \alpha } \nonumber\\
  &&+ \frac{1}{48} (\alpha_{1}{} - 3 \alpha_{2}{} - 4 \alpha_{3}{} + \alpha_{4}{} + \alpha_{5}{}) q_{\rho \tau \nu } q^{\rho \tau \nu } q_{\lambda \mu \alpha } q^{\lambda \mu \alpha } + \frac{1}{8} (- \alpha_{2}{} - 6 \alpha_{3}{} -  \alpha_{4}{} -  \alpha_{5}{}) q_{\alpha \tau \nu } q_{\lambda }{}^{\rho \tau } q^{\lambda \mu \alpha } q_{\mu \rho }{}^{\nu } \nonumber\\
  &&+ \frac{1}{2} (- \alpha_{2}{} + \alpha_{3}{}) (\nabla_{\alpha }q_{\lambda \mu \rho } -\nabla_{\rho }q_{\lambda \mu \alpha } )\nabla^{\rho }q^{\lambda \mu \alpha } -  \frac{1}{2} \alpha_{4}{} q_{\lambda \mu }{}^{\rho } q^{\lambda \mu \alpha } \nabla_{\tau }q_{\alpha \rho}{}^{\tau }   \,,\\
    16\pi\mathcal{L}_{\rm g}^{(q-W)}&=&\alpha_{4}{} q_{\alpha }{}^{\rho \tau } q_{\rho }{}^{\lambda \mu } 
q_{\tau \lambda \mu } W^{\alpha } + \frac{1}{4} (\alpha_{1}{} - 2 \alpha_{2}{} + 2 \alpha_{3}{}) q_{\rho \tau \lambda } q^{\rho \tau \lambda } W_{\alpha } W^{\alpha }- 2 \alpha_{4}{} q_{\alpha \rho \tau } W^{\alpha } W^{\rho } W^{\tau } \nonumber\\
&&+\Bigl(\frac{1}{2} \alpha_{2}{} - 4 \alpha_{3}{} - 2 \alpha_{4}{} + 2 \alpha_{5}{} + \alpha_{6}{}\Bigr) q_{\alpha }{}^{\tau \lambda } q_{\rho \tau \lambda } W^{\alpha } W^{\rho }  + \frac{1}{2} (- \alpha_{1}{} + 3 \alpha_{2}{} + 4 \alpha_{3}{}) q_{\rho \tau \lambda } q^{\rho \tau \lambda } \nabla_{\alpha }W^{\alpha } \nonumber\\
&&+ (- \alpha_{2}{} - 6 \alpha_{3}{}) q_{\alpha }{}^{\tau \lambda } q_{\rho \tau \lambda } \nabla^{\rho }W^{\alpha } + \alpha_{4}{} W^{\alpha } W^{\rho } \nabla_{\tau }q_{\alpha \rho }{}^{\tau } + 2 \alpha_{4}{} \nabla^{\rho }W^{\alpha } \nabla_{\tau }q_{\alpha \rho }{}^{\tau } - 4 \alpha_{4}{} q_{\alpha \rho \tau } W^{\alpha } \nabla^{\tau }W^{\rho } \nonumber\\
&&+ (- \alpha_{2}{} + \alpha_{3}{} + 2 \alpha_{4}{} - 2 \alpha_{5}{} -  \alpha_{6}{}) q_{\alpha }{}^{\rho \tau } W^{\alpha } \nabla_{\lambda }q_{\rho \tau }{}^{\lambda }\,,\\
 16\pi   \mathcal{L}_{\rm g}^{(q-\Lambda)}&=&\frac{1}{4} \bigl[- \alpha_{2}{} - 3 (2 \alpha_{3}{} + \alpha_{4}{})\bigr] q_{\alpha }{}^{\rho \tau } q_{\rho }{}^{\lambda \mu } q_{\tau \lambda \mu } \Lambda^{\alpha } + \frac{1}{64} (-3 \alpha_{1}{} + 14 \alpha_{2}{} + 42 \alpha_{3}{} + 12 \alpha_{4}{}) q_{\rho \tau \lambda } q^{\rho \tau \lambda } \Lambda_{\alpha } \Lambda^{\alpha } \nonumber\\
 &&+ \frac{9}{32} (\alpha_{2}{} - 8 \alpha_{3}{} - 4 \alpha_{4}{} + 4 \alpha_{5}{} + 2 \alpha_{6}{}) q_{\alpha }{}^{\tau \lambda } q_{\rho \tau \lambda } \Lambda^{\alpha } \Lambda^{\rho } + \frac{1}{32} (49 \alpha_{2}{} + 42 \alpha_{3}{} - 9 \alpha_{4}{} + 72 \alpha_{5}{} + 36 \alpha_{6}{}) q_{\alpha \rho \tau } \Lambda^{\alpha } \Lambda^{\rho } \Lambda^{\tau } \nonumber\\
 &&+ \frac{1}{8} \bigl[3 \alpha_{1}{} - 9 \alpha_{2}{} - 2 (6 \alpha_{3}{} + \alpha_{4}{})\bigr] q_{\rho \tau \lambda } q^{\rho \tau \lambda } \nabla_{\alpha }\Lambda^{\alpha } + \Bigl(\frac{3}{4} \alpha_{2}{} + \frac{9}{2} \alpha_{3}{} + \alpha_{4}{}\Bigr) q_{\alpha }{}^{\tau \lambda } q_{\rho \tau \lambda } \nabla^{\rho }\Lambda^{\alpha } \nonumber\\
 &&+ \frac{1}{16} (12 \alpha_{2}{} - 12 \alpha_{3}{} - 11 \alpha_{4}{} + 24 \alpha_{5}{} + 12 \alpha_{6}{}) \Lambda^{\alpha } \Lambda^{\rho } \nabla_{\tau }q_{\alpha \rho }{}^{\tau } + \Bigl(- \alpha_{2}{} + \alpha_{3}{} + \frac{1}{2} \alpha_{4}{} - 2 \alpha_{5}{} -  \alpha_{6}{}\Bigr) \nabla^{\rho }\Lambda^{\alpha } \nabla_{\tau }q_{\alpha \rho }{}^{\tau }  \nonumber\\
&&+ \frac{1}{4} (-9 \alpha_{2}{} - 12 \alpha_{3}{} -  \alpha_{4}{} - 12 \alpha_{5}{} - 6 \alpha_{6}{}) q_{\alpha \rho \tau } \Lambda^{\alpha } \nabla^{\tau
}\Lambda^{\rho }+ \frac{1}{4} (3 \alpha_{2}{} - 3 \alpha_{3}{} - 4 \alpha_{4}{} + 6 \alpha_{5}{} + 3 \alpha_{6}{}) q_{\alpha }{}^{\rho \tau } \Lambda^{\alpha } \nabla_{\lambda }q_{\rho \tau }{}^{\lambda }\,,\nonumber \\ \\
16\pi    \mathcal{L}_{\rm g}^{(q-T)}&=& \frac{16}{27} \alpha_{4}{} q_{\alpha \rho \tau } T^{\alpha } T^{\rho } T^{\tau }- \frac{2}{3} \alpha_{4}{} q_{\alpha }{}^{\rho \tau } q_{\rho }{}^{\nu \lambda } q_{\tau \nu \lambda } T^{\alpha } + \frac{4}{9} \alpha_{4}{} T^{\alpha } T^{\rho } \nabla_{\tau }q_{\alpha \rho }{}^{\tau } -  \frac{4}{3} \alpha_{4}{} \nabla^{\rho }T^{\alpha } \nabla_{\tau }q_{\alpha \rho }{}^{\tau } -  \frac{16}{9} \alpha_{4}{} q_{\alpha \rho \tau } T^{\alpha } \nabla^{\tau }T^{\rho }  \nonumber\\
&&+ \frac{2}{9} (\alpha_{2}{} - 8 \alpha_{3}{} - 4 \alpha_{4}{} + 4 \alpha_{5}{} + 2 \alpha_{6}{}) q_{\alpha }{}^{\tau \nu } q_{\rho \tau \nu } T^{\alpha } T^{\rho }+ \frac{2}{3} (\alpha_{2}{} -  \alpha_{3}{} - 2 \alpha_{4}{} + 2 \alpha_{5}{} + \alpha_{6}{}) q_{\alpha }{}^{\rho \tau } T^{\alpha } \nabla_{\nu }q_{\rho \tau }{}^{\nu } \nonumber\\
&&+ \frac{2}{3} (\alpha_{2}{} + 6 \alpha_{3}{}) q_{\alpha }{}^{\tau \nu } q_{\rho \tau \nu } \nabla^{\rho }T^{\alpha } + \frac{1}{9} (\alpha_{1}{} - 2 \alpha_{2}{} + 2 \alpha_{3}{}) q_{\rho \tau \nu } q^{\rho \tau \nu } T_{\alpha } T^{\alpha }+ \frac{1}{3} (\alpha_{1}{} - 3 \alpha_{2}{} - 4 \alpha_{3}{}) q_{\rho \tau \nu } q^{\rho \tau \nu } \nabla_{\alpha }T^{\alpha } \,,\nonumber\\
&&\\
 16\pi    \mathcal{L}_{\rm g}^{(q-S)}&=&-\,\frac{1}{144} \alpha_{1}{} q_{\rho \tau \nu } q^{\rho \tau \nu } S_{\alpha } S^{\alpha } + \frac{1}{144} (a_{5}{} + a_{6}{} + a_{7}{} + 4 \alpha_{2}{} + 10 \alpha_{3}{}) q_{\alpha }{}^{\tau \nu } q_{\rho \tau \nu } S^{\alpha } S^{\rho } -  \frac{1}{36} \alpha_{4}{} S^{\alpha } S^{\rho } \nabla_{\tau }q_{\alpha \rho }{}^{\tau } \nonumber\\
 &&+ \frac{1}{36} (a_{5}{} -  a_{7}{}) q_{\alpha \rho \tau } S^{\alpha } \nabla^{\tau }S^{\rho } + \frac{1}{6} (\alpha_{2}{} -  \alpha_{3}{}) \varepsilon_{\alpha \nu \lambda \mu } q^{\rho \tau \nu } S^{\alpha } \nabla^{\mu }q_{\rho \tau }{}^{\lambda }\,,\\
16\pi \mathcal{L}_{\rm g}^{(q-RTSW\Lambda)}&=&\frac{8}{3} \alpha_{4}{} q_{\alpha \rho \tau } (R^{\rho \tau }- T^{\rho } W^{\tau }) T^{\alpha } + 4 \alpha_{4}{} q_{\alpha \rho \tau } (T^{\alpha } W^{\rho } W^{\tau }-  R^{\rho \tau } W^{\alpha })+ \frac{8}{3} \alpha_{4}{} q_{\alpha \rho \tau }(W^{\alpha } \nabla^{\tau }T^{\rho }+ T^{\alpha } \nabla^{\tau }W^{\rho })\nonumber\\
&&
-  \frac{4}{3} \alpha_{4}{} T^{\alpha } W^{\rho } \nabla_{\tau }q_{\alpha
\rho }{}^{\tau }+ \frac{1}{54} (- a_{5}{} + a_{7}{} - 2 \alpha_{4}{}) q_{\alpha \rho \tau } S^{\alpha } S^{\rho } T^{\tau } + \frac{1}{3} (- \alpha_{1}{} + 2 \alpha_{2}{} - 2 \alpha_{3}{}) q_{\rho \tau \nu } q^{\rho \tau \nu } T^{\alpha } W_{\alpha } \nonumber\\
&& -  \frac{2}{3} (\alpha_{2}{} - 8 \alpha_{3}{} - 4 \alpha_{4}{} + 4 \alpha_{5}{} + 2 \alpha_{6}{}) q_{\alpha }{}^{\tau \nu } q_{\rho \tau \nu } T^{\alpha } W^{\rho } + \frac{1}{36} (a_{5}{} -  a_{7}{} + 2 \alpha_{4}{}) q_{\alpha \rho \tau } S^{\alpha } S^{\rho } W^{\tau } \nonumber\\
&&  + \frac{1}{12} (3 \alpha_{1}{} - 6 \alpha_{2}{} + 6 \alpha_{3}{} + 2 \alpha_{4}{}) q_{\rho \tau \nu } q^{\rho \tau \nu } T^{\alpha } \Lambda_{\alpha } + \frac{1}{3} (2 \alpha_{2}{} - 2 \alpha_{3}{} - \alpha_{4}{} + 4 \alpha_{5}{} + 2 \alpha_{6}{}) T^{\alpha } \Lambda^{\rho } \nabla_{\tau }q_{\alpha \rho }{}^{\tau } \nonumber\\
&&+ \frac{1}{8} \bigl[-3 \alpha_{1}{} + 6 \alpha_{2}{} - 2 (3 \alpha_{3}{} + \alpha_{4}{})\bigr] q_{\rho \tau \nu } q^{\rho \tau \nu } W^{\alpha } \Lambda_{\alpha } + (\alpha_{2}{} + 6 \alpha_{3}{} + 3 \alpha_{4}{}) q_{\alpha \rho \tau } R^{\rho \tau } \Lambda^{\alpha }  \nonumber\\
&&+\Bigl(\frac{1}{2} \alpha_{2}{} - 4 \alpha_{3}{} - 2 \alpha_{4}{} + 2 \alpha_{5}{} + \alpha_{6}{}\Bigr) q_{\alpha }{}^{\tau \nu } q_{\rho \tau \nu } T^{\alpha } \Lambda^{\rho } -  \frac{3}{4} (\alpha_{2}{} - 8 \alpha_{3}{} - 4 \alpha_{4}{} + 4 \alpha_{5}{} + 2 \alpha_{6}{}) q_{\alpha }{}^{\tau \nu } q_{\rho \tau \nu } W^{\alpha } \Lambda^{\rho }\nonumber\\
&&+ \frac{1}{144} (-7 a_{5}{} - 4 a_{6}{} -  a_{7}{} + 2 \alpha_{2}{} - 16 \alpha_{3}{} - 6 \alpha_{4}{}) q_{\alpha \rho \tau } S^{\alpha } S^{\rho } \Lambda^{\tau } + \frac{2}{9} (5 \alpha_{2}{} + 2 \alpha_{3}{} + \alpha_{4}{} + 8 \alpha_{5}{} + 4 \alpha_{6}{}) q_{\alpha \rho \tau } T^{\alpha } T^{\rho } \Lambda^{\tau } \nonumber\\
&&+\frac{1}{6}(5 \alpha_{2}{} + 2 \alpha_{3}{} + \alpha_{4}{} + 8 \alpha_{5}{} + 4 \alpha_{6}{})(3   W^{\alpha } -  4 T^{\alpha })q_{\alpha \rho \tau }W^{\rho }\Lambda^{\tau } + (2\alpha_{2}{} - 2 \alpha_{3}{} -  \alpha_{4}{} + 4 \alpha_{5}{} + 2\alpha_{6}{}) q_{\alpha \rho \tau } W^{\alpha } \nabla^{\tau}\Lambda^{\rho }\nonumber\\
&&+ \Bigl(\frac{5}{2} \alpha_{2}{} + \alpha_{3}{} -  \frac{3}{4} \alpha_{4}{} + 4 \alpha_{5}{} + 2 \alpha_{6}{}\Bigr) q_{\alpha \rho \tau } T^{\alpha } \Lambda^{\rho } \Lambda^{\tau } -  \frac{3}{8} (10 \alpha_{2}{} + 4 \alpha_{3}{} - 3 \alpha_{4}{} + 16 \alpha_{5}{} + 8 \alpha_{6}{}) q_{\alpha \rho \tau } W^{\alpha } \Lambda^{\rho } \Lambda^{\tau } \nonumber\\
&&+ \Bigl(- \alpha_{2}{} +\alpha_{3}{} + \frac{1}{2} \alpha_{4}{} - 2 \alpha_{5}{} - \alpha_{6}{}\Bigr) W^{\alpha } \Lambda^{\rho } \nabla_{\tau }q_{\alpha \rho }{}^{\tau }  -  \frac{2}{3} (\alpha_{2}{} + 6\alpha_{3}{} + 3 \alpha_{4}{}) q_{\alpha \rho \tau } \Lambda^{\alpha} \nabla^{\tau }T^{\rho }\nonumber\\
&& + (\alpha_{2}{} + 6\alpha_{3}{} + 3 \alpha_{4}{}) q_{\alpha \rho \tau } \Lambda^{\alpha} \nabla^{\tau }W^{\rho } -  \frac{2}{3} (2 \alpha_{2}{} - 2 \alpha_{3}{} -  \alpha_{4}{} + 4 \alpha_{5}{} + 2 \alpha_{6}{}) q_{\alpha \rho \tau } T^{\alpha } \nabla^{\tau }\Lambda^{\rho }\,,
\end{eqnarray}}\normalsize
\small{\begin{eqnarray}
    16\pi \mathcal{L}_{\rm g}^{(q-t)}&=&\frac{1}{4} ( a_{8}{} +  a_{9}{}-a_{10}{} - a_{11}{} ) \Big[4 \nabla_{\rho}t^{\alpha \rho \nu} \nabla_{\beta}q_{\alpha \nu}{}^{\beta} - t^{\alpha \rho \nu} ( - 2 t_{\rho \nu}{}^{\beta} \nabla_{\lambda }q_{\alpha \beta}{}^{\lambda } + t^{\beta}{}_{\rho \nu} \nabla_{\lambda }q_{\alpha \beta}{}^{\lambda }) +4q_{\alpha\beta\rho}t^{\alpha\gamma\beta}\nabla_{\lambda}t^{\rho}{}_{\gamma}{}^\lambda\Big]\nonumber\\
   &&  - \frac{1}{2} (a_{9}{}-a_{11}{} ) \bigl(q_{\alpha \lambda \mu } t^{\alpha \rho \nu} 
t_{\rho \nu}{}^{\beta} t^{\lambda }{}_{\beta}{}^{
\mu }+2t^{\alpha\rho\nu}q_{\alpha\nu\beta}\nabla_{\lambda}t^\lambda{}_\rho{}^\beta\bigr)+ \frac{1}{2} (a_{5}{} -  a_{7}{}) t^{\alpha \rho \nu} \bigl[q_{\alpha
\lambda \mu } t_{\rho}{}^{\beta \lambda } (t_{\beta \nu}{}^{\mu } + t^{\mu }{}_{\nu \beta}) \nonumber\\
&& + q_{\alpha \beta \lambda } (-2 \nabla_{\nu}t^{\beta}{}_{\rho}{}^{\lambda }+ \nabla^{\lambda }t^{\beta}{}_{\rho \nu})\bigr]+ \frac{1}{4} (  2 a_{5}{} - 2 a_{7}{}+a_{8}{} + a_{9}{}- a_{10}{} -  a_{11}{}  
) q_{\alpha \nu \mu } t^{\alpha \rho \nu} t_{\rho}{}^{\beta \lambda } t^{\mu }{}_{\beta \lambda } \nonumber\\
&&-\frac{1}{2} (a_{8}{}-a_{10}{} ) q_{\beta \lambda \mu } 
t^{\alpha \rho \nu} t_{\rho \nu}{}^{\beta} t^{\lambda }{}_{\alpha }{}^{\mu }- \frac{1}{2} \alpha_{4}{} q_{\alpha \rho }{}^{\nu } q_{\tau }{}^{\beta \lambda } q_{\nu \beta \lambda } t^{\alpha \rho \tau } + \frac{1}{24} (- \alpha_{1}{} + 3 \alpha_{2}{} + 4 \alpha_{3}{} -  \alpha_{4}{} -  \alpha_{5}{}) q_{\nu \beta \lambda } q^{\nu \beta \lambda } t_{\alpha \rho \tau } t^{\alpha \rho \tau } \nonumber\\
&&+ \frac{1}{2} (\alpha_{2}{} -  \alpha_{3}{}) q_{\tau }{}^{\beta \lambda } q_{\nu \beta \lambda } t_{\alpha \rho }{}^{\nu } t^{\alpha \rho \tau } + \frac{1}{4} (- \alpha_{2}{} - 6 \alpha_{3}{} -  \alpha_{4}{} -  \alpha_{5}{}) q_{\rho \nu }{}^{\lambda } q_{\tau \beta \lambda } t_{\alpha }{}^{\nu \beta } t^{\alpha \rho \tau } \nonumber\\
&&+ \frac{1}{12} (- \alpha_{1}{} + 3 \alpha_{2}{} + 4 \alpha_{3}{} -  \alpha_{4}{} - \alpha_{5}{}) q_{\nu \beta \lambda } q^{\nu \beta \lambda } t^{\alpha \rho \tau } t_{\rho \alpha \tau } + \frac{1}{2} (\alpha_{2}{} -  \alpha_{3}{}) q_{\tau }{}^{\beta \lambda } q_{\nu \beta \lambda } t^{\alpha \rho \tau } t_{\rho \alpha }{}^{\nu } \nonumber\\
&&+ \frac{1}{2} (- \alpha_{4}{} -  \alpha_{5}{}) q_{\alpha }{}^{\beta \lambda } q_{\nu \beta \lambda } t^{\alpha \rho \tau } t_{\rho \tau }{}^{\nu } + (- \alpha_{2}{} + \alpha_{3}{}) q_{\alpha \nu }{}^{\lambda } q_{\tau \beta \lambda } t^{\alpha \rho \tau } t_{\rho }{}^{\nu \beta } + \frac{1}{8} \beta_{20}{} q_{\alpha }{}^{\beta \lambda } q_{\nu \beta \lambda } t^{\alpha \rho \tau } t^{\nu }{}_{\rho \tau } \nonumber\\
&&+ \Bigl(\alpha_{2}{} + \frac{5}{2} \alpha_{3}{} + \alpha_{4}{} + \alpha_{5}{} -  \frac{1}{4} \beta_{20}{}\Bigr) q_{\alpha \beta }{}^{\lambda } q_{\tau \nu \lambda } t^{\alpha \rho \tau } t^{\nu }{}_{\rho }{}^{\beta } + \frac{1}{2} (- \alpha_{2}{} + 8 \alpha_{3}{} + \alpha_{4}{} + \alpha_{5}{}) q_{\alpha \nu }{}^{\lambda } q_{\tau \beta \lambda } t^{\alpha \rho \tau } t^{\nu }{}_{\rho }{}^{\beta } \nonumber\\
&&+ \frac{1}{16} \beta_{21}{} q_{\alpha \tau }{}^{\lambda } q_{\nu \beta \lambda } t^{\alpha \rho \tau } t^{\nu }{}_{\rho }{}^{\beta } + \frac{1}{16} (32 \alpha_{2}{} + 80 \alpha_{3}{} + 24 \alpha_{4}{} + 8 \alpha_{5}{} - 4 \alpha_{6}{} + \beta_{21}{}) q_{\alpha \rho \beta } q_{\tau \nu \lambda } t^{\alpha \rho \tau } t^{\nu \beta \lambda }\nonumber\\
&&+ (\alpha_{2}{} -  \alpha_{3}{}) q_{\rho }{}^{\nu \beta } t^{\alpha \rho \tau } \nabla_{\alpha }q_{\tau \nu \beta } + (- \alpha_{4}{} -  \alpha_{5}{}) q_{\alpha }{}^{\nu \beta } q_{\tau \nu \beta } \nabla_{\rho }t^{\alpha \rho \tau }+ (\alpha_{2}{} -  \alpha_{3}{}) q_{\rho }{}^{\nu \beta } t^{\alpha \rho \tau } \nabla_{\tau }q_{\alpha \nu \beta } \nonumber\\
&& + (- \alpha_{2}{} + \alpha_{3}{}) q_{\alpha }{}^{\nu \beta } t^{\alpha \rho \tau } \nabla_{\tau }q_{\rho \nu \beta } + 2 (\alpha_{2}{} + 6 \alpha_{3}{} + \alpha_{4}{} + \alpha_{5}{}) q_{\alpha \rho }{}^{\beta } q_{\tau \nu \beta } \nabla^{\nu }t^{\alpha \rho \tau } + (-2 \alpha_{2}{} + 2 \alpha_{3}{}) q_{\rho }{}^{\nu \beta } t^{\alpha \rho \tau } \nabla_{\beta }q_{\alpha \tau \nu } \nonumber\\
&&+ \Bigl(- \alpha_{4}{} + \alpha_{5}{} + \frac{1}{2} \alpha_{6}{}\Bigr) q_{\alpha \rho }{}^{\nu } t^{\alpha \rho \tau } \nabla_{\beta }q_{\tau \nu }{}^{\beta } \,,
\end{eqnarray}}
\small{\begin{eqnarray}
      16\pi  \mathcal{L}_{\rm g}^{(q-\Omega)}&=&\frac{1}{54} {\ast\Omega}^{\alpha \rho \gamma} \big[4 \beta_4{\ast\Omega}_{\alpha }{}^{\beta \lambda } {\ast\Omega}_{\rho \beta}{}^{\mu } q_{\gamma \lambda \mu } +\big(2 \beta_1 {\ast\Omega}_{\alpha \rho}{}^{\beta} + \beta_2{\ast\Omega}_{\rho \alpha }{}^{\beta}\big) {\ast\Omega}^{\lambda }{}_{\gamma}{}^{\mu } q_{\beta \lambda \mu } -  \beta_3 {\ast\Omega}_{\rho \gamma}{}^{\beta} ({\ast\Omega}^{\lambda }{}_{\beta}{}^{\mu } q_{\alpha \lambda \mu } + {\ast\Omega}^{\lambda }{}_{\alpha }{}^{\mu } q_{\beta \lambda \mu })\big]\nonumber\\
       && + \frac{1}{54} {\ast\Omega}_{\rho}{}^{\beta \lambda } \big\{\beta_3{\ast\Omega}^{\alpha }{}_{\beta \lambda } {\ast\Omega}^{\gamma \rho \mu } q_{\alpha \gamma \mu } + 2\beta_6{\ast\Omega}^{\alpha \rho \gamma} {\ast\Omega}^{\mu }{}_{\beta \gamma} q_{\lambda \alpha \mu }+ \frac{2}{3} {\ast\Omega}_{\beta}{}^{\alpha \gamma} \big[\big( 2 \beta_4-\beta_6\big){\ast\Omega}_{\alpha }{}^{\rho \mu } -3\beta_6{\ast\Omega}^{\mu \rho}{}_{\alpha }\big] q_{\lambda \gamma \mu }\big\}\nonumber\\
       &&+\frac{1}{9} \beta_9
{\ast\Omega}^{\alpha \rho \gamma} \bigl[{\ast\Omega}^{\beta}{}_{\rho}{}^{\lambda } (\nabla_{\beta}q_{\alpha \gamma \lambda } -  \nabla_{\lambda }q_{\alpha \gamma \beta}) - 2 {\ast\Omega}_{\rho}{}^{\beta \lambda } \nabla_{\lambda }q_{\alpha \gamma \beta}\bigr]+ \frac{1}{18} \beta_{10}{\ast\Omega}^{\alpha \rho \gamma} (2 {\ast\Omega}_{\rho}{}^{\beta}{}_{\gamma} + {\ast\Omega}^{\beta}{}_{\rho \gamma}) \nabla_{\lambda }q_{\alpha \beta}{}^{\lambda } \nonumber\\
&&+ \frac{1}{18} {\ast\Omega}^{\alpha \gamma \rho} \bigl(\beta_{11} {\ast\Omega}_{\alpha }{}^{\beta}{}_{\rho} + 2 \beta_{12}{\ast\Omega}_{\rho}{}^{\beta}{}_{\alpha }\bigr) \nabla_{\lambda }q_{\gamma \beta}{}^{\lambda }- \frac{1}{9} \beta_{13} {\ast\Omega}^{\alpha \rho \gamma} q_{\beta \lambda \gamma} \nabla_{\alpha }{\ast\Omega}^{\beta}{}_{\rho}{}^{\lambda } + \frac{1}{9} \beta_{14} {\ast\Omega}^{\alpha \rho \gamma} q_{\beta \lambda \gamma} \nabla_{\rho}{\ast\Omega}^{\beta}{}_{\alpha }{}^{\lambda }\nonumber\\
&&-  \frac{1}{9} \beta_{15} {\ast\Omega}^{\alpha \rho \gamma} q_{\lambda \alpha \gamma} \nabla_{\beta}{\ast\Omega}_{\rho}{}^{\beta \lambda } + \frac{1}{3} (- a_{10}{} + a_{11}{}) {\ast\Omega}^{\alpha \rho \gamma} q_{\lambda \alpha \gamma} \nabla_{\beta}{\ast\Omega}^{\beta}{}_{\rho}{}^{\lambda } + \frac{1}{6} \beta_{16} \nabla_{\rho}{\ast\Omega}^{\alpha \rho \gamma} \nabla_{\beta}q_{\alpha \gamma}{}^{\beta} \nonumber\\
&&+ \frac{1}{3} \beta_{9}\nabla_{\gamma}q_{\alpha \rho \beta} \nabla^{\beta}{\ast\Omega}^{\alpha \rho \gamma} + \frac{1}{9} \beta_{18} {\ast\Omega}^{\alpha \rho \gamma} q_{\beta \alpha \gamma} \nabla_{\lambda }{\ast\Omega}^{\beta}{}_{\rho}{}^{\lambda } -  \frac{1}{9} \beta_{19} {\ast\Omega}^{\alpha \rho \gamma} q_{\beta \lambda \gamma} \nabla^{\lambda }{\ast\Omega}_{\alpha \rho}{}^{\beta} + \frac{1}{9}\beta_{17} {\ast\Omega}^{\alpha \rho \gamma} q_{\beta \lambda \gamma} \nabla^{\lambda }{\ast\Omega}^{\beta}{}_{\alpha \rho} \nonumber\\
&&-  \frac{1}{9} {\ast\Omega}^{\alpha \rho \gamma} \big\{\beta_8 q_{\gamma \beta \lambda } \nabla^{\lambda }{\ast\Omega}_{\rho \alpha }{}^{\beta} + q_{\alpha \beta \lambda } \big[\beta_{5} \nabla_{\gamma}{\ast\Omega}^{\beta}{}_{\rho}{}^{\lambda } - \beta_{17}\big(\nabla^{\lambda }{\ast\Omega}_{\rho \gamma}{}^{\beta} -  \nabla^{\lambda }{\ast\Omega}^{\beta}{}_{\rho \gamma}\big)
\big]\big\}\nonumber\\
&&\frac{1}{432} \beta_{27}{} {\ast\Omega}^{\alpha \rho \tau } {\ast\Omega}^{\nu \beta \gamma } q_{\alpha \rho \beta } q_{\tau \nu \gamma } + \frac{1}{36} (\beta_{22}{} - 5 \beta_{23}{} -  \beta_{26}{}) {\ast\Omega}^{\alpha \rho \tau } {\ast\Omega}^{\nu }{}_{\rho }{}^{\beta } q_{\alpha \beta }{}^{\gamma } q_{\tau \nu \gamma } + \frac{1}{18} (\beta_{22}{} + \beta_{23}{}) {\ast\Omega}^{\alpha \rho \tau } {\ast\Omega}_{\rho }{}^{\nu \beta } q_{\alpha \nu }{}^{\gamma } q_{\tau \beta \gamma } \nonumber\\
&&+ \frac{1}{36} (2 \beta_{22}{} - 4 \beta_{23}{} -  \beta_{26}{}) {\ast\Omega}^{\alpha \rho \tau } {\ast\Omega}^{\nu }{}_{\rho }{}^{\beta } q_{\alpha \nu }{}^{\gamma } q_{\tau \beta \gamma } + \frac{1}{36} (\beta_{22}{} + \beta_{23}{}) {\ast\Omega}_{\alpha }{}^{\nu \beta } {\ast\Omega}^{\alpha \rho \tau } q_{\rho \nu }{}^{\gamma } q_{\tau \beta \gamma }\nonumber\\
&&+ \frac{1}{432} \beta_{28}{} {\ast\Omega}^{\alpha \rho \tau } {\ast\Omega}^{\nu }{}_{\rho }{}^{\beta } q_{\alpha \tau }{}^{\gamma } q_{\nu \beta \gamma } -  \frac{1}{54} \beta_{30}{} {\ast\Omega}^{\alpha \rho \tau } {\ast\Omega}_{\rho \tau }{}^{\nu } q_{\alpha }{}^{\beta \gamma } q_{\nu \beta \gamma }+ \frac{1}{108} \beta_{30}{} {\ast\Omega}^{\alpha \rho \tau } {\ast\Omega}^{\nu }{}_{\rho \tau } q_{\alpha }{}^{\beta \gamma } q_{\nu \beta \gamma } \nonumber\\
&& + \frac{1}{108} (-9 \beta_{22}{} + 3 \beta_{23}{} + \beta_{24}{} - 2 \beta_{25}{} + \beta_{26}{}) {\ast\Omega}_{\alpha \rho 
}{}^{\nu } {\ast\Omega}^{\alpha \rho \tau } q_{\tau }{}^{\beta \gamma } q_{\nu \beta \gamma }  + \frac{1}{3} \beta_{22}{} {\ast\Omega}^{\alpha \rho \tau } q_{\rho }{}^{\nu \beta } \nabla_{\alpha }q_{\tau \nu \beta }\nonumber\\
&&+ \frac{1}{108} (-6 \beta_{22}{} - 3 \beta_{23}{} - 3 \beta_{25}{} + \beta_{26}{}) {\ast\Omega}^{\alpha \rho \tau } {\ast\Omega}_{\rho \alpha }{}^{\nu } q_{\tau }{}^{\beta \gamma } q_{\nu \beta \gamma } + \frac{1}{72} (-6 \beta_{22}{} + \beta_{24}{} + 4 \beta_{25}{} - 4 \beta_{26}{}) {\ast\Omega}^{\alpha \rho \tau } q_{\alpha \rho }{}^{\nu } q_{\tau }{}^{\beta \gamma } q_{\nu \beta \gamma } \nonumber\\
&&-  \frac{1}{18} a_{1}{} {\ast\Omega}_{\alpha \rho \tau } {\ast\Omega}^{\alpha \rho \tau } q_{\nu \beta \gamma } q^{\nu \beta \gamma } -  \frac{1}{18} a_{1}{} {\ast\Omega}^{\alpha \rho \tau } {\ast\Omega}_{\rho \alpha \tau } q_{\nu \beta \gamma } q^{\nu \beta \gamma } + \frac{1}{12} \beta_{24}{} q_{\alpha }{}^{\nu \beta } q_{\tau \nu \beta } \nabla_{\rho }{\ast\Omega}^{\alpha \rho \tau }\nonumber\\
&&+ \frac{1}{6} \beta_{22}{} {\ast\Omega}^{\alpha \rho \tau } q_{\rho }{}^{\nu \beta } \nabla_{\tau }q_{\alpha \nu \beta } -  \frac{1}{6} \beta_{22}{} {\ast\Omega}^{\alpha \rho \tau } q_{\alpha }{}^{\nu \beta } \nabla_{\tau }q_{\rho \nu \beta } + \frac{1}{6} (3 \beta_{22}{} + \beta_{26}{}) q_{\alpha \rho }{}^{\beta } q_{\tau \nu \beta } \nabla^{\nu }{\ast\Omega}^{\alpha \rho \tau } -  \frac{1}{2} \beta_{22}{} {\ast\Omega}^{\alpha \rho \tau } q_{\rho }{}^{\nu \beta } \nabla_{\beta }q_{\alpha \tau \nu } \nonumber\\
&&+ \frac{1}{108} \beta_{29}{} {\ast\Omega}^{\alpha \rho \tau } q_{\alpha \rho }{}^{\nu } \nabla_{\beta }q_{\tau \nu }{}^{\beta }
\,,
\end{eqnarray}}
\small{\begin{eqnarray}
     16\pi \mathcal{L}_{\rm g}^{(q-RTStW\Lambda)}&=&\frac{1}{12} (a_{5}{} + a_{6}{} + a_{7}{} + 2 \alpha_{2}{} - 2 \alpha_{3}{}) \varepsilon_{\alpha \nu \lambda \mu } q_{\rho }{}^{\sigma \lambda } q_{\tau \sigma }{}^{\mu } S^{\alpha } t^{\rho \tau \nu } + \frac{1}{12} (- \alpha_{2}{} + 8 \alpha_{3}{} + \alpha_{4}{} + \alpha_{5}{}) \varepsilon_{\alpha \rho \lambda \mu } q_{\tau }{}^{\sigma \lambda } q_{\nu \sigma }{}^{\mu } S^{\alpha } t^{\rho \tau \nu } \nonumber\\
     &&+ \frac{1}{24} (a_{5}{} + a_{6}{} + a_{7}{} - 2 \alpha_{2}{} + 2 \alpha_{3}{}) \varepsilon_{\alpha \tau \nu \mu } q_{\rho }{}^{\sigma \lambda } q_{\sigma \lambda }{}^{\mu } S^{\alpha } t^{\rho \tau \nu } + \frac{1}{12} (- \alpha_{4}{} -  \alpha_{5}{}) \varepsilon_{\alpha \rho \nu \mu } q_{\tau }{}^{\sigma \lambda } q_{\sigma \lambda }{}^{\mu } S^{\alpha } t^{\rho \tau \nu } \nonumber\\
     &&+ \frac{1}{72} (- \alpha_{1}{} + 3 \alpha_{2}{} + 4 \alpha_{3}{} -  \alpha_{4}{} -  \alpha_{5}{}) \varepsilon_{\alpha \rho \tau \nu } q_{\sigma \lambda \mu } q^{\sigma \lambda \mu } S^{\alpha } t^{\rho \tau \nu } \nonumber\\
     &&+ \frac{1}{48} (-4 a_{10}{} - 4 a_{5}{} + 4 a_{7}{} - 4 a_{9}{} - 4 \alpha_{4}{} + \alpha_{6}{}) \varepsilon_{\alpha \sigma \lambda \mu } q_{\rho \nu }{}^{\mu } S^{\alpha } t^{\rho \tau \nu } t_{\tau }{}^{\sigma \lambda } + \frac{1}{12} (a_{5}{} -  a_{7}{}) \varepsilon_{\alpha \rho \lambda \mu } q_{\nu \sigma }{}^{\mu } S^{\alpha } t^{\rho \tau \nu } t_{\tau }{}^{\sigma \lambda }\nonumber\\
     &&- 2 \alpha_{4}{} q_{\rho \tau \nu } R^{\alpha \rho } t^{\tau }{}_{\alpha }{}^{\nu } + \frac{1}{72} (a_{5}{} -  a_{7}{} + 2 \alpha_{4}{}) q_{\rho \tau \nu } S^{\alpha } S^{\rho } t^{\tau }{}_{\alpha }{}^{\nu } + \frac{1}{48} (-4 a_{10}{} - 4 a_{9}{} + \alpha_{6}{}) \varepsilon_{\alpha \sigma \lambda \mu } q_{\rho \nu }{}^{\mu } S^{\alpha } t^{\rho \tau \nu } t^{\sigma }{}_{\tau }{}^{\lambda } \nonumber\\
     &&+ \frac{1}{48} (-4 a_{10}{} + 4 a_{5}{} - 4 a_{7}{} - 4 a_{9}{} - 4 \alpha_{4}{} + \alpha_{6}{}) \varepsilon_{\alpha \nu \lambda \mu } q_{\rho \tau \sigma } S^{\alpha } t^{\rho \tau \nu } t^{\sigma \lambda \mu } \nonumber\\
     &&+ \frac{1}{48} (-4 a_{10}{} - 4 a_{9}{} - 8 \alpha_{4}{} + \alpha_{6}{}) \varepsilon_{\alpha \nu \sigma \mu } q_{\rho \tau \lambda } S^{\alpha } t^{\rho \tau \nu } t^{\sigma \lambda \mu } + \frac{1}{3} (2 \alpha_{2}{} + 12 \alpha_{3}{} + \alpha_{4}{} + \alpha_{5}{}) q_{\rho }{}^{\nu \sigma } q_{\tau \nu \sigma } t^{\rho }{}_{\alpha }{}^{\tau } T^{\alpha }\nonumber\\
     &&+ \frac{2}{3} (7 \alpha_{3}{} + 3 \alpha_{4}{} -  \alpha_{5}{} -  \alpha_{6}{}) q_{\alpha \tau }{}^{\sigma } q_{\rho \nu \sigma } t^{\rho \tau \nu } T^{\alpha } + \frac{1}{3} (- a_{5}{} + a_{7}{} - 4 \alpha_{4}{}) q_{\alpha \rho \sigma } t^{\rho \tau \nu } t_{\tau \nu }{}^{\sigma } T^{\alpha }\nonumber\\
     &&+ \frac{1}{6} (-4 a_{10}{} - 2 a_{5}{} + 2 a_{7}{} - 4 a_{9}{} - 4 \alpha_{4}{} + \alpha_{6}{}) q_{\tau \nu \sigma } t_{\alpha }{}^{\rho \tau } t^{\nu }{}_{\rho }{}^{\sigma } T^{\alpha } + \frac{1}{12} (-4 a_{10}{} - 4 a_{9}{} + \alpha_{6}{}) q_{\tau \nu \sigma } t^{\rho }{}_{\alpha }{}^{\tau } t^{\nu }{}_{\rho }{}^{\sigma } T^{\alpha } \nonumber\\
     &&+ \frac{1}{12} (4 a_{10}{} + 4 a_{5}{} - 4 a_{7}{} + 4 a_{9}{} + 8 \alpha_{4}{} -  \alpha_{6}{}) q_{\rho \nu \sigma } t^{\rho }{}_{\alpha }{}^{\tau } t^{\nu }{}_{\tau }{}^{\sigma } T^{\alpha } + \frac{2}{3} \alpha_{4}{} q_{\alpha \rho \sigma } t^{\rho \tau \nu } t^{\sigma }{}_{\tau \nu } T^{\alpha } \nonumber\\
     &&+ \frac{1}{18} (a_{5}{} -  a_{7}{})
\varepsilon_{\alpha \nu \sigma \lambda } q_{\rho \tau }{}^{\lambda } S^{\alpha } t^{\tau \nu \sigma } T^{\rho } + \frac{1}{18} (- a_{5}{} + a_{7}{} - 4 \alpha_{4}{}) \varepsilon_{\alpha \tau \sigma \lambda }
q_{\rho \nu }{}^{\lambda } S^{\alpha } t^{\tau \nu \sigma } T^{\rho } -  \frac{4}{3} \alpha_{4}{} q_{\rho \tau \nu } t^{\tau }{}_{\alpha }{}^{\nu } T^{\alpha } T^{\rho } \nonumber\\
&&+ \Bigl(- \alpha_{2}{} - 6 \alpha_{3}{} -  \frac{1}{2} \alpha_{4}{} -  \frac{1}{2} \alpha_{5}{}\Bigr) q_{\rho }{}^{\nu \sigma } q_{\tau \nu \sigma } t^{\rho }{}_{\alpha }{}^{\tau } W^{\alpha } + (-7 \alpha_{3}{} - 3 \alpha_{4}{} + \alpha_{5}{} + \alpha_{6}{}) q_{\alpha \tau }{}^{\sigma } q_{\rho \nu \sigma } t^{\rho \tau \nu } W^{\alpha } \nonumber\\
&&+ \frac{1}{2} (a_{5}{} -  a_{7}{} + 4 \alpha_{4}{}) q_{\alpha \rho \sigma } t^{\rho \tau \nu } t_{\tau \nu }{}^{\sigma } W^{\alpha } + \Bigl(a_{10}{} + \frac{1}{2} a_{5}{} -  \frac{1}{2} a_{7}{} + a_{9}{} + \alpha_{4}{} -  \frac{1}{4} \alpha_{6}{}\Bigr) q_{\tau \nu \sigma } t_{\alpha }{}^{\rho \tau } t^{\nu }{}_{\rho }{}^{\sigma } W^{\alpha } \nonumber\\
&&+ \frac{1}{8} (4 a_{10}{} + 4 a_{9}{} -  \alpha_{6}{}) q_{\tau \nu \sigma } t^{\rho }{}_{\alpha }{}^{\tau } t^{\nu }{}_{\rho }{}^{\sigma } W^{\alpha } + \frac{1}{8} (-4 a_{10}{} - 4 a_{5}{} + 4 a_{7}{} - 4 a_{9}{} - 8 \alpha_{4}{} + \alpha_{6}{}) q_{\rho \nu \sigma } t^{\rho }{}_{\alpha }{}^{\tau } t^{\nu }{}_{\tau }{}^{\sigma } W^{\alpha } \nonumber\\
&&-  \alpha_{4}{} q_{\alpha \rho \sigma } t^{\rho \tau \nu } t^{\sigma }{}_{\tau \nu } W^{\alpha } + \frac{1}{12} (- a_{5}{} + a_{7}{}) \varepsilon_{\alpha \nu \sigma \lambda } q_{\rho \tau }{}^{\lambda } S^{\alpha } t^{\tau \nu \sigma } W^{\rho } + \frac{1}{12} (a_{5}{} -  a_{7}{} + 4 \alpha_{4}{}) \varepsilon_{\alpha \tau \sigma \lambda } q_{\rho \nu }{}^{\lambda } S^{\alpha } t^{\tau \nu \sigma } W^{\rho } \nonumber\\
&&+ 2 \alpha_{4}{} q_{\rho \tau \nu } t^{\tau }{}_{\alpha }{}^{\nu } T^{\alpha } W^{\rho } + 2 \alpha_{4}{} q_{\alpha \tau \nu } t^{\tau }{}_{\rho }{}^{\nu } T^{\alpha } W^{\rho } - 3 \alpha_{4}{} q_{\rho \tau \nu } t^{\tau }{}_{\alpha }{}^{\nu } W^{\alpha } W^{\rho } \nonumber\\
&&+ \frac{1}{8} (6 \alpha_{2}{} + 36 \alpha_{3}{} + 13 \alpha_{4}{} + 3 \alpha_{5}{}) q_{\rho }{}^{\nu \sigma } q_{\tau \nu \sigma } t^{\rho }{}_{\alpha }{}^{\tau } \Lambda^{\alpha } + \frac{1}{4} \bigl[21 \alpha_{3}{} + 7 \alpha_{4}{} - 3 \bigl(\alpha_{5}{} + \alpha_{6}{}\bigr)\bigr] q_{\alpha \tau }{}^{\sigma } q_{\rho \nu \sigma } t^{\rho \tau \nu } \Lambda^{\alpha } \nonumber\\
&&+ \frac{1}{2} (\alpha_{2}{} -  \alpha_{3}{}) q_{\alpha \nu \sigma } t_{\rho \tau }{}^{\sigma } t^{\rho \tau \nu } \Lambda^{\alpha } + \frac{1}{2} (\alpha_{2}{} -  \alpha_{3}{}) q_{\alpha \nu \sigma } t^{\rho \tau \nu } t_{\tau \rho }{}^{\sigma } \Lambda^{\alpha } \nonumber\\
&&+ \frac{1}{8} (-5 a_{5}{} - 2 a_{6}{} + a_{7}{} - 28 \alpha_{3}{} - 12 \alpha_{4}{}) q_{\alpha \rho \sigma } t^{\rho \tau \nu } t_{\tau \nu }{}^{\sigma } \Lambda^{\alpha } \nonumber\\
&&+ \frac{1}{16} (-28 a_{10}{} + 16 a_{12}{} - 2 a_{5}{} + 4 a_{6}{} + 10 a_{7}{} + 4 a_{9}{} + 4 \alpha_{4}{} + 16 \alpha_{5}{} + 7 \alpha_{6}{}) q_{\tau \nu \sigma } t_{\alpha }{}^{\rho \tau } t^{\nu }{}_{\rho }{}^{\sigma } \Lambda^{\alpha }\nonumber\\
&&+ \Bigl(- \frac{1}{8} a_{10}{} -  \frac{1}{4} a_{12}{} -  \frac{5}{8} a_{9}{} + \alpha_{2}{} -  \alpha_{3}{} -  \frac{3}{2} \alpha_{4}{} + \alpha_{5}{} + \frac{21}{32} \alpha_{6}{}\Bigr) q_{\tau \nu \sigma } t^{\rho }{}_{\alpha }{}^{\tau } t^{\nu }{}_{\rho }{}^{\sigma } \Lambda^{\alpha }\nonumber\\
&&+ \frac{1}{32} (4 a_{10}{} + 8 a_{12}{} + 4 a_{5}{} - 8 a_{6}{} - 20 a_{7}{} + 20 a_{9}{} + 32 \alpha_{2}{} - 32 \alpha_{3}{} - 8 \alpha_{4}{} + 48 \alpha_{5}{} + 19 \alpha_{6}{}) q_{\rho \nu \sigma } t^{\rho }{}_{\alpha }{}^{\tau } t^{\nu }{}_{\tau }{}^{\sigma } \Lambda^{\alpha }\nonumber\\
&&+ \frac{1}{4} (- a_{5}{} -  a_{6}{} -  a_{7}{} + \alpha_{2}{} + 6 \alpha_{3}{} + 3 \alpha_{4}{}) q_{\alpha \rho \sigma } t^{\rho \tau \nu } t^{\sigma }{}_{\tau \nu } \Lambda^{\alpha } \nonumber\\
&&+ \frac{1}{48} (9 a_{5}{} + 6 a_{6}{} + 3 a_{7}{} - 4 \alpha_{2}{} + 4 \alpha_{3}{}) \varepsilon_{\alpha \nu \sigma \lambda } q_{\rho \tau }{}^{\lambda } S^{\alpha } t^{\tau \nu \sigma } \Lambda^{\rho } + \frac{1}{12} (- a_{5}{} + a_{7}{}) \varepsilon_{\alpha \tau \nu \lambda } q_{\rho \sigma }{}^{\lambda } S^{\alpha } \nabla^{\sigma }t^{\rho \tau \nu } \nonumber\\
&&+ \frac{1}{48} (-5 a_{5}{} - 2 a_{6}{} + a_{7}{} - 28 \alpha_{3}{} - 12 \alpha_{4}{}) \varepsilon_{\alpha \tau \sigma \lambda } q_{\rho \nu }{}^{\lambda } S^{\alpha } t^{\tau \nu \sigma } \Lambda^{\rho } \nonumber\\
&&+ \frac{1}{16} (-4 a_{10}{} + 4 a_{12}{} + 4 a_{9}{} + 4 \alpha_{4}{} + 4 \alpha_{5}{} + \alpha_{6}{}) \varepsilon_{\alpha \rho \sigma \lambda } q_{\tau \nu }{}^{\lambda } S^{\alpha } t^{\tau \nu \sigma } \Lambda^{\rho } \nonumber\\
&&+ \frac{1}{6} (-4 \alpha_{2}{} - 10 \alpha_{3}{} - 5 \alpha_{4}{} - 4 \alpha_{5}{} - 2 \alpha_{6}{}) q_{\rho \tau \nu } t^{\tau }{}_{\alpha }{}^{\nu } T^{\alpha } \Lambda^{\rho } + \Bigl(-2 \alpha_{2}{} + 2 \alpha_{3}{} + \frac{5}{2} \alpha_{4}{} - 4 \alpha_{5}{} - 2 \alpha_{6}{}\Bigr) q_{\alpha \tau \nu } t^{\tau }{}_{\rho }{}^{\nu } T^{\alpha } \Lambda^{\rho } \nonumber\\
&&+ \Bigl(\alpha_{2}{} + \frac{5}{2} \alpha_{3}{} + \frac{5}{4} \alpha_{4}{} + \alpha_{5}{} + \frac{1}{2} \alpha_{6}{}\Bigr) q_{\rho \tau \nu } t^{\tau }{}_{\alpha }{}^{\nu } W^{\alpha } \Lambda^{\rho } + \Bigl(3 \alpha_{2}{} - 3 \alpha_{3}{} -  \frac{15}{4} \alpha_{4}{} + 6 \alpha_{5}{} + 3 \alpha_{6}{}\Bigr) q_{\alpha \tau \nu } t^{\tau }{}_{\rho }{}^{\nu } W^{\alpha } \Lambda^{\rho } \nonumber\\
&&-  \frac{3}{16} (16 \alpha_{2}{} - 2 \alpha_{3}{} - 11 \alpha_{4}{} + 28 \alpha_{5}{} + 14 \alpha_{6}{}) q_{\rho \tau \nu } t^{\tau }{}_{\alpha }{}^{\nu } \Lambda^{\alpha } \Lambda^{\rho } + \frac{1}{12} (- a_{5}{} + a_{7}{}) \varepsilon_{\alpha \nu \sigma \lambda } q_{\rho \tau }{}^{\lambda } t^{\tau \nu \sigma } \nabla^{\rho }S^{\alpha }\nonumber\\
&&+ \frac{1}{24} (-4 a_{10}{} - 4 a_{9}{} - 4
\alpha_{4}{} + \alpha_{6}{}) \varepsilon_{\alpha \rho \sigma \lambda } q_{\tau \nu }{}^{\lambda } t^{\tau \nu \sigma } \nabla^{\rho }S^{\alpha } \nonumber\\
&&+ \frac{1}{6} (-4 a_{10}{} - 2 a_{5}{} + 2 a_{7}{} - 4 a_{9}{} + \alpha_{6}{}) q_{\alpha \tau \nu } t^{\tau }{}_{\rho }{}^{\nu } \nabla^{\rho }T^{\alpha }+ \frac{1}{6} (- a_{5}{} + a_{7}{}) \varepsilon_{\alpha \nu \sigma \lambda } q_{\rho \tau }{}^{\lambda } S^{\alpha } \nabla^{\sigma }t^{\rho \tau \nu }\nonumber\\
&&+ \Bigl(- a_{12}{} + a_{15}{} -  a_{16}{} + \frac{1}{2}
a_{6}{} + a_{7}{} - 2 a_{9}{} - 3 \alpha_{4}{} -  \alpha_{5}{}\Bigr) q_{\rho \tau \nu } t^{\tau }{}_{\alpha }{}^{\nu } \nabla^{\rho }W^{\alpha } \nonumber\\
&&+ \Bigl(a_{12}{} -  a_{15}{} + a_{16}{} -  \frac{1}{2} a_{6}{} - a_{7}{} + 2 a_{9}{} + \alpha_{4}{} + \alpha_{5}{}\Bigr) q_{\alpha \tau \nu } t^{\tau }{}_{\rho }{}^{\nu } \nabla^{\rho }W^{\alpha } \nonumber\\
&&+ \frac{1}{8} (12 a_{10}{} - 6 a_{12}{} - 3 a_{6}{} - 6 a_{7}{} + 8 \alpha_{2}{} - 8 \alpha_{3}{} - 2 \alpha_{4}{} + 2 \alpha_{5}{} + \alpha_{6}{}) q_{\rho \tau \nu } t^{\tau }{}_{\alpha }{}^{\nu } \nabla^{\rho }\Lambda^{\alpha } \nonumber\\
&&+ \frac{1}{8} (-12 a_{10}{} + 6 a_{12}{} + 3 a_{6}{} + 6 a_{7}{} - 2 \alpha_{4}{} + 14 \alpha_{5}{} + 7 \alpha_{6}{}) q_{\alpha \tau \nu } t^{\tau }{}_{\rho }{}^{\nu } \nabla^{\rho }\Lambda^{\alpha } -  \frac{8}{3} \alpha_{4}{} q_{\alpha \rho \nu } T^{\alpha } \nabla_{\tau }t^{\rho \tau \nu } \nonumber\\
&&+ 4 \alpha_{4}{} q_{\alpha \rho \nu } W^{\alpha } \nabla_{\tau }t^{\rho \tau \nu } + \bigl[- \alpha_{2}{} - 3 \bigl(2 \alpha_{3}{} + \alpha_{4}{}\bigr)\bigr] q_{\alpha \rho \nu } \Lambda^{\alpha } \nabla_{\tau }t^{\rho \tau \nu } + (\alpha_{2}{} -  \alpha_{3}{}) t^{\rho \tau \nu } \Lambda^{\alpha } \nabla_{\nu }q_{\alpha \rho \tau } \nonumber\\
&&-  \frac{2}{3} \alpha_{4}{} t^{\rho }{}_{\alpha }{}^{\tau } T^{\alpha } \nabla_{\nu }q_{\rho \tau }{}^{\nu } + \alpha_{4}{} t^{\rho }{}_{\alpha }{}^{\tau } W^{\alpha } \nabla_{\nu }q_{\rho \tau }{}^{\nu } + \Bigl[- \alpha_{2}{} + \alpha_{3}{} + \frac{7}{4} \alpha_{4}{} -  \frac{5}{4} \bigl(2 \alpha_{5}{} + \alpha_{6}{}\bigr)\Bigr] t^{\rho }{}_{\alpha }{}^{\tau } \Lambda^{\alpha } \nabla_{\nu }q_{\rho \tau }{}^{\nu } \nonumber\\
&&-  \frac{1}{6} \alpha_{4}{} \varepsilon_{\alpha \rho \nu \lambda } S^{\alpha } t^{\rho \tau \nu } \nabla_{\sigma }q_{\tau }{}^{\sigma \lambda }+ \frac{1}{6} (4 a_{10}{} + 2 a_{5}{} - 2 a_{7}{} + 4 a_{9}{} + 8 \alpha_{4}{} -  \alpha_{6}{}) q_{\rho \tau \nu } t^{\tau }{}_{\alpha }{}^{\nu } \nabla^{\rho }T^{\alpha }   \,,
\end{eqnarray}}
\small{\begin{eqnarray}
     16\pi \mathcal{L}_{\rm g}^{(q-RTSW\Lambda\Omega)}&=&\frac{1}{3} (4 \alpha_{4}{} + \alpha_{5}{}) {\ast\Omega}^{\alpha \rho \tau } q_{\alpha \tau \nu } R_{\rho }{}^{\nu } -  \frac{2}{3} (\alpha_{2}{} + 6 \alpha_{3}{} + \alpha_{4}{} + \alpha_{5}{}) {\ast\Omega}^{\alpha \rho \tau } q_{\rho }{}^{\nu \sigma } R_{\alpha \nu \tau \sigma } \nonumber\\
&&
+ \frac{1}{108} (-2 a_{10}{} + a_{11}{} -  a_{12}{} + 2 a_{13}{} + 4 \
a_{2}{} - 2 a_{3}{} - 3 a_{4}{} + 2 a_{5}{} - 2 a_{7}{})( \varepsilon_{\alpha \nu \lambda \mu } q_{\rho \tau \sigma } +\varepsilon_{\alpha \nu \sigma \mu }  q_{\rho \tau \lambda }){\ast\Omega}^{\rho \tau \nu }{\ast\Omega}^{\sigma \lambda \mu }S^{\alpha }  \nonumber\\
 \nonumber\\
&&
+ \frac{1}{108} (-3 a_{10}{} - 3 a_{12}{} + 2 a_{6}{} + 4 a_{7}{} - 4 \alpha_{2}{} + 4 \alpha_{3}{} - 2 \alpha_{5}{}) ({\ast\Omega}_{\tau }{}^{\sigma \lambda } +{\ast\Omega}^{\sigma }{}_{\tau }{}^{\lambda } )\varepsilon_{\alpha \sigma \lambda \mu } {\ast\Omega}^{\rho \tau \nu }  q_{\rho \nu }{}^{\mu } S^{\alpha }\nonumber\\
&&  + \frac{1}{108} (2 a_{2}{} + 4 a_{3}{} + 2 a_{4}{} + a_{5}{} - 2 
a_{6}{} - 4 a_{7}{}) (\varepsilon_{\alpha \nu \lambda \mu } {\ast\Omega}_{\tau }{}^{\sigma \lambda }q_{\rho \sigma }{}^{\mu }+ \varepsilon_{\alpha \rho \lambda \mu } {\ast\Omega}_{\tau }{}^{\sigma \lambda } q_{\nu \sigma }{}^{\mu }  -\varepsilon_{\alpha \nu \sigma \mu } {\ast\Omega}^{\sigma }{}_{\tau }{}^{\lambda } q_{\rho \lambda }{}^{\mu } ){\ast\Omega}^{\rho \tau \nu }  S^{\alpha } \nonumber\\
&&
+ \frac{1}{72} \bigl[- a_{6}{} + 2 \bigl(- a_{7}{} - 2 \alpha_{2}{} + 9 \alpha_{3}{} + \alpha_{4}{} + \alpha_{5}{}\bigr)\bigr] (\varepsilon_{\alpha \nu \lambda \mu }  q_{\rho }{}^{\sigma \lambda } q_{\tau \sigma }{}^{\mu }  -\varepsilon_{\alpha \rho \lambda \mu } q_{\tau }{}^{\sigma \lambda } q_{\nu \sigma }{}^{\mu } ){\ast\Omega}^{\rho \tau \nu } S^{\alpha }\nonumber\\
&&
 + \frac{1}{72} (- a_{6}{} - 2 a_{7}{} + 2 \alpha_{2}{} - 2 \alpha_{3}{}
+ \alpha_{5}{}) (\varepsilon_{\alpha \tau \nu \mu }  q_{\rho }{}^{\sigma \lambda } +\varepsilon_{\alpha \rho \nu \mu }  q_{\tau }{}^{\sigma \lambda }  ){\ast\Omega}^{\rho \tau \nu } S^{\alpha }q_{\sigma \lambda }{}^{\mu } \nonumber\\
&&+ \frac{1}{432} (4 a_{5}{} + 7 a_{6}{} + 10 a_{7}{} - 8 \alpha_{2}{} + 22 \alpha_{3}{} - 6 \alpha_{4}{}) {\ast\Omega}^{\tau }{}_{\alpha }{}^{\nu } q_{\rho \tau \nu } S^{\alpha } S^{\rho }\nonumber\\
&& + \frac{1}{27} (2 a_{4}{} + 2 a_{7}{} + 2 \alpha_{2}{} - 9 \alpha_{3}{} - 9 \alpha_{4}{} + 3 \alpha_{5}{} + 2 \alpha_{6}{}) ( 2{\ast\Omega}_{\tau \nu }{}^{\sigma } - {\ast\Omega}^{\sigma }{}_{\tau \nu } ){\ast\Omega}^{\rho \tau \nu } T^{\alpha }q_{\alpha \rho \sigma } \nonumber\\
&& + \frac{1}{27} \bigl[- a_{4}{} -  a_{7}{} + 2 \bigl(4 \alpha_{2}{} + 3 \alpha_{3}{} + \alpha_{4}{} + 3 \alpha_{5}{}+ \alpha_{6}{}\bigr)\bigr] {\ast\Omega}_{\rho \tau }{}^{\sigma } {\ast\Omega}^{\rho \tau \nu } q_{\alpha \nu \sigma } T^{\alpha }  \nonumber\\
&&+ \frac{1}{27} (a_{4}{} + a_{7}{} + 10 \alpha_{2}{} - 3 \alpha_{3}{} - 7 \alpha_{4}{} + 9 \alpha_{5}{} + 4 \alpha_{6}{}) {\ast\Omega}^{\rho \tau \nu } {\ast\Omega}_{\tau \rho }{}^{\sigma } q_{\alpha \nu \sigma } T^{\alpha } \nonumber\\
&&+ \frac{1}{54} (18 a_{10}{} - 18 a_{4}{} - 18 a_{7}{} + 12 \alpha_{2}{} + 30 \alpha_{3}{} + 14 \alpha_{4}{} + 6 \alpha_{5}{} - 3 \alpha_{6}{}) {\ast\Omega}^{\rho }{}_{\alpha }{}^{\tau } {\ast\Omega}^{\nu }{}_{\tau }{}^{\sigma } q_{\rho \nu \sigma } T^{\alpha }\nonumber\\
&& + \frac{1}{9} \bigl[-21 \alpha_{3}{} - 11 \alpha_{4}{} + 3 \bigl(\alpha_{5}{} + \alpha_{6}{}\bigr)\bigr] {\ast\Omega}^{\rho \tau \nu } q_{\alpha \tau }{}^{\sigma } q_{\rho \nu \sigma } T^{\alpha }\nonumber\\
&&
+ \frac{1}{36} (-24 a_{10}{} + 12 a_{4}{} + 12 a_{7}{} - 8 \alpha_{2}{} - 48 \alpha_{3}{} - 8 \alpha_{4}{} - 8 \alpha_{5}{} + 3 \alpha_{6}{}) {\ast\Omega}_{\alpha }{}^{\rho \tau } {\ast\Omega}^{\nu }{}_{\rho }{}^{\sigma } q_{\tau \nu \sigma } T^{\alpha } \nonumber\\
&&+ \frac{1}{108} (-36 a_{10}{} - 84 \alpha_{3}{} + 4 \alpha_{4}{} - 12 \alpha_{5}{} + 3 \alpha_{6}{}) {\ast\Omega}^{\rho }{}_{\alpha }{}^{\tau } {\ast\Omega}^{\nu }{}_{\rho }{}^{\sigma } q_{\tau \nu \sigma } T^{\alpha } \nonumber\\
&&+ \frac{1}{18} (-6 \alpha_{2}{} - 36 \alpha_{3}{} - 2 \alpha_{4}{} - 3 \alpha_{5}{}) {\ast\Omega}^{\rho }{}_{\alpha }{}^{\tau } q_{\rho }{}^{\nu \sigma } q_{\tau \nu \sigma } T^{\alpha }+ \frac{1}{12} (6 \alpha_{2}{} + 36 \alpha_{3}{} + 2 \alpha_{4}{} + 3 \alpha_{5}{}) {\ast\Omega}^{\rho }{}_{\alpha }{}^{\tau } q_{\rho }{}^{\nu \sigma } q_{\tau \nu \sigma } W^{\alpha }  \nonumber\\
&&+ \frac{1}{108} \bigl[8 a_{4}{} - 3 a_{6}{} + 2 \bigl(a_{7}{} - 2 \alpha_{2}{} - 12 \alpha_{3}{} - 2 \alpha_{4}{} + \alpha_{6}{}\bigr)\bigr] (\varepsilon_{\alpha \nu \sigma \lambda }  q_{\rho \tau }{}^{\lambda }  +\varepsilon_{\alpha \tau \sigma \lambda }  q_{\rho \nu }{}^{\lambda }){\ast\Omega}^{\tau \nu \sigma } S^{\alpha } T^{\rho }\nonumber\\
&&+  \frac{1}{18} (-2 a_{4}{} + a_{6}{} + 7 \alpha_{3}{} + \alpha_{4}{} + \alpha_{5}{}) \varepsilon_{\alpha \rho \sigma \lambda } {\ast\Omega}^{\tau \nu \sigma } q_{\tau \nu }{}^{\lambda } S^{\alpha } T^{\rho }\nonumber\\
&&-  \frac{2}{27} (2 \alpha_{2}{} + 5 \alpha_{3}{} - 11 \alpha_{4}{} + 2 \alpha_{5}{} + \alpha_{6}{}) {\ast\Omega}^{\tau }{}_{\alpha }{}^{\nu } q_{\rho \tau \nu } T^{\alpha } T^{\rho }+ \frac{1}{6} \bigl[21 \alpha_{3}{} + 11 \alpha_{4}{} - 3 \big(\alpha_{5}{} + \alpha_{6}{}\big)\bigr] {\ast\Omega}^{\rho \tau \nu } q_{\alpha \tau }{}^{\sigma } q_{\rho \nu \sigma } W^{\alpha } \nonumber\\
&&
+ \frac{1}{18} (-2 a_{4}{} - 2 a_{7}{} - 2 \alpha_{2}{} + 9 \alpha_{3}{} + 9 \alpha_{4}{} - 3 \alpha_{5}{} - 2 \alpha_{6}{}) ( 2{\ast\Omega}_{\tau \nu }{}^{\sigma } - {\ast\Omega}^{\sigma }{}_{\tau \nu } ){\ast\Omega}^{\rho \tau \nu }q_{\alpha \rho \sigma } W^{\alpha }\nonumber\\
&& + \frac{1}{18} \bigl[a_{4}{} + a_{7}{} - 2 \bigl(4 \alpha_{2}{} + 3 \alpha_{3}{} + \alpha_{4}{} + 3 \alpha_{5}{} + \alpha_{6}{}\bigr)\bigr] {\ast\Omega}_{\rho \tau }{}^{\sigma } {\ast\Omega}^{\rho \tau \nu } q_{\alpha \nu \sigma } W^{\alpha } \nonumber\\
&&+ \frac{1}{18} (- a_{4}{} -  a_{7}{} - 10 \alpha_{2}{} + 3 \alpha_{3}{} + 7 \alpha_{4}{} - 9 \alpha_{5}{} - 4 \alpha_{6}{}) {\ast\Omega}^{\rho \tau \nu } {\ast\Omega}_{\tau \rho }{}^{\sigma } q_{\alpha \nu \sigma } W^{\alpha } \nonumber\\
&&
+ \frac{1}{36} (-18 a_{10}{} + 18 a_{4}{} + 18 a_{7}{} - 12 \alpha_{2}{} - 30 \alpha_{3}{} - 14 \alpha_{4}{} - 6 \alpha_{5}{} + 3 \alpha_{6}{}) {\ast\Omega}^{\rho }{}_{\alpha }{}^{\tau } {\ast\Omega}^{\nu }{}_{\tau }{}^{\sigma } q_{\rho \nu \sigma } W^{\alpha }\nonumber\\
&&
 + \frac{1}{24} (24 a_{10}{} - 12 a_{4}{} - 12 a_{7}{} + 8 \alpha_{2}{} + 48 \alpha_{3}{} + 8 \alpha_{4}{} + 8 \alpha_{5}{} - 3 \alpha_{6}{}) {\ast\Omega}_{\alpha }{}^{\rho \tau } {\ast\Omega}^{\nu }{}_{\rho }{}^{\sigma } q_{\tau \nu \sigma } W^{\alpha } \nonumber\\
&&+ \frac{1}{72} (36 a_{10}{} + 84 \alpha_{3}{} - 4 \alpha_{4}{} + 12 \alpha_{5}{} - 3 \alpha_{6}{}) {\ast\Omega}^{\rho }{}_{\alpha }{}^{\tau } {\ast\Omega}^{\nu }{}_{\rho }{}^{\sigma } q_{\tau \nu \sigma } W^{\alpha } \nonumber\\
&& + \frac{1}{72} (-8 a_{4}{} + 3 a_{6}{} - 2 a_{7}{} + 4 \alpha_{2}{} + 24 \alpha_{3}{} + 4 \alpha_{4}{} - 2 \alpha_{6}{}) (\varepsilon_{\alpha \nu \sigma \lambda }  q_{\rho \tau }{}^{\lambda }  +\varepsilon_{\alpha \tau \sigma \lambda } q_{\rho \nu }{}^{\lambda }){\ast\Omega}^{\tau \nu \sigma } S^{\alpha } W^{\rho }\nonumber\\
&&+ \frac{1}{12} (2 a_{4}{} -  a_{6}{} - 7 \alpha_{3}{} -  \alpha_{4}{} -  \alpha_{5}{}) \varepsilon_{\alpha \rho \sigma \lambda } {\ast\Omega}^{\tau \nu \sigma } q_{\tau \nu }{}^{\lambda } S^{\alpha } W^{\rho } \nonumber\\
&&+ \frac{1}{9} (2 \alpha_{2}{} + 5 \alpha_{3}{} - 11 \alpha_{4}{} + 2 \alpha_{5}{} + \alpha_{6}{}) ({\ast\Omega}^{\tau }{}_{\rho }{}^{\nu } q_{\alpha \tau \nu }  + {\ast\Omega}^{\tau }{}_{\alpha }{}^{\nu } q_{\rho \tau \nu } )T^{\alpha } W^{\rho } \nonumber\\
&&+ \frac{1}{6} (-2 \alpha_{2}{} - 5 \alpha_{3}{} + 11 \alpha_{4}{} - 2 
\alpha_{5}{} -  \alpha_{6}{}) {\ast\Omega}^{\tau }{}_{\alpha }{}^{\nu } q_{\rho \tau \nu } W^{\alpha } W^{\rho } \nonumber\\
&&+ \frac{1}{36} (10 a_{4}{} + 10 a_{7}{} + 2 \alpha_{2}{} - 65 \alpha_{3}{} - 25 \alpha_{4}{} + 7 
\alpha_{5}{} + 6 \alpha_{6}{}) ({\ast\Omega}_{\tau \nu }{}^{\sigma }  -\frac{1}{2} {\ast\Omega}^{\sigma }{}_{\tau \nu } ){\ast\Omega}^{\rho \tau \nu }\Lambda^{\alpha }q_{\alpha \rho \sigma }   \nonumber\\
&&+ \frac{1}{72} (-5 a_{4}{} - 5 a_{7}{} + 40 \alpha_{2}{} + 58 \alpha_{3}{} + 14 \alpha_{4}{} + 22 \alpha_{5}{}+ 6 \alpha_{6}{}) {\ast\Omega}_{\rho \tau }{}^{\sigma } {\ast\Omega}^{\rho \tau \nu } q_{\alpha \nu \sigma } \Lambda^{\alpha }  \nonumber\\
&&+ \frac{1}{72} (5 a_{4}{} + 5 a_{7}{} + 42 \alpha_{2}{} - 7 \alpha_{3}{} - 11 \alpha_{4}{} + 29 \alpha_{5}{} + 12 \alpha_{6}{}) {\ast\Omega}^{\rho \tau \nu } {\ast\Omega}_{\tau \rho }{}^{\sigma } q_{\alpha \nu \sigma } \Lambda^{\alpha } \nonumber\\
&&+ \frac{1}{144} (90 a_{10}{} - 90 a_{4}{} - 90 a_{7}{} + 92 \alpha_{2}{} + 174 \alpha_{3}{} - 18 \alpha_{4}{} + 78 \alpha_{5}{} + 9 \alpha_{6}{}) {\ast\Omega}^{\rho }{}_{\alpha }{}^{\tau } {\ast\Omega}^{\nu }{}_{\tau }{}^{\sigma } q_{\rho \nu \sigma } \Lambda^{\alpha } \nonumber\\
&&+ \Bigl[- \frac{1}{3} \alpha_{2}{} + \frac{1}{8} \bigl(-37 \alpha_{3}{} - 11 \alpha_{4}{} + \alpha_{5}{} + 3 \alpha_{6}{}\bigr)\Bigr] {\ast\Omega}^{\rho \tau \nu } q_{\alpha \tau }{}^{\sigma } q_{\rho \nu \sigma } \Lambda^{\alpha } \nonumber\\
&&-  \frac{5}{96} (24 a_{10}{} - 12 a_{4}{} - 12 a_{7}{} + 8 \alpha_{2}{} + 48 \alpha_{3}{} + 8 \alpha_{4}{} + 8 \alpha_{5}{} - 3 \alpha_{6}{}) {\ast\Omega}_{\alpha }{}^{\rho \tau } {\ast\Omega}^{\nu }{}_{\rho }{}^{\sigma } q_{\tau \nu \sigma } \Lambda^{\alpha } \nonumber\\
&&+ \frac{1}{288} (-180 a_{10}{} + 64 \alpha_{2}{} - 372 \alpha_{3}{} - 156 \alpha_{4}{} + 36 \alpha_{5}{} + 63 \alpha_{6}{}) {\ast\Omega}^{\rho }{}_{\alpha }{}^{\tau } {\ast\Omega}^{\nu }{}_{\rho }{}^{\sigma } q_{\tau \nu \sigma } 
\Lambda^{\alpha } \nonumber\\
&&+ \frac{1}{48} \bigl[-14 \alpha_{2}{} - 3 \bigl(28 \alpha_{3}{} + 14 \alpha_{4}{} + 5 \alpha_{5}{}\bigr)\bigr] {\ast\Omega}^{\rho }{}_{\alpha }{}^{\tau } q_{\rho }{}^{\nu \sigma } q_{\tau \nu \sigma } \Lambda^{\alpha } \nonumber\\
&&+ \frac{1}{288} \bigl[24 a_{4}{} - 19 a_{6}{} - 2 \bigl(7 a_{7}{} + 2 \alpha_{2}{} + 26 \alpha_{3}{} + 4 \alpha_{4}{} - 3 \alpha_{6}{}\bigr)\bigr] (\varepsilon_{\alpha \nu \sigma \lambda }  q_{\rho \tau }{}^{\lambda } +\varepsilon_{\alpha \tau \sigma \lambda } q_{\rho \nu }{}^{\lambda } )  {\ast\Omega}^{\tau \nu \sigma }S^{\alpha } \Lambda^{\rho }\nonumber\\
&&  + \frac{1}{16} (4 a_{10}{} - 2 a_{12}{} - 2 a_{4}{} + a_{6}{} + 7 \alpha_{3}{} + \alpha_{4}{} -  \alpha_{5}{} -  \alpha_{6}{}) \varepsilon_{\alpha \rho \sigma \lambda } {\ast\Omega}^{\tau \nu \sigma } q_{\tau \nu }{}^{\lambda } S^{\alpha } \Lambda^{\rho } \nonumber\\
&&+ \frac{1}{12} (10 \alpha_{2}{} - 17 \alpha_{3}{} - 21 \alpha_{4}{} + 22 \alpha_{5}{} + 11 \alpha_{6}{}) {\ast\Omega}^{\tau }{}_{\rho }{}^{\nu } q_{\alpha \tau \nu } T^{\alpha } \Lambda^{\rho } \nonumber\\
&&+ \frac{1}{12} (2 \alpha_{2}{} + 5 \alpha_{3}{} + 5 \alpha_{4}{} + 2 \alpha_{5}{} + \alpha_{6}{}) {\ast\Omega}^{\tau }{}_{\alpha }{}^{\nu } q_{\rho \tau \nu } T^{\alpha } \Lambda^{\rho } \nonumber\\
&&+ \frac{1}{8} (-10 \alpha_{2}{} + 17 \alpha_{3}{} + 21 \alpha_{4}{} - 22 \alpha_{5}{} - 11 \alpha_{6}{}) {\ast\Omega}^{\tau }{}_{\rho }{}^{\nu } q_{\alpha \tau \nu } W^{\alpha } \Lambda^{\rho } \nonumber\\
&& + \frac{1}{8} (-2 \alpha_{2}{} - 5 \alpha_{3}{} - 5 \alpha_{4}{} - 2 \alpha_{5}{} -  \alpha_{6}{}) {\ast\Omega}^{\tau }{}_{\alpha }{}^{\nu } q_{\rho \tau \nu } W^{\alpha } \Lambda^{\rho } \nonumber\\
&& + \frac{1}{32} (34 \alpha_{2}{} - 69 \alpha_{3}{} - 57 \alpha_{4}{} + 78 \alpha_{5}{} + 39 \alpha_{6}{}) {\ast\Omega}^{\tau }{}_{\alpha }{}^{\nu } q_{\rho \tau \nu } \Lambda^{\alpha } \Lambda^{\rho }  \nonumber\\
&& + \frac{1}{36} \bigl[-4 a_{4}{} + a_{6}{} + 2 \bigl(- a_{7}{} + \alpha_{2}{} + 6 \alpha_{3}{} + \alpha_{4}{} + \alpha_{5}{}\bigr)\bigr]( \varepsilon_{\alpha \nu \sigma \lambda }  q_{\rho \tau }{}^{\lambda } +\varepsilon_{\alpha \tau \sigma \lambda }  q_{\rho \nu }{}^{\lambda } ) {\ast\Omega}^{\tau \nu \sigma }\nabla^{\rho }S^{\alpha }\nonumber\\
&&  + \frac{1}{12} (2 a_{10}{} + a_{12}{} + \alpha_{5}{}) \varepsilon_{\alpha \rho \sigma \lambda } {\ast\Omega}^{\tau \nu \sigma } q_{\tau \nu }{}^{\lambda } \nabla^{\rho }S^{\alpha } \nonumber\\
&& + \frac{1}{18} (12 a_{10}{} + 6 a_{12}{} - 12 a_{4}{} + 3 a_{6}{} - 6 a_{7}{} + 8 \alpha_{2}{} + 48 \alpha_{3}{} + 12 \alpha_{5}{}) {\ast\Omega}^{\tau }{}_{\rho }{}^{\nu } q_{\alpha \tau \nu } \nabla^{\rho }T^{\alpha } \nonumber\\
&& + \frac{1}{18} (-12 a_{10}{} - 6 a_{12}{} + 12 a_{4}{} - 3 a_{6}{} + 6 a_{7}{} - 4 \alpha_{2}{} - 24 \alpha_{3}{} - 12 \alpha_{4}{} - 12 \alpha_{5}{}) {\ast\Omega}^{\tau }{}_{\alpha }{}^{\nu } q_{\rho \tau \nu } \nabla^{\rho }T^{\alpha }\nonumber\\
&&  + \Bigl(- a_{12}{} -  a_{16}{} + a_{4}{} + a_{7}{} -  \frac{2}{3} \alpha_{2}{} - 4 \alpha_{3}{} -  \frac{3}{2} \alpha_{5}{} -  \frac{1}{4} \alpha_{6}{}\Bigr) {\ast\Omega}^{\tau }{}_{\rho }{}^{\nu } q_{\alpha \tau \nu } \nabla^{\rho }W^{\alpha } \nonumber\\
&& + \Bigl(a_{12}{} + a_{16}{} -  a_{4}{} -  a_{7}{} + \frac{1}{3} \alpha_{2}{} + 2 \alpha_{3}{} + \alpha_{4}{} + \frac{3}{2} \alpha_{5}{} + \frac{1}{4} \alpha_{6}{}\Bigr) {\ast\Omega}^{\tau }{}_{\alpha }{}^{\nu } q_{\rho \tau \nu } \nabla^{\rho }W^{\alpha } \nonumber\\
&& + \frac{1}{48} (72 a_{10}{} - 36 a_{4}{} - 36 a_{7}{} + 24 \alpha_{2}{} + 144 \alpha_{3}{} + 32 \alpha_{4}{} - 6 \alpha_{5}{} - 21 \alpha_{6}{}) {\ast\Omega}^{\tau }{}_{\rho }{}^{\nu } q_{\alpha \tau \nu } \nabla^{\rho }\Lambda^{\alpha } \nonumber\\
&& + \frac{1}{48} (-72 a_{10}{} + 36 a_{4}{} + 36 a_{7}{} - 36 \alpha_{2}{} - 48 \alpha_{3}{} - 4 \alpha_{4}{} - 42 \alpha_{5}{} - 3 \alpha_{6}{}) {\ast\Omega}^{\tau }{}_{\alpha }{}^{\nu } q_{\rho \tau \nu } \nabla^{\rho }\Lambda^{\alpha }\nonumber\\
&&  +\frac{1}{9}(\alpha_{2}{} -  \alpha_{3}{} - 8 \alpha_{4}{} + 2 \alpha_{5}{} + \alpha_{6}{})( - 2 T^{\alpha }  +  3W^{\alpha } )  q_{\alpha \rho \nu }\nabla_{\tau }{\ast\Omega}^{\rho \tau \nu }\nonumber\\
&& + \frac{1}{12} (3 \alpha_{2}{} + 39 \alpha_{3}{} + 22 \alpha_{4}{} - 6 \alpha_{5}{} - 3 \alpha_{6}{}) q_{\alpha \rho \nu } \Lambda^{\alpha } \nabla_{\tau }{\ast\Omega}^{\rho \tau \nu } +\frac{1}{3}(\alpha_{2}{} -  \alpha_{3}{}) (-2 T^{\alpha }  + 3W^{\alpha }  )\nabla_{\nu } q_{\alpha \rho \tau }{\ast\Omega}^{\rho \tau \nu }\nonumber\\
&&-  \frac{5}{4} (\alpha_{2}{} -  \alpha_{3}{}) {\ast\Omega}^{\rho \tau \nu } \Lambda^{\alpha } \nabla_{\nu }q_{\alpha \rho \tau } + \frac{1}{36} (-4 \alpha_{2}{} + 4 \alpha_{3}{} + 8 \alpha_{4}{} - 2 \alpha_{5}{} -  \alpha_{6}{}) (2T^{\alpha } -3W^{\alpha } ){\ast\Omega}^{\rho }{}_{\alpha }{}^{\tau } \nabla_{\nu }q_{\rho \tau }{}^{\nu } \nonumber\\
&&  + \frac{1}{48} (12 \alpha_{2}{} - 12 \alpha_{3}{} - 56 \alpha_{4}{} + 54 \alpha_{5}{} + 27 \alpha_{6}{}) {\ast\Omega}^{\rho }{}_{\alpha }{}^{\tau } \Lambda^{\alpha } \nabla_{\nu }q_{\rho \tau }{}^{\nu } \nonumber\\
&&+ \frac{1}{72} (2 \alpha_{5}{} + \alpha_{6}{})( \varepsilon_{\alpha \tau \nu \lambda } \nabla_{\sigma }q_{\rho }{}^{\sigma \lambda }+\varepsilon_{\alpha \rho \nu \lambda } \nabla_{\sigma }q_{\tau }{}^{\sigma \lambda } ){\ast\Omega}^{\rho \tau \nu } S^{\alpha }+ \frac{1}{18} (\alpha_{2}{} - \alpha_{3}{}) \varepsilon_{\alpha \nu \sigma \lambda } {\ast\Omega}^{\rho \tau \nu } S^{\alpha } \nabla^{\lambda }q_{\rho \tau }{}^{\sigma }  \nonumber\\
&&  + \frac{1}{36} (-3 a_{6}{} - 6 a_{7}{} + 2 \alpha_{2}{} - 2 \alpha_{3}{}) \varepsilon_{\alpha \nu \sigma \lambda } q_{\rho \tau }{}^{\lambda } S^{\alpha } \nabla^{\sigma }{\ast\Omega}^{\rho \tau \nu } \nonumber\\
&&+ \frac{1}{36} \bigl[- a_{6}{} - 2 \bigl(a_{7}{} -  \alpha_{2}{} + \alpha_{3}{}\bigr)\bigr] (\varepsilon_{\alpha \tau \nu \lambda } q_{\rho \sigma }{}^{\lambda }  +\varepsilon_{\alpha \rho \nu \lambda } q_{\tau \sigma }{}^{\lambda } )S^{\alpha } \nabla^{\sigma }{\ast\Omega}^{\rho \tau \nu }  \,,
\end{eqnarray}}
\small{\begin{eqnarray}
     16\pi \mathcal{L}_{\rm g}^{(q-TStW\Lambda\Omega)}&=&\frac{1}{18} (\alpha_{1}{} - 3 \alpha_{2}{} - 4 \alpha_{3}{} + \alpha_{4}{} + \alpha_{5}{}) {\ast\Omega}^{\alpha \rho \tau } q_{\nu \sigma \lambda } q^{\nu \sigma \lambda } t_{\alpha \rho \tau } + \frac{1}{12} (-6 \alpha_{2}{} + 6 \alpha_{3}{} -  \alpha_{5}{}) {\ast\Omega}^{\alpha \rho \tau } q_{\tau }{}^{\sigma \lambda } q_{\nu \sigma \lambda } t_{\alpha \rho }{}^{\nu }\nonumber\\
     &&+ \frac{1}{72} (-12 a_{10}{} + 12 a_{12}{} - 84 \alpha_{3}{} - 12 \alpha_{4}{} - 4 \alpha_{5}{} + 3 \alpha_{6}{}) {\ast\Omega}^{\alpha \rho \tau } {\ast\Omega}^{\nu \sigma \lambda } q_{\tau \nu \lambda } t_{\alpha \rho \sigma } \nonumber\\&&
     + \frac{1}{36} (-6 a_{10}{} + 6 a_{12}{} + 8 \alpha_{5}{} + 3 \alpha_{6}{}) {\ast\Omega}^{\alpha \rho \tau } {\ast\Omega}^{\nu }{}_{\rho }{}^{\sigma } q_{\nu \sigma \lambda } t_{\alpha \tau }{}^{\lambda }  \nonumber\\&&
     + \frac{1}{18} (a_{4}{} -  a_{6}{} -  a_{7}{} - 6 \alpha_{2}{} -  \alpha_{3}{} -  \alpha_{4}{} -  \alpha_{5}{}) {\ast\Omega}^{\alpha \rho \tau } {\ast\Omega}_{\rho }{}^{\nu \sigma } q_{\tau \sigma \lambda } t_{\alpha \nu }{}^{\lambda } \nonumber\\&&+ \frac{1}{18} (2 a_{4}{} + a_{6}{} + 4 a_{7}{} - 8 \alpha_{2}{} + 8 \alpha_{3}{}) {\ast\Omega}^{\alpha \rho \tau } {\ast\Omega}^{\nu }{}_{\rho }{}^{\sigma } q_{\tau \sigma \lambda } t_{\alpha \nu }{}^{\lambda } + \frac{1}{6} (2 \alpha_{2}{} + 5 \alpha_{3}{} + \alpha_{4}{} + \alpha_{5}{}) {\ast\Omega}^{\alpha \rho \tau } q_{\rho \nu }{}^{\lambda } q_{\tau \sigma \lambda } t_{\alpha }{}^{\nu \sigma }  \nonumber\\&&
     + \frac{1}{18} (a_{4}{} + 2 a_{6}{} + 5 a_{7}{} - 2 \alpha_{2}{} + 9 \alpha_{3}{} + \alpha_{4}{} + \alpha_{5}{}) {\ast\Omega}^{\alpha \rho \tau } {\ast\Omega}^{\nu }{}_{\rho }{}^{\sigma } q_{\tau \nu \lambda } t_{\alpha \sigma }{}^{\lambda }  \nonumber\\&&+ \frac{1}{18} (\alpha_{1}{} - 3 \alpha_{2}{} - 4 \alpha_{3}{} + \alpha_{4}{} + \alpha_{5}{}) {\ast\Omega}^{\alpha \rho \tau } q_{\nu \sigma \lambda } q^{\nu \sigma \lambda } t_{\rho \alpha \tau } + \frac{1}{12} (-6 \alpha_{2}{} + 6 \alpha_{3}{} + 4 \alpha_{4}{} + \alpha_{5}{}) {\ast\Omega}^{\alpha \rho \tau } q_{\tau }{}^{\sigma \lambda } q_{\nu \sigma \lambda } t_{\rho \alpha }{}^{\nu }  \nonumber\\&&
     + \frac{1}{72} (-12 
a_{10}{} - 84 \alpha_{3}{} + 4 \alpha_{4}{} - 20 \alpha_{5}{} - 3 \alpha_{6}{}) {\ast\Omega}^{\alpha \rho \tau } {\ast\Omega}^{\nu \sigma \lambda } q_{\tau \nu \lambda } t_{\rho \alpha \sigma } + \frac{1}{6} (2 
\alpha_{4}{} + \alpha_{5}{}) {\ast\Omega}^{\alpha \rho \tau } q_{\alpha }{}^{\sigma \lambda } q_{\nu \sigma \lambda } t_{\rho \tau }{}^{\nu }  \nonumber\\&&
+ \frac{1}{36} (-6 a_{12}{} + 8 \alpha_{4}{} - 8 \alpha_{5}{} - 3
\alpha_{6}{}) {\ast\Omega}^{\alpha \rho \tau } {\ast\Omega}^{\nu \sigma \lambda } q_{\alpha \nu \lambda } t_{\rho \tau \sigma }  \nonumber\\&&
+ \frac{1}{72} (12 a_{10}{} + 24 a_{4}{} - 4 a_{6}{} + 16 a_{7}{} - 16 \alpha_{2}{} - 68 \alpha_{3}{} + 4 \alpha_{4}{} - 12 \alpha_{5}{} - 3 \alpha_{6}{}) {\ast\Omega}^{\alpha \rho \tau } {\ast\Omega}^{\nu \sigma \lambda } q_{\alpha \tau \lambda } t_{\rho \nu \sigma }  \nonumber\\&&
+ \frac{1}{18} \bigl[2 a_{4}{} -  a_{6}{} - 2 \bigl(2 \alpha_{2}{} + 5 \alpha_{3}{} + \alpha_{4}{} + \alpha_{5}{}\bigr)\bigr] {\ast\Omega}_{\alpha }{}^{\nu \sigma } {\ast\Omega}^{\alpha \rho \tau } q_{\tau \sigma \lambda } t_{\rho \nu }{}^{\lambda }  \nonumber\\&&
+ \frac{1}{6} (2 \alpha_{2}{} + 5 \alpha_{3}{} + \alpha_{4}{} + \alpha_{5}{}) ({\ast\Omega}^{\alpha \rho \tau } q_{\tau \sigma \lambda } t_{\alpha }{}^{\nu \sigma } t_{\rho \nu }{}^{\lambda } +  {\ast\Omega}^{\alpha \rho \tau } q_{\alpha \nu }{}^{\lambda } q_{\tau \sigma \lambda } t_{\rho }{}^{\nu \sigma }) \nonumber\\&&
+ \frac{1}{72} (-12 a_{10}{} - 24 a_{4}{} + 4 a_{6}{} - 16 a_{7}{} + 16 \alpha_{2}{} + 68 \alpha_{3}{} - 4 \alpha_{4}{} + 12 \alpha_{5}{} + 3 \alpha_{6}{}) {\ast\Omega}^{\alpha \rho \tau } {\ast\Omega}^{\nu \sigma \lambda } q_{\alpha \tau \nu } t_{\rho \sigma \lambda } \nonumber\\&&
+ \frac{1}{72} (-4 a_{4}{} + 4 a_{5}{} + 5 a_{6}{} + 2 a_{7}{} - 4 \alpha_{2}{} + 18 \alpha_{3}{} + 2 \alpha_{4}{} + 2 \alpha_{5}{}) \varepsilon_{\alpha \nu \lambda \mu } {\ast\Omega}^{\rho \tau \nu } q_{\tau \sigma }{}^{\mu } S^{\alpha } t_{\rho }{}^{\sigma \lambda }  \nonumber\\&&
+ \frac{1}{36} (-6 a_{10}{} - 8 \alpha_{4}{} + 4 \alpha_{5}{} + 3 \alpha_{6}{}) {\ast\Omega}^{\alpha \rho \tau } {\ast\Omega}^{\nu }{}_{\rho }{}^{\sigma } q_{\nu \sigma \lambda } t_{\tau \alpha }{}^{\lambda } \nonumber\\&&
+ \frac{1}{18} (- a_{4}{} -  a_{7}{} - 2 \alpha_{2}{} + 9 \alpha_{3}{} + \alpha_{4}{} + \alpha_{5}{}) {\ast\Omega}^{\alpha \rho \tau } {\ast\Omega}_{\rho }{}^{\nu \sigma } q_{\alpha \sigma \lambda } t_{\tau \nu }{}^{\lambda }  \nonumber\\&&
+ \frac{1}{18} (7 a_{4}{} + 7 a_{7}{} - 2 \alpha_{2}{} - 19 \alpha_{3}{} - 3 \alpha_{4}{} - 3 \alpha_{5}{}) {\ast\Omega}^{\alpha \rho \tau } {\ast\Omega}^{\nu }{}_{\rho }{}^{\sigma } q_{\alpha \sigma \lambda } t_{\tau \nu }{}^{\lambda } \nonumber\\&&
+ \frac{2}{9} (2 a_{4}{} + 2 a_{7}{} - 7 \alpha_{3}{} -  \alpha_{4}{} -  \alpha_{5}{}) {\ast\Omega}^{\alpha \rho \tau } {\ast\Omega}^{\nu }{}_{\rho }{}^{\sigma } q_{\alpha \nu \lambda } t_{\tau \sigma }{}^{\lambda } \nonumber\\&&
+ \frac{1}{72} (6 a_{10}{} + 3 a_{12}{} + 2 a_{5}{} -  a_{6}{} - 4 a_{7}{} + 2 \alpha_{2}{} - 2 \alpha_{3}{} + 3 \alpha_{5}{}) \varepsilon_{\alpha \sigma \lambda \mu } {\ast\Omega}^{\rho \tau \nu } q_{\rho \nu }{}^{\mu } S^{\alpha } t_{\tau }{}^{\sigma \lambda }  \nonumber\\&&
+ \frac{1}{72} (-8 a_{4}{} + 5 a_{6}{} + 2 a_{7}{} + 28 \alpha_{3}{} + 4 \alpha_{4}{} + 4 \alpha_{5}{}) \varepsilon_{\alpha \nu \lambda \mu } {\ast\Omega}^{\rho \tau \nu } q_{\rho \sigma }{}^{\mu } S^{\alpha } t_{\tau }{}^{\sigma \lambda }  \nonumber\\&&
+ \frac{1}{36} (-2 a_{4}{} - 2 a_{5}{} + 2 \alpha_{2}{} + 5 \alpha_{3}{} + \alpha_{4}{} + \alpha_{5}{}) \varepsilon_{\alpha \rho \lambda \mu } {\ast\Omega}^{\rho \tau \nu } q_{\nu \sigma }{}^{\mu } S^{\alpha } t_{\tau }{}^{\sigma \lambda }  \nonumber\\&&
+ \frac{1}{18} \bigl[-5 a_{4}{} + a_{6}{} + 3 \bigl(- a_{7}{} - 2 \alpha_{2}{} + 9 \alpha_{3}{} + \alpha_{4}{} + \alpha_{5}{}\bigr)\bigr] {\ast\Omega}^{\alpha \rho \tau } {\ast\Omega}_{\rho }{}^{\nu \sigma } q_{\tau \sigma \lambda } t_{\nu \alpha }{}^{\lambda }  \nonumber\\&&
+ \frac{1}{12} \bigl[4 a_{4}{} -  a_{6}{} + 2 \bigl(a_{7}{} + 2 \alpha_{2}{} - 9 \alpha_{3}{} -  \alpha_{4}{} -  \alpha_{5}{}\bigr)\bigr] {\ast\Omega}^{\alpha \rho \tau } q_{\tau \sigma \lambda } t_{\rho }{}^{\nu \sigma } t_{\nu \alpha }{}^{\lambda } \nonumber\\&&
+ \frac{1}{12} \bigl[-4 a_{4}{} + a_{6}{} + 2 \bigl(- a_{7}{} + 2 \alpha_{2}{} + 5 \alpha_{3}{} + \alpha_{4}{} + \alpha_{5}{}\bigr)\bigr] {\ast\Omega}^{\alpha \rho \tau } q_{\tau \sigma \lambda } t_{\alpha }{}^{\nu \sigma } t_{\nu \rho }{}^{\lambda }  \nonumber\\&&
+ \frac{1}{18} (-7 a_{4}{} + 2 a_{6}{} - 3 a_{7}{} - 2 \alpha_{2}{} + 37 \alpha_{3}{} + 5 \alpha_{4}{} + 5 \alpha_{5}{}) {\ast\Omega}^{\alpha \rho \tau } {\ast\Omega}_{\rho }{}^{\nu \sigma } q_{\alpha \sigma \lambda } t_{\nu \tau }{}^{\lambda }  \nonumber\\&&
+ \frac{1}{6} \bigl[4 a_{4}{} -  a_{6}{} + 2 \bigl(a_{7}{} - 7 \alpha_{3}{} -  \alpha_{4}{} -  \alpha_{5}{}\bigr)\bigr] {\ast\Omega}^{\alpha \rho \tau } q_{\alpha \sigma \lambda } t_{\rho }{}^{\nu \sigma } t_{\nu \tau }{}^{\lambda } + \frac{1}{18} (3 a_{12}{} + 4 \alpha_{4}{} + 2 \alpha_{5}{}) {\ast\Omega}^{\alpha \rho \tau } {\ast\Omega}_{\rho }{}^{\nu \sigma } q_{\alpha \tau \lambda } t_{\nu \sigma }{}^{\lambda }  \nonumber\\&&
+ \frac{1}{12} \bigl[-6 a_{10}{} - 3 a_{12}{} - 4 \bigl(\alpha_{4}{} + \alpha_{5}{}\bigr)\bigr] {\ast\Omega}^{\alpha \rho \tau } q_{\alpha \tau \lambda } t_{\rho }{}^{\nu \sigma } t_{\nu \sigma }{}^{\lambda } + \frac{1}{6} (2 \alpha_{2}{} - 9 \alpha_{3}{} -  \alpha_{4}{} -  \alpha_{5}{}) {\ast\Omega}^{\alpha \rho \tau } q_{\tau \sigma \lambda } t_{\nu \rho }{}^{\lambda } t^{\nu }{}_{\alpha }{}^{\sigma }  \nonumber\\&&
+ \frac{1}{12} (- a_{6}{} - 2 a_{7}{} + 2 \alpha_{2}{} - 2 \alpha_{3}{} + 4 \alpha_{4}{} + 3 \alpha_{5}{}) {\ast\Omega}^{\alpha \rho \tau } q_{\tau }{}^{\sigma \lambda } q_{\nu \sigma \lambda } t^{\nu }{}_{\alpha \rho }\nonumber\\
&&+ \frac{1}{12} (a_{6}{} + 2 a_{7}{} + 8 \alpha_{2}{} - 8 \alpha_{3}{}) {\ast\Omega}^{\alpha \rho \tau } q_{\rho \nu }{}^{\lambda } q_{\tau \sigma \lambda } t^{\nu }{}_{\alpha }{}^{\sigma } \nonumber\\&&
+ \frac{1}{12} (a_{6}{} + 2 a_{7}{} - 2 \alpha_{2}{} + 2 \alpha_{3}{} - 4 \alpha_{4}{} - 3 \alpha_{5}{}) {\ast\Omega}^{\alpha \rho \tau } q_{\alpha }{}^{\sigma \lambda } q_{\nu \sigma \lambda } t^{\nu }{}_{\rho \tau } \nonumber\\&&
+ \frac{1}{6} (- a_{6}{} - 2 a_{7}{} - 2 \alpha_{2}{} - 19 \alpha_{3}{} - 3 \alpha_{4}{} - 3 \alpha_{5}{}) {\ast\Omega}^{\alpha \rho \tau } q_{\alpha \sigma }{}^{\lambda } q_{\tau \nu \lambda } t^{\nu }{}_{\rho }{}^{\sigma }  \nonumber\\&&
+ \frac{1}{12} \bigl[- a_{6}{} - 2 \bigl(a_{7}{} - 2 \alpha_{2}{} + 23 \alpha_{3}{} + 3 \alpha_{4}{} + 3 \alpha_{5}{}\bigr)\bigr] {\ast\Omega}^{\alpha \rho \tau } q_{\alpha \nu }{}^{\lambda } q_{\tau \sigma \lambda } t^{\nu }{}_{\rho }{}^{\sigma } \nonumber\\&&
+ \Bigl(- \frac{1}{2} a_{10}{} + \frac{1}{4} a_{12}{} + \alpha_{2}{} + 6 \alpha_{3}{} + \frac{4}{3} \alpha_{4}{} + \alpha_{5}{}\Bigr) {\ast\Omega}^{\alpha \rho \tau } q_{\alpha \tau }{}^{\lambda } q_{\nu \sigma \lambda } t^{\nu }{}_{\rho }{}^{\sigma } + \frac{1}{3} (\alpha_{2}{} -  \alpha_{3}{}) {\ast\Omega}^{\alpha \rho \tau } q_{\alpha \tau \lambda } t_{\nu \sigma }{}^{\lambda } t^{\nu }{}_{\rho }{}^{\sigma }  \nonumber\\&&
+ \frac{1}{12} (-6 a_{10}{} + 3 a_{12}{} - 12 \alpha_{2}{} + 12 \alpha_{3}{} - 4 \alpha_{4}{} + 6 \alpha_{5}{} + 3 \alpha_{6}{}) {\ast\Omega}^{\alpha \rho \tau } q_{\alpha \rho \sigma } q_{\tau \nu \lambda } t^{\nu \sigma \lambda } \nonumber\\&&
+ \frac{1}{6} (2 a_{4}{} -  a_{6}{} + 2 \alpha_{2}{} - 9 \alpha_{3}{} -  \alpha_{4}{} - \alpha_{5}{}) {\ast\Omega}^{\alpha \rho \tau } q_{\tau \nu \lambda } t^{\nu }{}_{\alpha }{}^{\sigma } t_{\sigma \rho }{}^{\lambda }  \nonumber\\&&
+ \frac{1}{12} (-8 a_{4}{} + a_{6}{} - 6 a_{7}{} + 28 \alpha_{3}{} + 4 \alpha_{4}{} + 4 \alpha_{5}{}) {\ast\Omega}^{\alpha \rho \tau } q_{\alpha \nu \lambda } t^{\nu }{}_{\rho }{}^{\sigma } t_{\sigma \tau }{}^{\lambda } + \frac{1}{3} (\alpha_{2}{} -  \alpha_{3}{}) {\ast\Omega}^{\alpha \rho \tau } q_{\alpha \tau \lambda } t^{\nu }{}_{\rho }{}^{\sigma } t_{\sigma \nu }{}^{\lambda }  \nonumber\\&&
+ \frac{1}{18} (4 a_{4}{} + 4 a_{7}{} - 4 \alpha_{2}{} - 10 \alpha_{3}{} + 2 \alpha_{4}{} - 4 \alpha_{5}{} -  \alpha_{6}{}) ( {\ast\Omega}_{\rho \tau }{}^{\nu }  - {\ast\Omega}^{\nu }{}_{\rho \tau } ){\ast\Omega}^{\alpha \rho \tau }q_{\nu \sigma \lambda } t^{\sigma }{}_{\alpha }{}^{\lambda } \nonumber\\&&
+ \frac{1}{12} (-2 a_{12}{} - 4 a_{9}{} - 8 \alpha_{4}{} + \alpha_{6}{}) {\ast\Omega}^{\alpha \rho \tau } q_{\nu \sigma \lambda } t_{\rho \tau }{}^{\nu } t^{\sigma }{}_{\alpha }{}^{\lambda }+ \frac{1}{12} (2 a_{12}{} + 4 a_{9}{} + 4 \alpha_{5}{} + \alpha_{6}{})
{\ast\Omega}^{\alpha \rho \tau } q_{\alpha \sigma \lambda } t_{\rho \tau }{}^{\nu } t^{\sigma }{}_{\nu }{}^{\lambda }   \nonumber\\&&
+ \frac{1}{36} (-2 a_{4}{} - 2 a_{5}{} -  a_{6}{} - 2 a_{7}{} + 2 \alpha_{2}{} + 5 \alpha_{3}{} + \alpha_{4}{} + \alpha_{5}{}) \varepsilon_{\alpha \nu \lambda \mu } {\ast\Omega}^{\rho \tau \nu } q_{\tau \sigma }{}^{\mu } S^{\alpha } t^{\sigma }{}_{\rho }{}^{\lambda }  \nonumber\\&&
+ \frac{1}{36} (-a_{6}{} - 2 a_{7}{} + 2 \alpha_{2}{} + 5 \alpha_{3}{} + \alpha_{4}{} + \alpha_{5}{}) \varepsilon_{\alpha \nu \sigma \mu } {\ast\Omega}^{\rho \tau \nu } q_{\tau \lambda }{}^{\mu } S^{\alpha } t^{\sigma }{}_{\rho }{}^{\lambda }  \nonumber\\&&
+ \frac{1}{72} (2 a_{10}{} + a_{12}{} - 2 a_{6}{} - 4 a_{7}{} + 4 a_{9}{} + 4 \alpha_{2}{} - 4 \alpha_{3}{} + 4 \alpha_{4}{} -  \alpha_{6}{}) \varepsilon_{\alpha \tau \nu \mu } {\ast\Omega}^{\rho \tau \nu } q_{\sigma \lambda }{}^{\mu } S^{\alpha } t^{\sigma }{}_{\rho }{}^{\lambda } \nonumber\\&&
+ \frac{1}{18} (-2 a_{4}{} - 2 a_{7}{} - 4 \alpha_{2}{} - 10 \alpha_{3}{} - 2 \alpha_{4}{} - 4 \alpha_{5}{} -  \alpha_{6}{}) {\ast\Omega}_{\alpha \rho }{}^{\nu } {\ast\Omega}^{\alpha \rho \tau } q_{\nu \sigma \lambda } t^{\sigma }{}_{\tau }{}^{\lambda }  \nonumber\\&&
+ \frac{1}{9} (a_{4}{} + a_{7}{} - 4 \alpha_{2}{} - 10 \alpha_{3}{} - 4 \alpha_{5}{} -  \alpha_{6}{}) {\ast\Omega}^{\alpha \rho \tau } {\ast\Omega}_{\rho \alpha }{}^{\nu } q_{\nu \sigma \lambda } t^{\sigma }{}_{\tau }{}^{\lambda } \nonumber\\&&
+ \frac{1}{72} (6 a_{10}{} + 3 a_{12}{} - 4 \alpha_{4}{} + 2 \alpha_{5}{}) \varepsilon_{\alpha \sigma \lambda \mu } {\ast\Omega}^{\rho \tau \nu } q_{\rho \nu }{}^{\mu } S^{\alpha } t^{\sigma }{}_{\tau }{}^{\lambda } \nonumber\\
&&+ \frac{1}{18} (-2 a_{4}{} + a_{6}{} + 7 \alpha_{3}{} + \alpha_{4}{} + \alpha_{5}{}) \varepsilon_{\alpha \nu \lambda \mu } {\ast\Omega}^{\rho \tau \nu } q_{\rho \sigma }{}^{\mu } S^{\alpha } t^{\sigma }{}_{\tau }{}^{\lambda } - \frac{1}{72} (a_{6}{} + 2 a_{7}{}) \varepsilon_{\alpha \nu \sigma \mu } {\ast\Omega}^{\rho \tau \nu } q_{\rho \lambda }{}^{\mu } S^{\alpha } t^{\sigma }{}_{\tau }{}^{\lambda } \nonumber\\&&
 + \frac{1}{12} \bigl\{2 a_{10}{} -  a_{12}{} - 2 \bigl[a_{6}{} + 2 \bigl(a_{7}{} -  \alpha_{2}{} + \alpha_{3}{} + \alpha_{4}{}\bigr)\bigr]\bigr\}  {\ast\Omega}^{\alpha \rho \tau } q_{\nu \sigma \lambda } t^{\nu }{}_{\alpha \rho } t^{\sigma }{}_{\tau }{}^{\lambda }  \nonumber\\&&
+ \frac{1}{36} (-2 a_{4}{} + 2 a_{5}{} + 3 a_{6}{} + 2 a_{7}{} - 2 \alpha_{2}{} + 9 \alpha_{3}{} + \alpha_{4}{} + \alpha_{5}{}) \varepsilon_{\alpha \rho \lambda \mu } 
{\ast\Omega}^{\rho \tau \nu } q_{\nu \sigma }{}^{\mu } S^{\alpha } t^{\sigma }{}_{\tau }{}^{\lambda }  \nonumber\\&&
+ \frac{1}{72} \bigl[a_{6}{} + 2 \bigl(a_{7}{} - 2 \alpha_{2}{} - 5 \alpha_{3}{} -  \alpha_{4}{} -  \alpha_{5}{}\bigr)\bigr] \varepsilon_{\alpha \rho \sigma \mu } {\ast\Omega}^{\rho \tau \nu } q_{\nu \lambda }{}^{\mu } S^{\alpha } t^{\sigma }{}_{\tau }{}^{\lambda } \nonumber\\&&
+ \frac{1}{72} (2 a_{10}{} + a_{12}{} - 2 a_{6}{} - 4 a_{7}{} + 4 a_{9}{} + 4 \alpha_{2}{} - 4 \alpha_{3}{} + 4 \alpha_{4}{} -  \alpha_{6}{}) \varepsilon_{\alpha \rho \nu \mu } {\ast\Omega}^{\rho \tau \nu } q_{\sigma \lambda }{}^{\mu } S^{\alpha } t^{\sigma }{}_{\tau }{}^{\lambda } \nonumber\\&&
+ \frac{1}{12} (-2 a_{10}{} + 3 a_{12}{} + 4 a_{9}{} + 4 \alpha_{2}{} - 4 \alpha_{3}{} + 4 \alpha_{4}{} + 4 \alpha_{5}{} + \alpha_{6}{}) {\ast\Omega}^{\alpha \rho \tau } q_{\nu \sigma \lambda } t_{\alpha \rho }{}^{\nu } t^{\sigma }{}_{\tau }{}^{\lambda } \nonumber\\&&
+ \frac{1}{12} (-2 a_{10}{} + a_{12}{} + 4 \alpha_{2}{} - 4 \alpha_{3}{} - 4 \alpha_{4}{} + 4 \alpha_{5}{} + 2 \alpha_{6}{}) {\ast\Omega}^{\alpha \rho \tau } q_{\nu \sigma \lambda } t_{\rho \alpha }{}^{\nu } t^{\sigma }{}_{\tau }{}^{\lambda }  \nonumber\\&&
+ \frac{1}{18} (4 a_{4}{} + 4 a_{7}{} - 4 \alpha_{2}{} - 10 \alpha_{3}{} + 2 \alpha_{4}{} - 4 \alpha_{5}{} -  \alpha_{6}{}) {\ast\Omega}^{\alpha \rho \tau } {\ast\Omega}_{\rho \tau }{}^{\nu } q_{\alpha \sigma \lambda } t^{\sigma }{}_{\nu }{}^{\lambda } \nonumber\\&&
+ \frac{1}{12} (2 a_{10}{} - 3 a_{12}{} - 4 a_{9}{} + 4 \alpha_{2}{} + 24 \alpha_{3}{} + 2 \alpha_{5}{}) {\ast\Omega}^{\alpha \rho \tau } q_{\tau \sigma \lambda } t_{\alpha \rho }{}^{\nu } t^{\sigma }{}_{\nu }{}^{\lambda } \nonumber\\&&
+ \frac{1}{12} (2 a_{10}{} - a_{12}{} + 4 \alpha_{2}{} + 24 \alpha_{3}{} + 6 \alpha_{5}{} + 
\alpha_{6}{}) {\ast\Omega}^{\alpha \rho \tau } q_{\tau \sigma \lambda } t_{\rho \alpha }{}^{\nu } t^{\sigma }{}_{\nu }{}^{\lambda } \nonumber\\&&
+ \frac{1}{12} (-2 a_{10}{} + a_{12}{} - 8 a_{4}{} + 2 a_{6}{} - 4 a_{7}{} + 4 \alpha_{2}{} + 24 \alpha_{3}{} + 6 \alpha_{5}{} + \alpha_{6}{}) (q_{\tau \sigma \lambda } t^{\nu }{}_{\alpha \rho } - q_{\alpha \sigma \lambda } t^{\nu }{}_{\rho \tau } ){\ast\Omega}^{\alpha \rho \tau } t^{\sigma }{}_{\nu }{}^{\lambda } \nonumber\\&&
+ \frac{1}{72} (6 a_{10}{} + 3 a_{12}{} - 2 a_{5}{} + a_{6}{} + 4 a_{7}{} - 2 \alpha_{2}{} + 2 \alpha_{3}{} + 3 \alpha_{5}{}) \varepsilon_{\alpha \nu \lambda \mu } {\ast\Omega}^{\rho \tau \nu } q_{\rho \tau \sigma } S^{\alpha } t^{\sigma \lambda \mu } \nonumber\\&&
+ \frac{1}{72} \bigl[6 a_{10}{} + 3 a_{12}{} + 4 \bigl(\alpha_{4}{} + \alpha_{5}{}\bigr)\bigr] \varepsilon_{\alpha \nu \sigma \mu } {\ast\Omega}^{\rho \tau \nu } q_{\rho \tau \lambda } S^{\alpha } t^{\sigma \lambda \mu } \nonumber\\&&
+ \frac{1}{72} (2 a_{10}{} + a_{12}{} + 8 a_{4}{} - 2 a_{6}{} + 4 a_{7}{} + 4 a_{9}{} - 4 \alpha_{2}{} - 24 \alpha_{3}{} - 2 \alpha_{5}{}) (\varepsilon_{\alpha \tau \nu \mu }  q_{\rho \sigma \lambda } +\varepsilon_{\alpha \rho \nu \mu } q_{\tau \sigma \lambda }  ){\ast\Omega}^{\rho \tau \nu } S^{\alpha } t^{\sigma \lambda \mu }\nonumber\\&&
+ \frac{1}{18} \bigl\{3 a_{10}{} - 2 \bigl[a_{6}{} + 2 \bigl(a_{7}{} -  \alpha_{2}{} + \alpha_{3}{} + \alpha_{4}{}\bigr)\bigr]\bigr\} {\ast\Omega}^{\alpha \rho \tau } {\ast\Omega}^{\nu }{}_{\rho }{}^{\sigma } q_{\nu \sigma \lambda } t^{\lambda }{}_{\alpha \tau }  \nonumber\\&&
+ \frac{1}{18} (-3 a_{4}{} - 2 a_{6}{} - 7 a_{7}{} + 2 \alpha_{2}{} + 19 \alpha_{3}{} + 3 \alpha_{4}{} + 3 \alpha_{5}{}) {\ast\Omega}^{\alpha \rho \tau } {\ast\Omega}_{\rho }{}^{\nu \sigma } q_{\tau \sigma \lambda } t^{\lambda }{}_{\alpha \nu } \nonumber\\&&
+ \frac{1}{12} \bigl[-4 a_{5}{} -  a_{6}{} + 2 \bigl(a_{7}{} + 2 \alpha_{2}{} - 9 \alpha_{3}{} -  \alpha_{4}{} -  \alpha_{5}{}\bigr)\bigr] 
{\ast\Omega}^{\alpha \rho \tau } q_{\tau \sigma \lambda } t_{\rho }{}^{\nu \sigma } t^{\lambda }{}_{\alpha \nu }   \nonumber\\&&
+\frac{1}{18} (3 a_{4}{} + 2 a_{6}{} + 7 a_{7}{} - 2 \alpha_{2}{} - 19 \alpha_{3}{} - 3 \alpha_{4}{} - 3 \alpha_{5}{}) {\ast\Omega}^{\alpha \rho \tau } {\ast\Omega}^{\nu }{}_{\rho }{}^{\sigma } q_{\tau \nu \lambda } t^{\lambda }{}_{\alpha \sigma }  \nonumber\\&&
+ \frac{1}{12} (-2 a_{10}{} + a_{12}{} + 2 a_{6}{} + 4 a_{7}{} - 4 \alpha_{2}{} + 4 \alpha_{3}{} + 4 \alpha_{4}{}) {\ast\Omega}^{\alpha \rho \tau } q_{\nu \sigma \lambda } t^{\nu }{}_{\alpha }{}^{\sigma } t^{\lambda }{}_{\rho \tau } \nonumber\\&&
+ \frac{1}{6} (-2 a_{5}{} - 2 a_{6}{} - 2 a_{7}{} + 2 \alpha_{2}{} - 9 \alpha_{3}{} -  \alpha_{4}{} -  \alpha_{5}{}) {\ast\Omega}^{\alpha \rho \tau } q_{\tau \sigma \lambda } t_{\alpha }{}^{\nu \sigma } t^{\lambda }{}_{\rho \nu } \nonumber\\&&
+ \frac{1}{12} \bigl[-4 a_{4}{} -  a_{6}{} + 2 \bigl(-3 a_{7}{} + 2 \alpha_{2}{} + 5 \alpha_{3}{} + \alpha_{4}{} + \alpha_{5}{}\bigr)\bigr] {\ast\Omega}^{\alpha \rho \tau } q_{\tau \sigma \lambda } t^{\nu }{}_{\alpha }{}^{\sigma } t^{\lambda }{}_{\rho \nu }  \nonumber\\&&
+ \frac{1}{12}\bigl[- 3 a_{6}{} + 2 \bigl(-3 a_{7}{} + 2 \alpha_{2}{} + 5 \alpha_{3}{} + \alpha_{4}{} + \alpha_{5}{}\bigr)\bigr] {\ast\Omega}^{\alpha \rho \tau } q_{\tau \nu \lambda } t^{\nu }{}_{\alpha }{}^{\sigma } t^{\lambda }{}_{\rho \sigma } + \frac{1}{4} (a_{6}{} + 2 a_{7}{}) {\ast\Omega}^{\alpha \rho \tau } q_{\alpha \sigma \lambda } t_{\rho }{}^{\nu \sigma } t^{\lambda }{}_{\tau \nu }  \nonumber\\&&
+ \frac{1}{18} (-3 a_{4}{} - 2 a_{6}{} - 7 a_{7}{} + 2 \alpha_{2}{} + 19 \alpha_{3}{} + 3 \alpha_{4}{} + 3 \alpha_{5}{}) {\ast\Omega}^{\alpha \rho \tau } {\ast\Omega}_{\rho }{}^{\nu \sigma } q_{\alpha \sigma \lambda } t^{\lambda }{}_{\tau \nu }  \nonumber\\&&
+ \frac{1}{18} \bigl\{3 a_{10}{} - 2 \bigl[a_{6}{} + 2 \bigl(a_{7}{} -  \alpha_{2}{} + \alpha_{3}{} +\alpha_{4}{}\bigr)\bigr]\bigr\} {\ast\Omega}^{\alpha \rho \tau } {\ast\Omega}_{\rho }{}^{\nu \sigma } q_{\alpha \tau \lambda } t^{\lambda }{}_{\nu \sigma }  \nonumber\\&&
+ \frac{1}{12} (-2 a_{5}{} + a_{6}{} + 4 a_{7}{} - 2 \alpha_{2}{} + 2 \alpha_{3}{} + 4 \alpha_{4}{} + \alpha_{5}{}) {\ast\Omega}^{\alpha \rho \tau } q_{\alpha \tau \lambda } t_{\rho }{}^{\nu \sigma } t^{\lambda }{}_{\nu \sigma } \nonumber\\&&
+ \frac{1}{12} (-6 a_{10}{} - 3 a_{12}{} + 4 \alpha_{4}{} - 2 \alpha_{5}{}) {\ast\Omega}^{\alpha \rho \tau } q_{\alpha \tau \lambda } t^{\nu }{}_{\rho }{}^{\sigma } t^{\lambda }{}_{\nu \sigma } \nonumber\\&&
+ \frac{1}{6} (4 a_{10}{} + 2 a_{12}{} - 4 a_{4}{} + a_{6}{} - 2 a_{7}{} + 14 \alpha_{3}{} + 2 \alpha_{4}{} + 4 \alpha_{5}{}) {\ast\Omega}^{\rho \tau \nu } q_{\rho \nu \sigma } t_{\alpha \tau }{}^{\sigma } T^{\alpha }  \nonumber\\&&
+ \frac{1}{18} \bigl[4 a_{4}{} -  a_{6}{} + 2 \bigl(a_{7}{} - 6 \alpha_{2}{} - 8 \alpha_{3}{} - 2 \alpha_{4}{} - 4 \alpha_{5}{} -  \alpha_{6}{}\bigr)\bigr] {\ast\Omega}^{\rho \tau \nu } q_{\alpha \nu \sigma } t_{\rho \tau }{}^{\sigma } T^{\alpha }  \nonumber\\&&
+ \frac{1}{18} (6 a_{10}{} + 3 a_{12}{} - 4 \alpha_{2}{} + 18 \alpha_{3}{} - 2 \alpha_{4}{} + 4 \alpha_{5}{}) {\ast\Omega}^{\rho \tau \nu } q_{\rho \nu \sigma } t_{\tau \alpha }{}^{\sigma } T^{\alpha }  \nonumber\\&&
+ \frac{1}{18} \bigl[-4 a_{4}{} + a_{6}{} - 2 \bigl(a_{7}{} + 6 \alpha_{2}{} 
- 6 \alpha_{3}{} - 8 \alpha_{4}{} + 6 \alpha_{5}{} + 3 \alpha_{6}{}\bigr)\bigr] {\ast\Omega}^{\rho \tau \nu } q_{\alpha \nu \sigma } t_{\tau \rho }{}^{\sigma } T^{\alpha } \nonumber\\&&
+ \frac{1}{9} \bigl[-4 
a_{4}{} + a_{6}{} - 2 \bigl(a_{7}{} - 7 \alpha_{3}{} - 5 \alpha_{4}{} + \alpha_{5}{} + \alpha_{6}{}\bigr)\bigr] {\ast\Omega}^{\rho \tau \nu } q_{\alpha \rho \sigma } t_{\tau \nu }{}^{\sigma } T^{\alpha }  \nonumber\\&&
+ \frac{1}{18} 
(12 a_{10}{} - 6 a_{12}{} - 4 a_{4}{} -  a_{6}{} - 6 a_{7}{} + 8 \alpha_{2}{} + 20 \alpha_{3}{} + 4 \alpha_{4}{} - 2 \alpha_{5}{} - 3 \alpha_{6}{}) {\ast\Omega}_{\alpha }{}^{\rho \tau } q_{\tau \nu \sigma }
t^{\nu }{}_{\rho }{}^{\sigma } T^{\alpha }  \nonumber\\&&
+ \frac{1}{18} (6 a_{10}{} - 3 a_{12}{} + 4 a_{4}{} - 2 a_{6}{} + 4 \alpha_{2}{} - 4 \alpha_{3}{} - 4 \alpha_{4}{} - 2 \alpha_{5}{} -  \alpha_{6}{}) {\ast\Omega}^{\rho }{}_{\alpha }{}^{\tau } q_{\tau \nu \sigma } t^{\nu }{}_{\rho }{}^{\sigma } T^{\alpha }  \nonumber\\&&
+ \frac{1}{18} (-6 a_{10}{} + 3 a_{12}{} + 8 a_{4}{} -  a_{6}{} + 6 a_{7}{} - 4 \alpha_{2}{} - 24 \alpha_{3}{} - 8 \alpha_{4}{} + 2 \alpha_{6}{}) {\ast\Omega}^{\rho }{}_{\alpha }{}^{\tau } q_{\rho \nu \sigma } t^{\nu }{}_{\tau }{}^{\sigma } T^{\alpha } \nonumber\\&&
+ \frac{1}{18} (-6 a_{10}{} - 3 a_{12}{} + 3 a_{6}{} + 6 a_{7}{} - 4 \alpha_{2}{} + 18 \alpha_{3}{} - 2 \alpha_{4}{} - 2 \alpha_{5}{}) {\ast\Omega}^{\rho \tau \nu } q_{\rho \nu \sigma } t^{\sigma }{}_{\alpha \tau } T^{\alpha }  \nonumber\\&&
+ \frac{1}{18} \bigl[- a_{6}{} - 2 \bigl(a_{7}{} + 2 \alpha_{2}{} - 2 \alpha_{3}{} - 8 \alpha_{4}{} + 2 \alpha_{5}{} + \alpha_{6}{}\bigr)\bigr]( q_{\alpha \nu \sigma } t^{\sigma }{}_{\rho \tau } - q_{\alpha \rho \sigma } t^{\sigma }{}_{\tau \nu } ){\ast\Omega}^{\rho \tau \nu }T^{\alpha } \nonumber\\&&
+ \frac{1}{4} (-4 a_{10}{} - 2 a_{12}{} + 4 a_{4}{} -  a_{6}{} + 2 a_{7}{} - 14 \alpha_{3}{} - 2 \alpha_{4}{} - 4 \alpha_{5}{}) {\ast\Omega}^{\rho \tau \nu } q_{\rho \nu \sigma } t_{\alpha \tau }{}^{\sigma } W^{\alpha }  \nonumber\\&&
+ \frac{1}{12} \bigl[-4 a_{4}{} + a_{6}{} + 2 \bigl(- a_{7}{} + 6 \alpha_{2}{} + 8 \alpha_{3}{} + 2 \alpha_{4}{} + 4 \alpha_{5}{} + \alpha_{6}{}\bigr)\bigr] {\ast\Omega}^{\rho \tau \nu } q_{\alpha \nu \sigma } t_{\rho \tau }{}^{\sigma } W^{\alpha }  \nonumber\\&&
+ \frac{1}{12} \bigl[-6 a_{10}{} - 3 a_{12}{} + 2 \bigl(2 \alpha_{2}{} - 9 \alpha_{3}{} + \alpha_{4}{} - 2 \alpha_{5}{}\bigr)\bigr] {\ast\Omega}^{\rho \tau \nu } q_{\rho \nu \sigma } t_{\tau \alpha }{}^{\sigma } W^{\alpha }  \nonumber\\&&
+ \Bigl(\frac{1}{3} a_{4}{} -  \frac{1}{12} a_{6}{} + \frac{1}{6} a_{7}{} + \alpha_{2}{} -  \alpha_{3}{} -  \frac{4}{3} \alpha_{4}{} + \alpha_{5}{} + \frac{1}{2} \alpha_{6}{}\Bigr) {\ast\Omega}^{\rho \tau \nu } q_{\alpha \nu \sigma } t_{\tau \rho }{}^{\sigma } W^{\alpha } \nonumber\\&&
+ \frac{1}{6} \bigl[4 a_{4}{} -  a_{6}{} + 2 \bigl(a_{7}{} - 7 \alpha_{3}{} - 5 \alpha_{4}{} + \alpha_{5}{} + \alpha_{6}{}\bigr)\bigr] {\ast\Omega}^{\rho \tau \nu } q_{\alpha \rho \sigma } t_{\tau \nu }{}^{\sigma } W^{\alpha }  \nonumber\\&&
+ \frac{1}{12} (-12 a_{10}{} + 6 a_{12}{} + 4 a_{4}{} + a_{6}{} + 6 a_{7}{} - 8 \alpha_{2}{} - 20 \alpha_{3}{} - 4 \alpha_{4}{} + 2 \alpha_{5}{} + 3 \alpha_{6}{}) {\ast\Omega}_{\alpha }{}^{\rho \tau } q_{\tau \nu \sigma } t^{\nu }{}_{\rho }{}^{\sigma } W^{\alpha } \nonumber\\&&
+ \frac{1}{12} (-6 a_{10}{} + 3 a_{12}{} - 4 a_{4}{} + 2 a_{6}{} - 4 \alpha_{2}{} + 4 \alpha_{3}{} + 4 \alpha_{4}{} + 2 \alpha_{5}{} + \alpha_{6}{}) {\ast\Omega}^{\rho }{}_{\alpha }{}^{\tau } q_{\tau \nu \sigma } t^{\nu }{}_{\rho }{}^{\sigma } W^{\alpha }  \nonumber\\&&
+ \frac{1}{12} (6 a_{10}{} - 3 a_{12}{} - 8 a_{4}{} + a_{6}{} - 6 a_{7}{} + 4 \alpha_{2}{} + 24 \alpha_{3}{} + 8 \alpha_{4}{} - 2 \alpha_{6}{}) {\ast\Omega}^{\rho }{}_{\alpha }{}^{\tau } q_{\rho \nu \sigma } t^{\nu }{}_{\tau }{}^{\sigma } W^{\alpha }  \nonumber\\&&
+ \frac{1}{12} (6 a_{10}{} + 3 a_{12}{} - 3 a_{6}{} - 6 a_{7}{} + 4 \alpha_{2}{} - 18 \alpha_{3}{} + 2 \alpha_{4}{} + 2 \alpha_{5}{}) {\ast\Omega}^{\rho \tau \nu } q_{\rho \nu \sigma } t^{\sigma }{}_{\alpha \tau } W^{\alpha }  \nonumber\\&&
+ \frac{1}{12} \bigl[a_{6}{} + 2 \bigl(a_{7}{} + 2 \alpha_{2}{} - 2 \alpha_{3}{} - 8 \alpha_{4}{} + 2 \alpha_{5}{} + \alpha_{6}{}\bigr)\bigr] (q_{\alpha \nu \sigma } t^{\sigma }{}_{\rho \tau } -  q_{\alpha \rho \sigma } t^{\sigma }{}_{\tau \nu } ) {\ast\Omega}^{\rho \tau \nu }W^{\alpha } \nonumber\\&&
+ \frac{1}{16} (28 a_{10}{} - 2 a_{12}{} - 12 a_{4}{} + a_{6}{} - 10 a_{7}{} + 42 \alpha_{3}{} + 6 \alpha_{4}{} + 4 \alpha_{5}{} - 4 \alpha_{6}{}) {\ast\Omega}^{\rho \tau \nu } q_{\rho \nu \sigma } t_{\alpha \tau }{}^{\sigma } \Lambda^{\alpha }  \nonumber\\&&
+ \frac{1}{48} \bigl[12 a_{4}{} - 5 a_{6}{} + 2 \bigl(a_{7}{} - 30 \alpha_{2}{} - 26 \alpha_{3}{} - 8 \alpha_{4}{} - 12 \alpha_{5}{} - 3 \alpha_{6}{}\bigr)\bigr] {\ast\Omega}^{\rho \tau \nu } q_{\alpha \nu \sigma } t_{\rho \tau }{}^{\sigma } \Lambda^{\alpha }  \nonumber\\&&
+ \frac{1}{48} (6 a_{10}{} + 15 a_{12}{} - 36 \alpha_{2}{} + 78 \alpha_{3}{} + 34 \alpha_{4}{} - 12 \alpha_{5}{} - 12 \alpha_{6}{}) {\ast\Omega}^{\rho \tau \nu } q_{\rho \nu \sigma } t_{\tau \alpha }{}^{\sigma } \Lambda^{\alpha }  \nonumber\\&&
+ \frac{1}{48} \bigl[-12 a_{4}{} + 5 a_{6}{} - 2 \bigl(a_{7}{} + 30 \alpha_{2}{} - 44 \alpha_{3}{} - 18 \alpha_{4}{} + 18 \alpha_{5}{} + 9 \alpha_{6}{}\bigr)\bigr] {\ast\Omega}^{\rho \tau \nu } q_{\alpha \nu \sigma } t_{\tau \rho }{}^{\sigma } \Lambda^{\alpha }  \nonumber\\&&
+ \frac{1}{24} (-12 a_{4}{} + 5 a_{6}{} - 2 a_{7}{} + 70 \alpha_{3}{} + 26 \alpha_{4}{} - 6 \alpha_{5}{} - 6 \alpha_{6}{}) {\ast\Omega}^{\rho \tau \nu } q_{\alpha \rho \sigma } t_{\tau \nu }{}^{\sigma } \Lambda^{\alpha } \nonumber\\&&
+ \frac{1}{16} (20 a_{10}{} - 10 a_{12}{} - 4 a_{4}{} - 3 a_{6}{} - 10 a_{7}{} + 8 \alpha_{2}{} + 20 \alpha_{3}{} + 4 \alpha_{4}{} - 6 \alpha_{5}{} - 5 \alpha_{6}{}) {\ast\Omega}_{\alpha }{}^{\rho \tau } q_{\tau \nu \sigma } t^{\nu }{}_{\rho }{}^{\sigma } \Lambda^{\alpha }  \nonumber\\&&
+ \frac{1}{48} (30 a_{10}{} - 15 a_{12}{} + 12 a_{4}{} - 6 a_{6}{} - 12 \alpha_{2}{} + 12 \alpha_{3}{} + 28 \alpha_{4}{} - 42 \alpha_{5}{} - 21 \alpha_{6}{}) {\ast\Omega}^{\rho }{}_{\alpha }{}^{\tau } q_{\tau \nu \sigma } t^{\nu }{}_{\rho }{}^{\sigma } \Lambda^{\alpha }  \nonumber\\&&
+ \frac{1}{48} (-30 a_{10}{} + 15 a_{12}{} + 24 a_{4}{} + 3 a_{6}{} + 30 a_{7}{} - 36 \alpha_{2}{} - 48 \alpha_{3}{} + 16 \alpha_{4}{} - 24 \alpha_{5}{} - 6 \alpha_{6}{}) {\ast\Omega}^{\rho }{}_{\alpha }{}^{\tau } q_{\rho \nu \sigma } t^{\nu }{}_{\tau }{}^{\sigma } \Lambda^{\alpha }  \nonumber\\&&
+ \frac{1}{48} (-6 a_{10}{} - 15 a_{12}{} + 15 a_{6}{} + 30 a_{7}{} - 36 \alpha_{2}{} + 78 \alpha_{3}{} + 34 \alpha_{4}{} - 42 \alpha_{5}{} - 18 \alpha_{6}{})  {\ast\Omega}^{\rho \tau \nu } q_{\rho \nu \sigma } t^{\sigma }{}_{\alpha \tau } \Lambda^{\alpha } \nonumber\\&&
+ \frac{1}{48} \bigl[-9 a_{6}{} - 2 \bigl(9 a_{7}{} + 2 \alpha_{2}{} - 44 \alpha_{3}{} - 22 \alpha_{4}{} + 6 \alpha_{5}{} + 3 \alpha_{6}{}\bigr)\bigr] {\ast\Omega}^{\rho \tau \nu } q_{\alpha \nu \sigma } t^{\sigma }{}_{\rho \tau } \Lambda^{\alpha } \nonumber\\&&
+ \frac{1}{48} (9 a_{6}{} + 18 a_{7}{} + 4 \alpha_{2}{} - 88 \alpha_{3}{} - 44 \alpha_{4}{} + 12 \alpha_{5}{} + 6 \alpha_{6}{}) {\ast\Omega}^{\rho \tau \nu } q_{\alpha \rho \sigma } t^{\sigma }{}_{\tau \nu } \Lambda^{\alpha } \nonumber\\&&
+ \frac{1}{6} (-4 a_{4}{} + a_{6}{} - 2 
a_{7}{}) {\ast\Omega}^{\alpha \rho \tau } q_{\tau \nu \sigma } \nabla_{\alpha }t^{\nu }{}_{\rho }{}^{\sigma } + \frac{1}{3} (\alpha_{2}{} -  \alpha_{3}{}) (\nabla_{\rho }{\ast\Omega}^{\nu }{}_{\alpha }{}^{\sigma }  + \nabla_{\alpha }{\ast\Omega}^{\nu }{}_{\rho }{}^{\sigma } )q_{\tau \nu \sigma } t^{\alpha \rho \tau } \nonumber\\&&
+ \frac{1}{6} (4 a_{4}{} -  a_{6}{} + 2 a_{7}{} - 4 \alpha_{2}{} - 24 \alpha_{3}{} - 4 \alpha_{4}{} - 4 \alpha_{5}{}) {\ast\Omega}^{\alpha \rho \tau } q_{\tau \nu \sigma } \nabla_{\rho }t^{\nu }{}_{\alpha }{}^{\sigma } \nonumber\\&&
+ \frac{1}{6} (-3 a_{6}{} - 6 a_{7}{} + 2 \alpha_{2}{} - 2 \alpha_{3}{}) q_{\alpha \nu \sigma } t^{\alpha \rho \tau } \nabla_{\tau }{\ast\Omega}^{\nu }{}_{\rho }{}^{\sigma } + \frac{2}{3} (\alpha_{2}{} -  \alpha_{3}{}) ( t^{\nu }{}_{\rho }{}^{\sigma } \nabla_{\tau }q_{\alpha \nu \sigma }+  t^{\nu }{}_{\alpha }{}^{\sigma } \nabla_{\tau }q_{\rho \nu \sigma })  {\ast\Omega}^{\alpha \rho \tau } \nonumber\\&&
+ \frac{1}{3} \bigl[-4 a_{4}{} + a_{6}{} + 2 \bigl(- a_{7}{} + \alpha_{2}{} + 6 \alpha_{3}{} + \alpha_{4}{} + \alpha_{5}{}\bigr)\bigr] {\ast\Omega}^{\alpha \rho \tau } q_{\alpha \nu \sigma } \nabla_{\tau }t^{\nu }{}_{\rho }{}^{\sigma }  \nonumber\\&&
+ \frac{1}{6} (2 a_{10}{} -  a_{12}{} - 4 \alpha_{4}{}) q_{\alpha \tau \sigma } t^{\alpha \rho \tau } \nabla_{\nu }{\ast\Omega}_{\rho }{}^{\nu \sigma } + \frac{1}{6} (4 a_{10}{} - 2 a_{12}{} - 2 \alpha_{5}{} -  \alpha_{6}{}) q_{\alpha \tau \sigma } t^{\alpha \rho \tau } \nabla_{\nu }{\ast\Omega}^{\nu }{}_{\rho }{}^{\sigma } \nonumber\\&&
+ \frac{1}{3} (- \alpha_{2}{} + \alpha_{3}{}) {\ast\Omega}^{\alpha \rho \tau } t^{\nu }{}_{\rho }{}^{\sigma } \nabla_{\nu }q_{\alpha \tau \sigma } + \frac{1}{6} (-4 \alpha_{4}{} -  \alpha_{5}{}) {\ast\Omega}^{\alpha \rho \tau } q_{\alpha \tau \sigma } \nabla_{\nu }t_{\rho }{}^{\nu \sigma }  \nonumber\\&&
+ \frac{1}{2} (2 a_{10}{} + a_{12}{} + \alpha_{5}{}) {\ast\Omega}^{\alpha \rho \tau } q_{\alpha \tau \sigma } \nabla_{\nu }t^{\nu }{}_{\rho }{}^{\sigma } + \frac{1}{6} (2 a_{10}{} -  a_{12}{} + 4 \alpha_{4}{} - 2 \alpha_{5}{} -  \alpha_{6}{}) q_{\alpha \tau \nu } t^{\alpha \rho \tau } \nabla_{\sigma }{\ast\Omega}^{\nu }{}_{\rho }{}^{\sigma } \nonumber\\&&
+ \frac{1}{3} (\alpha_{2}{} -  \alpha_{3}{}) {\ast\Omega}^{\alpha \rho \tau } t_{\rho }{}^{\nu \sigma } \nabla_{\sigma }q_{\alpha \tau \nu } + \frac{1}{3} (- \alpha_{2}{} + \alpha_{3}{}) {\ast\Omega}^{\alpha \rho \tau } t^{\nu }{}_{\rho }{}^{\sigma } \nabla_{\sigma }q_{\alpha \tau \nu }  \nonumber\\&&
+ \frac{1}{12} (-8 \alpha_{4}{} + 2 \alpha_{5}{} + \alpha_{6}{}) {\ast\Omega}^{\alpha \rho \tau } t^{\nu }{}_{\rho \tau } \nabla_{\sigma }q_{\alpha \nu }{}^{\sigma } + \frac{1}{12} (-2 \alpha_{5}{} -  \alpha_{6}{}) {\ast\Omega}^{\alpha \rho \tau } t_{\alpha \rho }{}^{\nu } \nabla_{\sigma }q_{\tau \nu }{}^{\sigma }
 \nonumber\\&&
 + \frac{1}{12} \bigl[8 \alpha_{4}{} - 3 \bigl(2 \alpha_{5}{} + \alpha_{6}{}\bigr)\bigr] {\ast\Omega}^{\alpha \rho \tau } t_{\rho \alpha }{}^{\nu } \nabla_{\sigma }q_{\tau \nu }{}^{\sigma } + \frac{1}{12} (8 \alpha_{4}{} - 2 \alpha_{5}{} -  \alpha_{6}{}) {\ast\Omega}^{\alpha \rho \tau } t^{\nu }{}_{\alpha \rho } \nabla_{\sigma }q_{\tau \nu }{}^{\sigma }  \nonumber\\&&
 + \frac{1}{6} (4 \alpha_{4}{} + \alpha_{5}{}) 
{\ast\Omega}^{\alpha \rho \tau } q_{\alpha \tau \nu } \nabla_{\sigma }t^{\nu }{}_{\rho }{}^{\sigma } + \frac{1}{3} (\alpha_{2}{} -  \alpha_{3}{}) ( \nabla^{\sigma }{\ast\Omega}_{\alpha \rho }{}^{\nu } +  \nabla^{\sigma }{\ast\Omega}_{\rho \alpha }{}^{\nu }) q_{\tau \nu \sigma } t^{\alpha \rho \tau } \nonumber\\&&
+ \frac{1}{6} \bigl[- a_{6}{} - 2 \bigl(a_{7}{} -  \alpha_{2}{} + \alpha_{3}{}\bigr)\bigr] ( \nabla^{\sigma }{\ast\Omega}_{\rho \tau }{}^{\nu } - \nabla^{\sigma }{\ast\Omega}^{\nu }{}_{\rho \tau }) q_{\alpha \nu \sigma } t^{\alpha \rho \tau }\nonumber\\&&
+ \frac{1}{3} (\alpha_{2}{} + 6 \alpha_{3}{} + \alpha_{4}{} + \alpha_{5}{})(\nabla^{\sigma }t_{\alpha \rho }{}^{\nu } + \nabla^{\sigma }t_{\rho \alpha }{}^\nu){\ast\Omega}^{\alpha \rho \tau } q_{\tau \nu \sigma }+ \frac{1}{6} (4 \alpha_{4}{} - 2 \alpha_{5}{} -  \alpha_{6}{}) {\ast\Omega}^{\alpha \rho \tau } t_{\rho \tau }{}^{\nu } \nabla_{\sigma }q_{\alpha \nu }{}^{\sigma }   \nonumber\\&&
+ \frac{1}{6} \bigl[-4 a_{4}{} + a_{6}{} + 2 \bigl(- a_{7}{} + \alpha_{2}{} + 6 \alpha_{3}{} + \alpha_{4}{} + \alpha_{5}{}\bigr)\bigr](  q_{\tau \nu \sigma } \nabla^{\sigma }t^{\nu }{}_{\alpha \rho } - q_{\alpha \nu \sigma } \nabla^{\sigma }t^{\nu }{}_{\rho \tau }){\ast\Omega}^{\alpha \rho \tau } \,,
\end{eqnarray}}\normalsize
having defined ${\ast\Omega}_{\lambda\mu\nu}=\frac{1}{2}\varepsilon^{\rho\sigma}{}_{\mu\nu}\Omega_{\lambda\rho\sigma} $ and the following combinations:
\small{\begin{eqnarray}
    \beta_1&=&(a_{10}{} + a_{11}{} -  a_{12}{} -  a_{13}{} - 2 a_{3}{} + 3 a_{4}{} -  a_{6}{} + 2 a_{7}{})\,,\\
 \beta_2&=&   (2 a_{10}{} + 2 a_{11}{} - 3 a_{12}{} - 3 a_{13}{} + 8 a_{2}{} - 4 a_{3}{} + 2 a_{4}{} + 4 a_{5}{} - 2 a_{6}{} + 4 a_{8}{} + 4 a_{9}{}) \,,\\
 \beta_3&=&2\beta_1-\beta_2\,,\\
 \beta_4&=&(4 a_{2}{} + 2 a_{3}{} + 2 a_{5}{} + a_{6}{} + a_{7}{}) \,,\\
\beta_{5}&=&(4 a_{2}{} - 2 a_{3}{} - 3 a_{4}{} + 2 a_{5}{} -  a_{6}{} - 4 a_{7}{})\,,\\
 \beta_6&=& (2 a_{2}{} - 8 a_{3}{} - 3 a_{4}{} + a_{5}{} - 4 a_{6}{} - 7 a_{7}{}) \,,\\
 \beta_7&=&(2 a_{2}{} + 4 a_{3}{} + a_{4}{} + a_{5}{} + 2 a_{6}{} + 3 a_{7}{}) \,,\\
 \beta_8&=&(4 a_{2}{} 
+ a_{4}{} + 2 a_{5}{} + a_{7}{})\,,\\
 \beta_9&=&(2 a_{2}{} + 2 a_{3}{} + a_{5}{} + a_{6}{} + a_{7}{}) \,,\\
 \beta_{10}&=&\bigl(a_{12}{} + a_{13}{} - 2 (a_{8}{} + a_{9}{})\bigr) \,,\\
 \beta_{11}&=&(2 a_{10}{} + 2 a_{11}{} -  a_{12}{} -  a_{13}{})\,,\\
 \beta_{12}&=&(a_{10}{} + a_{11}{} -  a_{12}{} -  a_{13}{} + a_{8}{} + a_{9}{}) \,,\\
 \beta_{13}&=&(2 a_{2}{} + 2 a_{3}{} - 3 a_{4}{} + a_{5}{} + a_{6}{} - 2 a_{7}{})\,,\\
\beta_{14}&=& (2 a_{2}{} - 4 a_{3}{} + a_{5}{} - 2 a_{6}{} - 2 a_{7}{})\,,\\
\beta_{15}&=&a_{10}{} + 2 (-a_{11}{} + a_{8}{} + a_{9}{})\,,\\
\beta_{16}&=&(2 a_{10}{} + 2 a_{11}{} + a_{12}{} + a_{13}{} - 4 a_{8}{} - 4 a_{9}{})\,,\\
\beta_{17}&=& (-2 a_{2}{} + a_{4}{} -  a_{5}{} + a_{7}{})\,,\\
\beta_{18}&=&-2 a_{10}{} + a_{11}{} + 2 (a_{8}{} + a_{9}{})\,,\\
\beta_{19}&=&(2 a_{2}{} + 2 a_{4}{} + a_{5}{} + 2 a_{7}{})\,,\\
\beta_{20}&=&- a_{5}{} -  a_{6}{} -  a_{7}{} + 2 \alpha_{2}{} - 2 \alpha_{3}{} + 2 
\alpha_{4}{} + 2 \alpha_{5}{}\,,\\
\beta_{21}&=&4 a_{10}{} - 4 a_{12}{} - 4 a_{9}{} - 16 \alpha_{2}{} - 96 
\alpha_{3}{} - 24 \alpha_{4}{} - 16 \alpha_{5}{} + \alpha_{6}{}\,,\\
\beta_{22}&=&-2 a_{2}{} - 2 a_{3}{} -  a_{5}{} -  a_{6}{} -  a_{7}{}\,,\\
\beta_{23}&=&-2 a_{3}{} -  a_{6}{} -  a_{7}{}\,,\\
\beta_{24}&=&-2 a_{10}{} - 2 a_{11}{} - 3 a_{12}{} - 3 a_{13}{} - 4 a_{8}{} - 4 
a_{9}{}\,,\\
\beta_{25}&=&2 a_{10}{} + 2 a_{11}{} + 6 a_{3}{} - 3 a_{4}{} + 3 a_{6}{} - 2 
a_{8}{} - 2 a_{9}{}\,,\\
\beta_{26}&=&6 (2 a_{3}{} -  a_{4}{} + a_{6}{})\,,\\
\beta_{27}&=&12 (4 a_{10}{} - 5 a_{11}{} + a_{12}{} + a_{13}{} + 8 a_{2}{} + 10 
a_{3}{} -  a_{4}{} + 4 a_{5}{} + 5 a_{6}{} + 4 a_{7}{} - 2 a_{8}{} - 
2 a_{9}{})\,,\\
\beta_{28}&=&12 (5 a_{10}{} - 4 a_{11}{} -  a_{12}{} -  a_{13}{} - 8 a_{2}{} + 8 
a_{3}{} - 8 a_{4}{} - 4 a_{5}{} + 4 a_{6}{} - 4 a_{7}{} + 2 a_{8}{} + 
2 a_{9}{})\,,\\
\beta_{29}&=&96 \beta_{22}{} -  \beta_{24}{} - 4 \beta_{25}{} + 16 \beta_{26}{} + 
\beta_{27}{} -  \beta_{28}{}\,,\quad \beta_{30}=-3 \beta_{22}{} + 6 \beta_{23}{} + \beta_{24}{} + \beta_{25}{}\,.
\end{eqnarray}}\normalsize

Note that the other tensor modes $\Omega_{\lambda\mu\nu}$ and $t_{\lambda\mu\nu}$ of torsion and nonmetricity are vanishing at the background level, i.e. in FLRW cosmology, which means that all the terms appearing in the second line of Expression~\eqref{Lagterm} do not contribute to the spin-3 field perturbations around the FLRW background up to second order in the gravitational action.

\bibliographystyle{utphys}
\bibliography{references}

\end{document}